\documentclass[lettersize,journal]{IEEEtran}
\usepackage{amsmath,amsfonts}
\usepackage{algorithmic}
\usepackage{algorithm}
\usepackage{array}
\usepackage[caption=false,font=normalsize,labelfont=sf,textfont=sf]{subfig}
\usepackage[colorlinks,linkcolor=black,anchorcolor=black,citecolor=black,urlcolor=black]{hyperref}
\usepackage{textcomp}
\usepackage{stfloats}
\usepackage{url}
\usepackage{verbatim}
\usepackage{graphicx}
\usepackage{cite}
\usepackage{amssymb}
\usepackage{graphicx}
\usepackage{booktabs}
\usepackage{multirow}
\begin{document}
\title{Generalizable CT-Free PET Attenuation and Scatter Correction for Pediatric Patients}
\author{Jia-Mian Wu\textsuperscript{\#}, Jun Liu\textsuperscript{\#}, Siqi Li, Xiaoya Wang, Shibai Yin, Huanyu Luo, Lingling Zheng, Qiang Gao, Jigang 
Yang\textsuperscript{\dag}, and Tai-Xiang Jiang\textsuperscript{\dag}
\thanks{\# Equal contribution. \dag Corresponding authors.}
\thanks{Jia-Mian Wu, Shibai Yin, Qiang Gao, and Tai-Xiang Jiang are with the School of Computing and Artificial Intelligence, Southwestern University of Finance and Economics, Chengdu 611130, China. e-mail: taixiangjiang@gmail.com}
\thanks{Jun Liu, Siqi Li, Xiaoya Wang, Lingling Zheng, and Jigang Yang are with the Department of Nuclear Medicine, Beijing Friendship Hospital, Capital Medical University, Beijing 100050, China. e-mail: yangjigang@ccmu.edu.cn}
\thanks{Huanyu Luo is with the Department of Radiology, Beijing Children's Hospital, Capital Medical University, National Center for Children's Health, Beijing 100045, China.}
}
\maketitle
\begin{abstract}
Computed tomography (CT)-based attenuation and scatter correction improves quantitative PET but adds radiation exposure that is particularly undesirable in pediatric imaging.
Existing CT-free methods are commonly trained in homogeneous settings and often degrade under scanner or radiotracer shifts, which limits their clinical utility.
We propose the Generalizable PET Correction Network (GPCN), a dual-domain network for domain-robust CT-free PET attenuation and scatter correction.
GPCN combines a multi-band contextual refinement module, which models pediatric anatomical variability through wavelet-based multiscale decomposition and long-range spatial context modeling, with a frequency-aware spectral decoupling module, which performs coordinate-conditioned amplitude/phase refinement in the Fourier domain. By synergizing multi-band spatial contextual modeling with asymmetric frequency-spectrum decoupling, the network explicitly separates invariant topological structures from domain-specific noise, thereby achieving precise quantitative recovery of both anatomical organs and focal lesions.
This design aims to separate anatomy-dominant structures from domain-sensitive spectral residuals and to improve robustness across heterogeneous imaging conditions. We train and evaluate the method on 1085 pediatric whole-body PET scans acquired with two scanners and five radiotracers. In both joint training and zero-shot cross-domain evaluation, GPCN outperforms representative baselines and maintains stable quantitative accuracy on unseen scanner-tracer combinations. The method is further supported by ablation, region-wise quantitative analysis, and downstream segmentation experiments. In our cohort, the CT component of the conventional protocol corresponded to an average effective dose of 10.8 mSv, indicating the potential clinical value of reliable CT-free correction for pediatric PET. The source code are available at \url{https://github.com/wujm7/GPCN-PET}.
\end{abstract}


\section{Introduction}
\label{sec:introduction}
The `as low as reasonably achievable' (ALARA) principle is critical in pediatric PET/CT, a cornerstone of care for oncology and inflammatory diseases, because children's heightened radiosensitivity and longer lifespan increase their lifetime risk of radiation-induced malignancies \cite{fahey2012minimizing, brenner2007computed, pierce2000radiation}. In PET/CT, the CT scan provides an attenuation map ($\mu$-map) essential for quantitative reconstruction (e.g., OSEM, TOF) \cite{rezaei2012simultaneous, 876305, 7429780}. Despite low-dose CT protocols \cite{prieto2021ultra, xia2011ultra}, the cumulative radiation from repeated scans remains a major concern, making the elimination of the CT component a key goal in pediatric imaging. However, existing CT-free approaches have been developed almost exclusively on adult data, leaving a critical gap for the pediatric population.

\begin{figure*}[t]
\centering
\includegraphics[width=0.95\textwidth]{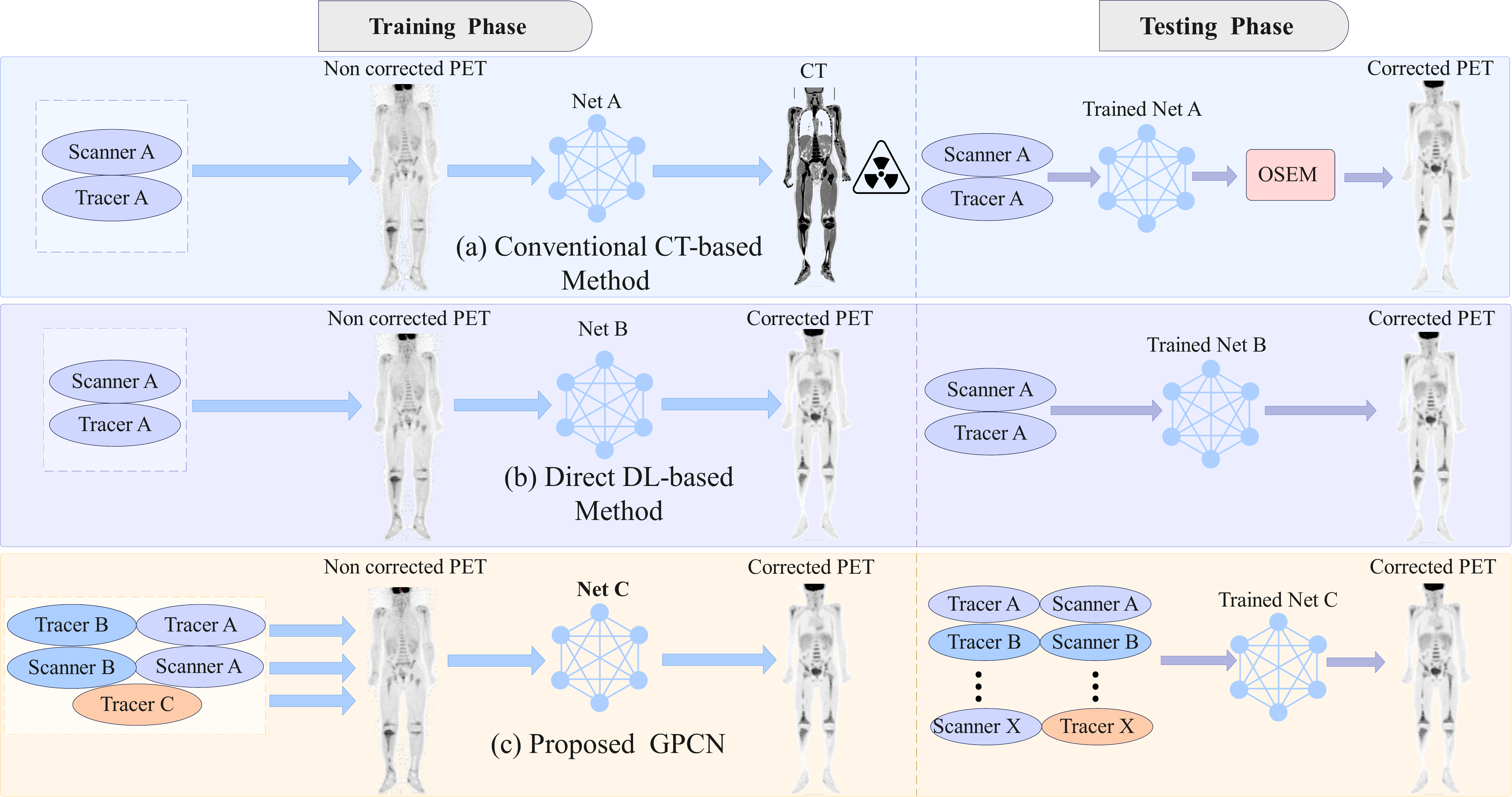}
\caption{Conceptual overview of PET correction paradigms.
Comparison of (a) conventional CT-based correction, (b) standard direct DL approach with limited generalization, and (c) the proposed GPCN trained on heterogeneous data for robust cross-scanner and cross-tracer performance.
}
\label{fig:1}

\end{figure*}
The pursuit of CT-free PET has progressed from early statistical methods, limited by high noise and crosstalk artifacts \cite{rezaei2012simultaneous,10315027}, to powerful deep learning (DL) approaches that frame attenuation correction as a data-driven image-to-image translation task \cite{shiri2020deep,li2022eliminating,wen2024high}. DL-based attenuation correction can be broadly grouped into two strategies \cite{chen2023deep}. Indirect methods first synthesize a pseudo-CT from non-attenuation and non-scatter corrected (NASC) PET, which is then used for a separate reconstruction, but the reliance on an intermediate anatomical map makes them vulnerable to error propagation \cite{liu2018deep,hwang2022comparison,8804223,spuhler2019synthesis}. In contrast, end-to-end strategies directly map NASC to attenuation- and scatter-corrected (ASC) PET, typically using U-Net or GAN architectures, thereby avoiding cumulative errors of multi-stage pipelines while implicitly learning complex physical effects that are difficult to model explicitly \cite{shiri2019direct,isola2017image,ronneberger2015u}.

As shown in Fig. ~\ref{fig:1}, a central difficulty in CT-free pediatric PET correction is domain heterogeneity rather than image translation alone. In practice, scanner differences alter noise texture and effective resolution, radiotracers change uptake contrast and lesion conspicuity, and the pediatric population introduces pronounced anatomical scale variation across age groups. Models \cite{van2019deep, laurent2025evaluation} optimized on homogeneous cohorts can therefore achieve low average reconstruction error by relying mainly on shared low-frequency anatomy while under-modeling domain-dependent local residuals \cite{ayyoubzadeh2021high}. This mismatch becomes clinically relevant \cite{finlayson2021clinician} when the same model is deployed across scanners, tracers, or age ranges not represented during training. 
We therefore formulate CT-free pediatric PET correction as a domain-robust restoration problem under coupled scanner, tracer, and anatomical shifts.

To address this problem, we propose the Generalizable PET Correction Network (GPCN), a dual-domain architecture for domain-robust PET correction. 
GPCN is built on the observation that heterogeneity in pediatric PET manifests across both spatial scale and spectral representation. The image-domain branch,
i.e., the multi-band contextual refinement (MBCR) module, performs wavelet-based multiscale decomposition and long-range context modeling to stabilize anatomical structure across large body-size variations. The frequency-domain branch, i.e., the frequency-aware spectral decoupling (FASD) module, performs coordinate-conditioned refinement of amplitude and phase to model domain-sensitive residuals related to contrast, noise texture, and boundary sharpness. By coupling these two branches in an iterative residual framework, GPCN learns to correct NASC PET with improved robustness to scanner- and tracer-induced distribution shifts.

The main contributions of this work are threefold.
First, we formulate CT-free pediatric PET correction as a domain-robust restoration problem and evaluate it under both joint multi-domain training and strict zero-shot cross-domain testing. Second, we propose a dual-domain architecture that combines wavelet-guided multiscale context modeling with frequency-coordinate-conditioned amplitude/phase refinement. Third, we validate the method on a heterogeneous cohort of 1085 pediatric scans from two scanners and five radiotracers, using image quality metrics, quantitative PET measures, physical-behavior analyses, and a downstream segmentation task.
\section{Proposed Method}
\begin{figure*}[!t]
\centering
\includegraphics[width=0.99\textwidth]{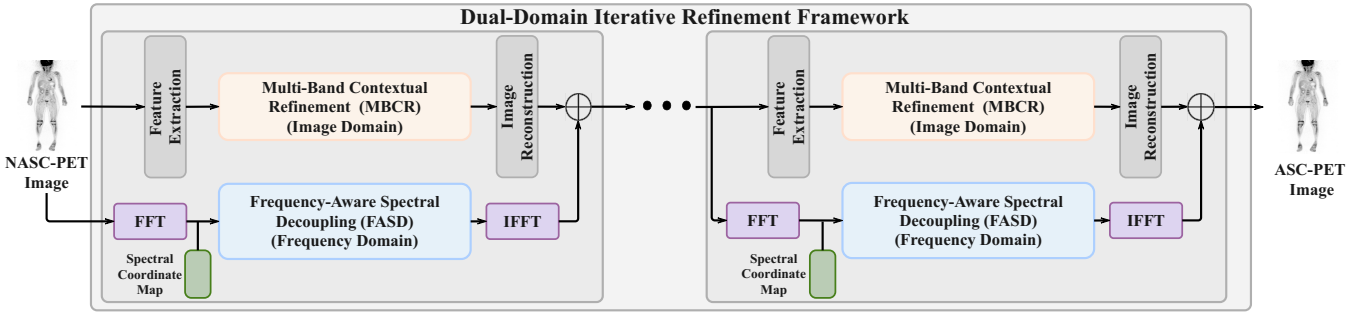}
\caption{Overview of the proposed generalizable PET correction network (GPCN). MBCR is the multi-band contextual refinement. FASD is the frequency-aware spectral decoupling. A spectral coordinate map provides coordinate information to guide the FASD. Here, $\boldsymbol{\bigoplus}$ denotes the element-wise addition.}
\label{fig:network}
\end{figure*}
 \subsection{Network Architecture}
 
As illustrated in Fig. \ref{fig:network}, the proposed generalizable PET correction network (GPCN) estimates ASC PET from NASC PET through three cascaded residual refinement blocks. Each block contains two complementary operators. The MBCR branch operates in the image domain to model multiscale anatomical context, whereas the FASD branch operates in the Fourier domain to correct domain-sensitive spectral residuals. Their outputs are fused to produce an incremental correction at each stage. This design separates two major sources of difficulty in pediatric PET correction: large structural variation across subjects and scanner-/tracer-dependent changes in local texture and contrast.

\paragraph{Multi-Band Contextual Refinement}
PET images contain coupled structural information at multiple spatial scales. In pediatric whole-body scans, coarse-scale anatomy varies markedly with age and body size, whereas fine-scale components contain lesion boundaries, tracer-dependent texture, and noise. The multi-band contextual refinement (MBCR)  addresses this heterogeneity by first decomposing features into wavelet subbands and then applying shared residual VMamba block (RVMB) \cite{liu2024vmamba} operators within each subband. This allows the network to model long-range context in low-frequency components while refining detail-dominant high-frequency components with reduced interference across scales \cite{sun2025spatial,jiang2025ph}.
\begin{figure}[!t]
\centering
\includegraphics[width=0.49\textwidth]{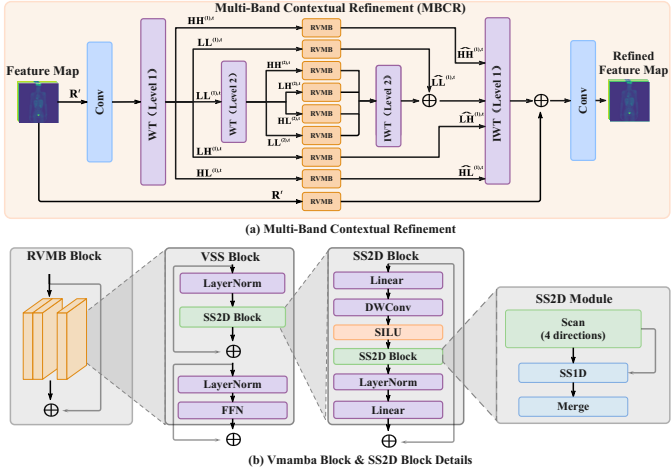}
\caption{Structure of the proposed MBCR. (a) Overview of the multi-band contextual refinement (MBCR). (b) The architecture of VMamba. (c) The details of the VSS Block. It is composed of an SS2D block and a feed-forward network (FFN), each preceded by LayerNorm. (d) The details of the SS2D Block. It consists of a Linear layer, DWConv, SiLU activation, an SS2D module, LayerNorm (LN), and a final Linear layer.}
\label{fig:WMM}
\end{figure}

The architecture of the MBCR is illustrated in Fig.~\ref{fig:WMM}. It employs RVMBs \cite{liu2024vmamba} to process each decoupled sub-band. By leveraging the 2D Selective Scan (SS2D) mechanism \cite{gu2023mamba}, VMamba efficiently models long-range dependencies with linear complexity, capturing global spatial continuity within each sub-band more effectively than standard convolutions.
The MBCR is a multi-scale architecture that uses a wavelet transform (WT) to separate and accurately reconstruct multi-scale features. This WT-based decoupling directly benefits the VMamba blocks. It allows them to focus on the specific characteristics of each sub-band, efficiently modeling long-range dependencies in the dense, low-frequency component for global context, while refining the sparse, high-frequency details for local texture and noise preservation.

As shown in Fig.~\ref{fig:WMM}, the input feature map $\mathbf{X}^t$ of the $t$-th MBCR block is initially processed through a 2D convolutional layer to produce the feature map $\mathbf{R}^t$. This map is then decomposed via a WT into four frequency components ($\mathbf{LH}^{(1),t}, \mathbf{HL}^{(1),t}, \mathbf{HH}^{(1),t}, \mathbf{LL}^{(1),t}$). Notably, the low-frequency component ($\mathbf{LL}^{(1),t}$) is recursively processed through a second WT, further splitting it into four sub-band components ($\mathbf{LH}^{(2),t}, \mathbf{HL}^{(2),t}, \mathbf{HH}^{(2),t}, \mathbf{LL}^{(2),t}$). Each of these eight sub-bands is then independently refined by a shared VMamba block:
\begin{equation}
\begin{aligned}
\small
&\{\mathbf{LL}^{(1),t},\mathbf{LH}^{(1),t},\mathbf{HL}^{(1),t},\mathbf{HH}^{(1),t}\} = \text{WT}(\mathbf{R}^t),\\
\{&\mathbf{LL}^{(2),t},\mathbf{LH}^{(2),t},\mathbf{HL}^{(2),t},\mathbf{HH}^{(2),t}\} = \text{WT}(\mathbf{LL}^{(1),t}),\\
&\widehat{\mathbf{LL}}^{(i),t} = \text{RVMB}(\mathbf{LL}^{(i),t}),\widehat{\mathbf{LH}}^{(i),t} = \text{RVMB}(\mathbf{LH}^{(i),t}),\\
&\widehat{\mathbf{HL}}^{(i),t} = \text{RVMB}(\mathbf{HL}^{(i),t}),\widehat{\mathbf{HH}}^{(i),t} = \text{RVMB}(\mathbf{HH}^{(i),t}),
\end{aligned}
\end{equation}
where $i\ (i=1,2)$ denotes the WT decomposition level. This design enables an effective hierarchical decomposition of the image, allowing for specialized processing of each component through the VMamba blocks.

Following the feature enhancement across all paths, the refined sub-bands are progressively merged. The level-2 components are first combined through an inverse wavelet transform (IWT), and the result is fused with the processed level-1 low-frequency component via element-wise addition. This combined map is then merged with the remaining level-1 components through a final IWT. The output is added back to the initial feature map $\mathbf{R}^{t}$ in a residual connection to produce the final refined output $\widehat{\mathbf{F}}^t$:
\begin{equation}
\begin{aligned}
\widehat{\widehat{\mathbf{LL}}}^{(1),t}&\!=\!\text{IWT}(\!\widehat{\mathbf{LL}}^{(2),t}\!,\!\widehat{\mathbf{LH}}^{(2),t}\!,\!\widehat{\mathbf{HL}}^{(2),t}\!,\!\widehat{\mathbf{HH}}^{(2),t}\!)\!+\!\widehat{\mathbf{LL}}^{(1),t},\\
\widehat{\mathbf{F}}^t&\!=\!\text{IWT}(\widehat{\widehat{\mathbf{LL}}}^{(1),t}, \widehat{\mathbf{LH}}^{(1),t},\widehat{\mathbf{HL}}^{(1),t},\widehat{\mathbf{HH}}^{(1),t})+\mathbf{R}^{t}.
\end{aligned}
\end{equation}
\begin{figure}[!t]
  \centering
  \includegraphics[width=0.49\textwidth]{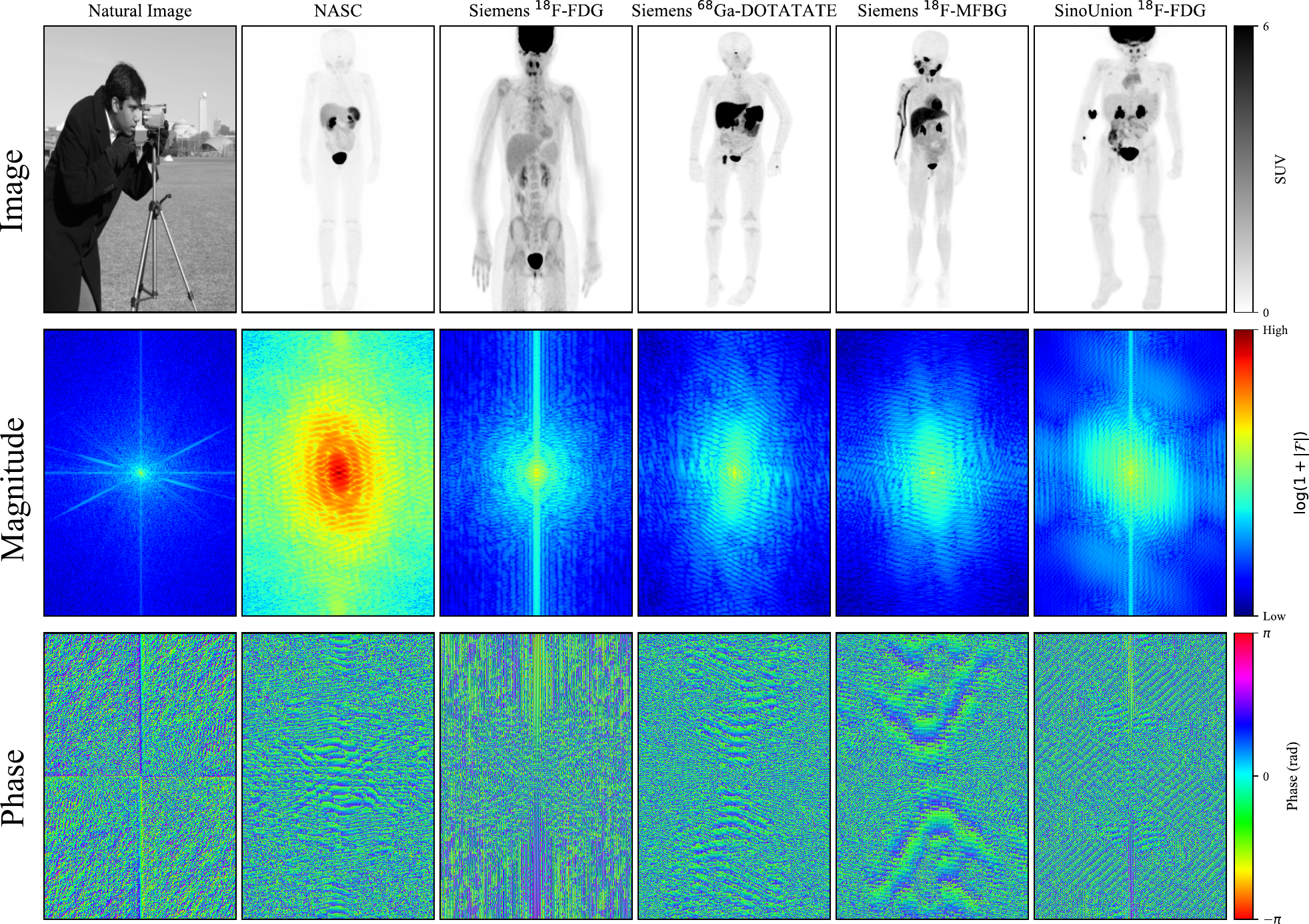}
  \caption{Frequency-domain characteristics of PET images across scanners and tracers. Rows show the spatial-domain MIP image, log-magnitude spectrum
           $\log(1+|\mathcal{F}|)$, and phase map $\angle\mathcal{F}$.
           Left group: natural image and NASC image.
           Right group: ASC MIPs for four representative scanner/tracer
           combinations, illustrating diverse frequency-domain signatures.
           }
  \label{fig:freq_combined}
\end{figure}

\paragraph{Frequency-Aware Spectral Decoupling}
We introduce the frequency-aware spectral decoupling (FASD) module to model correction residuals that are difficult to express purely in the image domain.
As shown in Fig.~\ref{fig:freq_combined}, both natural and PET images are low-frequency dominated, but the PET examples exhibit a more compact central spectrum and scanner-/tracer-dependent directional residual patterns. 
These observations suggest that part of the NASC-to-ASC correction problem lies in domain-dependent spectral residuals, which motivates an explicit frequency-domain refinement branch. Within this branch, we separately refine amplitude and phase so that contrast-related correction and structure-related correction can be handled with different parameterizations.
\begin{figure}[!t]
\centering
\includegraphics[width=0.49\textwidth,height=0.15\textwidth]{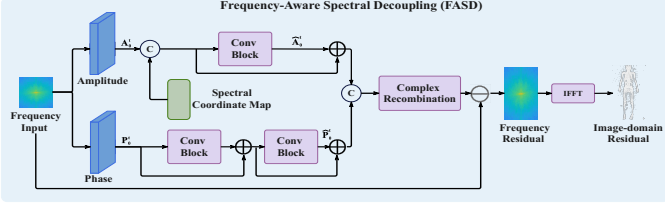}
\caption{frequency-aware spectral decoupling (FASD). (a) The dual-branch architecture of the FASD. The module operates in the frequency domain, decoupling the signal into amplitude (A) and phase (P) for asymmetric processing.
(b) Details of the Conv block. $\oplus$: residual connection, $\ominus$: subtraction operation, $\textcircled{c}$: channel-wise concatenation.}
\label{fig:PACM}
\end{figure}

The architecture of the FASD is shown in Fig.~\ref{fig:PACM}. 
Its design explicitly acknowledges the dual nature of spectral learning. 
First, to enable frequency-adaptive processing, we introduce a Spectral Coordinate Map (SCM). 
Specifically, this map is constructed by concatenating normalized horizontal and vertical coordinate grids with a radial distance map relative to the DC center. 
This composition is crucial: the radial component acts as a direct proxy for frequency magnitude, explicitly informing the network whether it is processing the anatomical backbone (low radial distance) or fine details (high radial distance), while the Cartesian grids capture directional spectral properties. 
By conditioning the spectral features on the SCM, we break the translational invariance, allowing the network to learn position-dependent kernels that adaptively regularize the spectrum. 
Second, to respect the distinct spectral roles of the signal, we employ an asymmetric decoupling strategy. 
We separate the processing of amplitude and phase, dedicating greater network capacity to the phase branch. 
This ensures that the high-frequency structural details encoded in the phase are meticulously recovered, while the amplitude branch focuses on restoring quantitative contrast.

We denote the frequency-domain input of the $t$-th FASD block as $\mathbf{K}^t \in \mathbb{C}^{H \times W}$, which is obtained by applying a fast Fourier transform (FFT) to the PET images.  We define $\mathcal{R}(\cdot)$ and $\mathcal {I}(\cdot)$ as the operators that extract the real and imaginary parts of the complex data, respectively. The amplitude and phase of $\mathbf{K}^t$ are then computed using the following equation:
\begin{equation}
\begin{aligned}
\small
\mathcal{A}(\mathbf{K}^t(i,j)) \!&=\!\sqrt{\mathcal{R}(\mathbf{K}^t(i,j))^2\!+\!\mathcal{I}(\mathbf{K}^t(i,j))^2},\!\quad\!\\
\mathcal{P}(\mathbf{K}^t(i,j))\!&=\!\operatorname{arctan}\left(\frac{\mathcal{I}(\mathbf{K}^t(i,j))}{\mathcal{R}(\mathbf{K}^t(i,j))}\right),
\end{aligned}
\end{equation}
where $\mathcal{A}(\cdot)$ and $\mathcal{P}(\cdot)$ represent the operators for calculating the amplitude and phase, respectively.

Next, the amplitude and phase are passed through separate subnetworks composed of Conv Block, as detailed in Fig.~\ref{fig:PACM}. We denote the initial amplitude of $\mathbf{K}^t$ as $\mathbf{A}^t_0$ and the phase as $\mathbf{P}^t_0$. To prioritize the preservation of detailed information encoded in the phase, we double the depth of the phase subnetwork compared to the amplitude subnetwork. To compensate for the translation invariance of frequency operations, we explicitly introduce a spectral coordinate map $\mathbf{M} \in \mathbb{R}^{H \times W \times 3}$, constructed by concatenating the normalized coordinate vector $\mathbf{c}_{uv} = (\frac{2u}{H}-1, \frac{2v}{W}-1)$ and its radial distance $\|\mathbf{c}_{uv}\|_2$. The computation process is as follows:
\begin{equation}
\begin{aligned}
\small
\mathbf{M}_{uv} &= \text{Concat}(\mathbf{c}_{uv}, \|\mathbf{c}_{uv}\|_2), \\
\widehat{\mathbf{A}}^t &= \text{ConvBlock}(\text{Concat}(\mathbf{A}^t_0,\mathbf{M})), \\
\mathbf{P}^t_1 &= \text{ConvBlock}(\mathbf{P}^t_0),\quad 
\widehat{\mathbf{P}}^t\!= \text{ConvBlock}(\mathbf{P}^t_1).
\end{aligned}
\end{equation}
Here, $\widehat{\mathbf{A}}^t$ and $\widehat{\mathbf{P}}^t$ are the refined amplitude and phase components, respectively.

Finally, the refined amplitude $\widehat{\mathbf{A}}^t$ and phase $\widehat{\mathbf{P}}^t$ are recombined to form the corrected spectrum $\widehat{\mathbf{K}}^t$. We then define the frequency residual as
\begin{equation}
\Delta \mathbf{K}^t =\widehat{\mathbf{K}}^t-\mathbf{K}^t
\end{equation}
and obtain the image-domain residual by
\begin{equation}
 \Delta \mathbf{I}^t = \text{IFFT}(\Delta \mathbf{K}^t).
\end{equation}
The block output is produced by adding $\Delta \mathbf{I}^t$ to the current image-domain estimate.
This residual formulation makes the role of FASD explicit: the branch predicts only the spectral correction needed to transform NASC PET toward ASC PET.

\subsubsection{Loss Function}
The optimization of GPCN is guided by a hybrid loss function that enforces fidelity in both the image and frequency domains. This dual-domain supervision is crucial for learning the complex domain-invariant mapping required for PET correction. The total loss, $\mathcal{L}_{\text{total}}$, is a weighted sum of an image-domain loss, $\mathcal{L}_{\text{img}}$, and a frequency-domain loss, $\mathcal{L}_{\text{freq}}$.

The individual loss components are defined as:
\begin{equation}
\label{eq:loss_components}
\begin{aligned}
\mathcal{L}_{\text{img}} &= \|\mathbf{X}_{\text{GPCN}} - \mathbf{X}_{\text{ASC}}\|_1 \\
\mathcal{L}_{\text{freq}} &= \|\mathcal{F}(\mathbf{X}_{\text{GPCN}}) - \mathcal{F}(\mathbf{X}_{\text{ASC}})\|_1
\end{aligned}
\end{equation}
where $\mathbf{X}_{\text{GPCN}}$ is the final corrected image generated by our network, $\mathbf{X}_{\text{ASC}}$ is the ground-truth CT-based attenuation-corrected PET image, and $\mathcal{F}$ denotes the fast Fourier transform.

The total loss is then formulated as:
\begin{equation}
\label{eq:total_loss}
\mathcal{L}_{\text{total}} = \mathcal{L}_{\text{img}} + \lambda \mathcal{L}_{\text{freq}}
\end{equation}
where the hyperparameter $\lambda$ balances the contribution of the two terms. In our experiments, we set $\lambda=1.0$ based on empirical validation.
We deliberately chose the L1 norm for both loss components due to its demonstrated superiority in preserving high-frequency details, which is of paramount importance in pediatric PET imaging. 
\subsection{Patient Data}
\begin{table*}[!t]
\small
\renewcommand\arraystretch{1}
\setlength{\tabcolsep}{1.1pt}
\centering
\caption{Information on our datasets' demographics.}
\label{tab:my_dataset_info}
\begin{tabular}{lccccccc}
\toprule
& \multicolumn{5}{c}{\textbf{Siemens Healthineers}} & \multicolumn{2}{c}{\textbf{SinoUnion Healthcare}} \\
\cmidrule(lr){2-6} \cmidrule(lr){7-8}
\textbf{Attribute} & \textbf{$^{18}$F-FDG} & \textbf{$^{68}$Ga-DOTATATE} & \textbf{$^{18}$F-MFBG} & \textbf{$^{18}$F-FAPI} & \textbf{$^{18}$F-DOPA} & \textbf{$^{18}$F-FDG} & \textbf{$^{18}$F-MFBG} \\
\midrule
    Number of patients                       & 622 & 86 & 205 & 3 & 2 & 103 & 64 \\
    Gender (Male/Female)                     & 332/290 & 54/32 & 109/96 & 2/1 & 0/2 & 62/41 & 32/32 \\
    Age (Year)                               & 7.2 $\pm$ 4.5 & 6.4 $\pm$ 4.3 & 4.9 $\pm$ 3.2 & 13.0 $\pm$ 2.2 & 0.3 $\pm$ 0.1 & 6.8 $\pm$ 4.1 & 6.4 $\pm$ 4.0 \\
    Height (cm)                              & 124.3 $\pm$ 29.9 & 118.4 $\pm$ 27.7 & 108.0 $\pm$ 23.6 & 168.7 $\pm$ 8.2 & 63.0 $\pm$ 1.0 & 121.6 $\pm$ 26.4 & 117.5 $\pm$ 30.1 \\
    Weight (kg)                              & 29.8 $\pm$ 19.6 & 25.8 $\pm$ 19.3 & 19.3 $\pm$ 12.0 & 62.3 $\pm$ 7.1 & 7.9 $\pm$ 1.0 & 26.4 $\pm$ 16.9 & 24.7 $\pm$ 16.5 \\
    Total dose (MBq)                         & 188.6 $\pm$ 74.0 & 74.3 $\pm$ 29.1 & 114.2 $\pm$ 62.5 & 284.9 $\pm$ 50.8 & 72.2 $\pm$ 20.4 & 171.1 $\pm$ 74.8 & 132.3 $\pm$ 73.1 \\
    Post-injection time (min)                & 74.3 $\pm$ 26.6 & 78.1 $\pm$ 16.8 & 92.3 $\pm$ 37.3 & 63.7 $\pm$ 10.7 & 78.0 $\pm$ 24.4 & 78.7 $\pm$ 13.7 & 108.6 $\pm$ 24.4 \\
\bottomrule
\end{tabular}
\end{table*}
\begin{table*}[htbp]
    \centering
    \caption{Comparison of Experimental Settings. Note that our method covers the most diverse clinical scenarios.}
    \label{tab:pet_comparison_final}
    \setlength{\tabcolsep}{5pt} 
    \begin{tabular}{lcccccccc}
        \toprule
        {Method} & {CycleGAN}\cite{dong2020deep} &{DBDL\cite{guo2022using}} & {IVNAC}\cite{guan2024synthetic} & {Pix2pix}\cite{li2022eliminating} & {DLSE \cite{laurent2025evaluation}} & {Ours (GPCN)} \\ 
        \midrule
        Dataset Size & 35 & {735} & 41 & 28 & 32  & \textbf{1085} \\
        Radiotracer & $^{18}$F-FDG &{Multiple ($N=5$)} & $^{18}$F-FDG & $^{18}$F-FDG & {Multiple ($N=5$)} &  \textbf{Multiple ($N=5$)} \\
        Scanner & Single &{Multiple ($N=4$)}& Single & Single & Single & \textbf{Multiple ($N=2$)} \\
        Region & Whole-body & Whole-body & Brain & Whole-body & Whole-body &\textbf{Whole-body} \\
        \bottomrule
    \end{tabular}
\end{table*}
In compliance with the Consent to Participate declaration, this study was granted a formal waiver of informed consent by the approving ethics committee. Specifically, the Institutional Review Board of Beijing Friendship Hospital, Capital Medical University, approved this retrospective study and waived the requirement for patient informed consent due to the retrospective analysis of anonymized data (Approval No. 2025-P2-415-01).  The study included 1085 pediatric patients who underwent PET/CT for various clinical assessments.
The dataset reflects significant clinical heterogeneity, with scans acquired from two PET/CT systems (Siemens Biograph mCT and SinoUnion PoleStar Flight).
A diverse range of five radiotracers was used across these systems based on clinical availability: $^{18}$F-FDG, $^{18}$F-MFBG, $^{68}$Ga-DOTATATE, $^{18}$F-FAPI, and $^{18}$F-DOPA.
A low-dose CT was acquired for attenuation correction prior to PET emission scanning.
For this study, both the resulting non-attenuation
 and non-scatter corrected (NASC) images, serving as model input, and the CT-based attenuation and scatter corrected (ASC) images, serving as ground truth, were archived.
All image volumes were standardized to a 200 $\times$ 200 matrix size, yielding approximately 400,000 axial slices in total. The images were reconstructed with a voxel size of 4.07 $\times$ 4.07 $\times$ 3.0 $\mathrm{mm}^3$.

As summarized in Table \ref{tab:my_dataset_info} and \ref{tab:pet_comparison_final}, our dataset represents a significant advancement in terms of both scale and diversity compared to prior works. Whereas previous methods were often validated on limited cohorts ranging from 28 to 41 scans [21], [29], [31], our study utilizes what is, to the best of our knowledge, the largest pediatric cohort for PET attenuation correction, comprising 1085 whole-body scans. Furthermore, unlike methods confined to homogeneous imaging protocols, our dataset explicitly addresses clinical heterogeneity by including five different radiotracers and data from multiple scanner manufacturers, establishing a new benchmark for generalizable PET correction.
\section{Experiments}
\subsection{Training and Implementation}
The proposed GPCN was implemented using the PyTorch framework. For the core VMamba blocks, the latent state dimension of the state space model was set to 16, and the internal expansion ratio was set to 2.0. With these configurations, GPCN achieves a highly compact and efficient architecture. 
To rigorously evaluate its performance, we employed two distinct training strategies. The first, a joint training strategy, utilized a heterogeneous dataset aggregating data from all scanners and tracers to assess overall performance. The second, a single-source strategy, was designed for generalization analysis, where the model was trained exclusively on data from the Siemens-FDG cohort (560 patients). Both training routines were conducted on an NVIDIA V100 GPU (16 GB VRAM) using the Adam optimizer with an initial learning rate of $2 \times 10^{-4}$. Each model was trained with a batch size of 2 for 400,000 iterations. The final performance of both trained models was then evaluated on a test set comprising 129 patients from all domains.

This study was conducted on a large and diverse retrospective cohort of 1085 pediatric subjects from Beijing Friendship Hospital. To rigorously evaluate the proposed GPCN under different clinical scenarios, the entire dataset was first partitioned into a global training pool and a unified test set by independently splitting each device-tracer cohort at a 9:1 ratio. The unified test set comprises samples from all seven device-tracer protocols and was kept strictly identical across all evaluations. Based on this partition, we designed two distinct experimental setups:

\begin{itemize}
    \item \textbf{Joint Training:} To assess the overall performance and robustness, models were trained on a heterogeneous subset of the training pool, which includes subjects scanned with $^{18}$F-FDG, $^{68}$Ga-DOTATATE, and $^{18}$F-MFBG on the Siemens system, along with $^{18}$F-FDG and $^{18}$F-MFBG on the SinoUnion Healthcare system. 
    
    \item \textbf{Zero-Shot Generalization:} To explicitly validate cross-domain robustness, models were trained \textit{exclusively} on the Siemens $^{18}$F-FDG subset from the training pool. These models were subsequently evaluated directly on the unified test set to strictly quantify their zero-shot generalization capabilities across the remaining six unseen scanner-tracer domains.
\end{itemize}

The performance of our proposed GPCN is compared with five   state-of-the-art methods: Pix2pix \cite{li2022eliminating}, CycleGAN \cite{dong2020deep}, IVNAC \cite{guan2024synthetic}, DBDL \cite{guo2022using} and MambaIR \cite{guo2024mambair}. The uncorrected NASC images were also included as a baseline reference.
\begin{table}[!t]
    \centering
    \footnotesize
    \renewcommand\arraystretch{1}
    \setlength{\tabcolsep}{1pt} 
    \caption{ASC performance via different methods on the joint training. The best values for each metric are highlighted in bold. NASC refers to the uncorrected input images.}
    \begin{tabular}{lccccc}
      \toprule
      {Method} & {\# Params$\downarrow$} & {PSNR (dB)$\uparrow$} & {SSIM$\uparrow$}& {nMAE$\downarrow$} & {NMSE$\downarrow$}  \\
      \midrule
NASC                          & N/A               & 22.18$\pm$6.70 & 0.813$\pm$0.15 & N/A & N/A\\
Pix2pix \cite{li2022eliminating} & 29.23M            & 36.00$\pm$4.40 & 0.989$\pm$0.01 & 2.46$\pm$2.38 & 0.04$\pm$0.03 \\
CycleGAN \cite{dong2020deep}  & 7.82M             & 33.20$\pm$4.14 & 0.985$\pm$0.01 & 3.26$\pm$2.83 & 0.07$\pm$0.05 \\
IVNAC \cite{guan2024synthetic}   & 2.12M             & 32.67$\pm$4.53 & 0.972$\pm$0.02 & 3.71$\pm$2.92 & 0.08$\pm$0.05 \\
DBDL \cite{guo2022using}      & 31.90M            & 35.65$\pm$4.03 & 0.991$\pm$0.01 & 2.30$\pm$2.17 & 0.04$\pm$0.03 \\
MambaIR \cite{guo2024mambair} & {1.60M}    & 37.96$\pm$4.19 & 0.991$\pm$0.01 & 2.05$\pm$1.62 & 0.03$\pm$0.03 \\
\textbf{GPCN}                 & {2.01M} & \textbf{38.75$\pm$4.11} & \textbf{0.993$\pm$0.01} & \textbf{1.80$\pm$1.53} & \textbf{0.02$\pm$0.02} \\
\bottomrule
    \end{tabular}
    \label{tab:quantitative_results}
\end{table}

\begin{figure}[htbp]
    \centering
    \includegraphics[width=\linewidth]{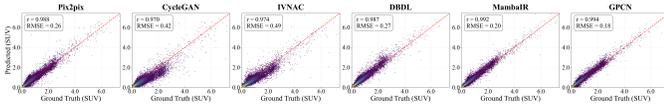}
    \caption{Joint histogram analysis demonstrating voxel-wise correlation between the predicted PET images and the ground-truth ASC PET in SUV units.}
    \label{fig:joint_hist}
\end{figure}

\begin{figure*}[t]
\centering
\includegraphics[width=0.9\textwidth]{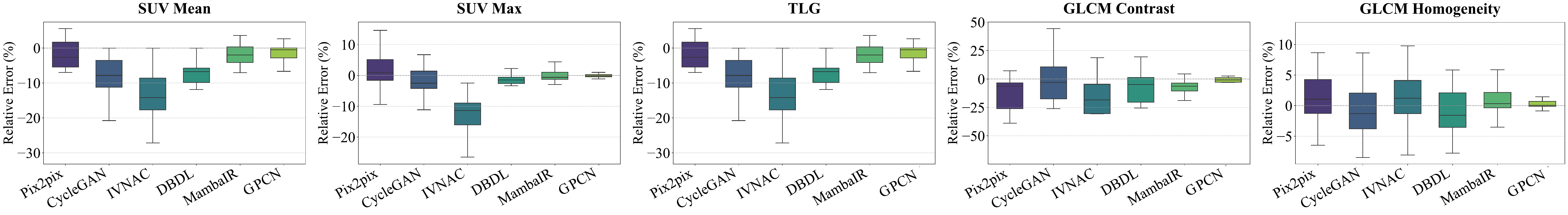}
\caption{Quantitative analysis of clinical and radiomic metrics under joint training. Box plots show error of VOI-based $\text{SUV}{\text{max}}$, $\text{SUV}{\text{mean}}$, GLCM Contrast, GLCM Homogeneity, and total lesion glycolysis (TLG) in major organs.}
\label{fig:suv_large}
\end{figure*}

\begin{figure*}[!t]
\centering
\includegraphics[width=0.82\textwidth]{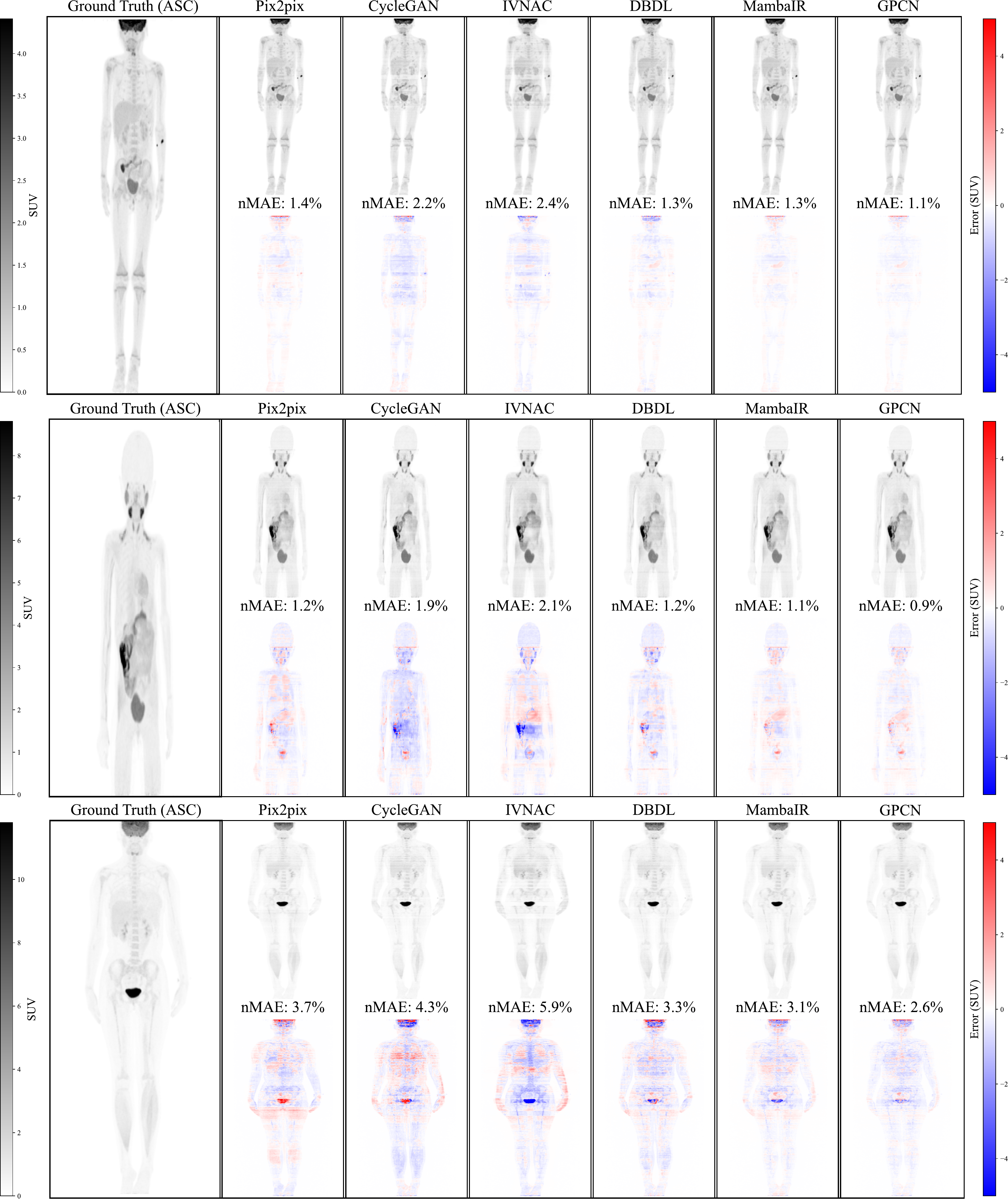}
\caption{Qualitative visual results of the joint training strategy on the heterogeneous test set. Each patient case is presented as a two-row block, comprising the predicted PET images displayed via maximum intensity projection (MIP) (top) and their corresponding spatial error maps (bottom).}
\label{fig:results_siemens}
\end{figure*}
\subsection{Performance on the joint training}
As summarized in Table~\ref{tab:quantitative_results}, GPCN consistently outperformed all competing methods, achieving a PSNR of 38.75~dB, SSIM of 0.993, nMAE of 1.8, and NMSE of 0.02, indicating superior image quality and fidelity. 
To evaluate voxel-wise quantitative accuracy, we performed a joint histogram analysis using the clinically standard Standardized Uptake Value (SUV). Fig. \ref{fig:joint_hist} illustrates the pixel-wise correlation between ground-truth ASC PET and model predictions for a representative case. While CycleGAN and IVNAC exhibit substantial scattering from the identity line with higher RMSE, GPCN maintains a compact distribution tightly aligned with the reference. Achieving the highest Pearson correlation ($r = 0.994$) and lowest error, this high voxel-level agreement confirms GPCN captures the underlying physical attenuation mapping rather than performing superficial image translation, effectively preserving the radiotracer's true biodistribution. 

We also evaluated key clinical metrics within volumes of interest (VOIs) \cite{presotto2018pet,li2019non,lohmann2017radiation}. As shown in Fig.~\ref{fig:suv_large}, GPCN achieves the lowest median error and a markedly more compact distribution for $\text{SUV}_{\text{max}}$, $\text{SUV}_{\text{mean}}$, total lesion glycolysis (TLG), and GLCM Contrast, which is critical for preserving quantitative values used in clinical diagnosis and response assessment.

To complement the quantitative metrics, Fig.~\ref{fig:results_siemens} shows the coronal MIP of corrected PET images for three representative patients, together with error maps for each method. Although most methods appear visually plausible, their error maps reveal substantial residual errors, particularly for CycleGAN and IVNAC, which exhibit widespread, high-magnitude errors throughout the torso. By contrast, GPCN closely matches the ground truth, with a markedly cleaner and less structured error map, indicating that its superior quantitative performance translates into a more accurate and reliable restoration of the true anatomical biodistribution.
\begin{figure*}[t]
\centering
\includegraphics[width=0.8\textwidth,height=0.16\textwidth]{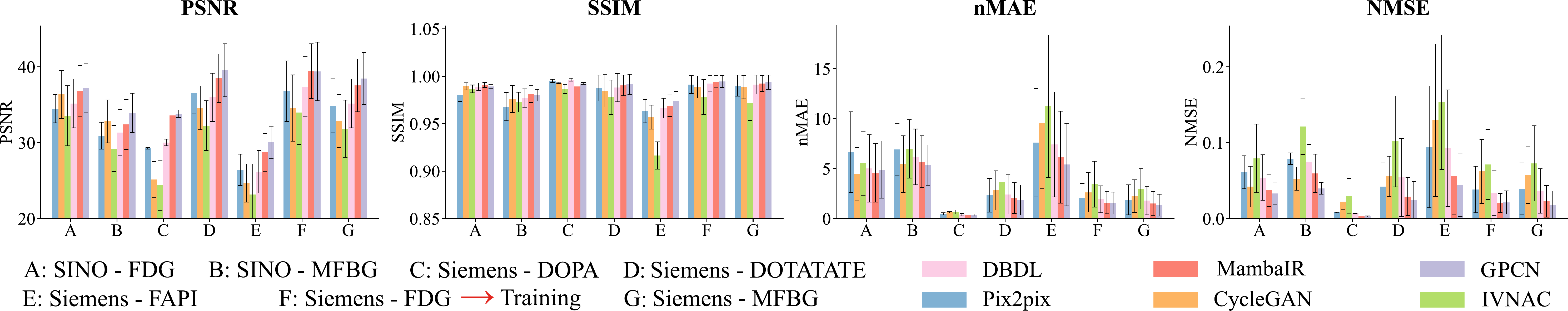}
\caption{Quantitative generalization on external scanners and tracers. Bar plots show PSNR, SSIM, nMAE, and NMSE for methods trained only on Siemens $^{18}$F-FDG and evaluated on six external scanner–tracer domains.}
\label{fig:crosspsnr}
\end{figure*}

\begin{figure*}[t]
\centering
\includegraphics[width=0.9\textwidth]{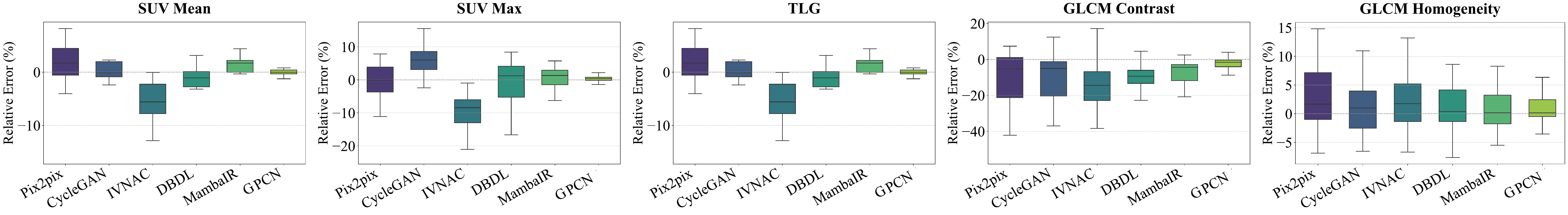}
\caption{Quantitative generalization analysis of clinical and radiomic metrics on external datasets. Box plots display the relative error (\%) of VOI-based $\text{SUV}_{\text{max}}$, $\text{SUV}_{\text{mean}}$, total lesion glycolysis (TLG), and GLCM features (Contrast and Homogeneity) across unseen scanners and radiotracers.}
\label{fig:suv_cross}
\end{figure*}

\begin{figure}[t]
\centering
\includegraphics[width=0.95\linewidth]{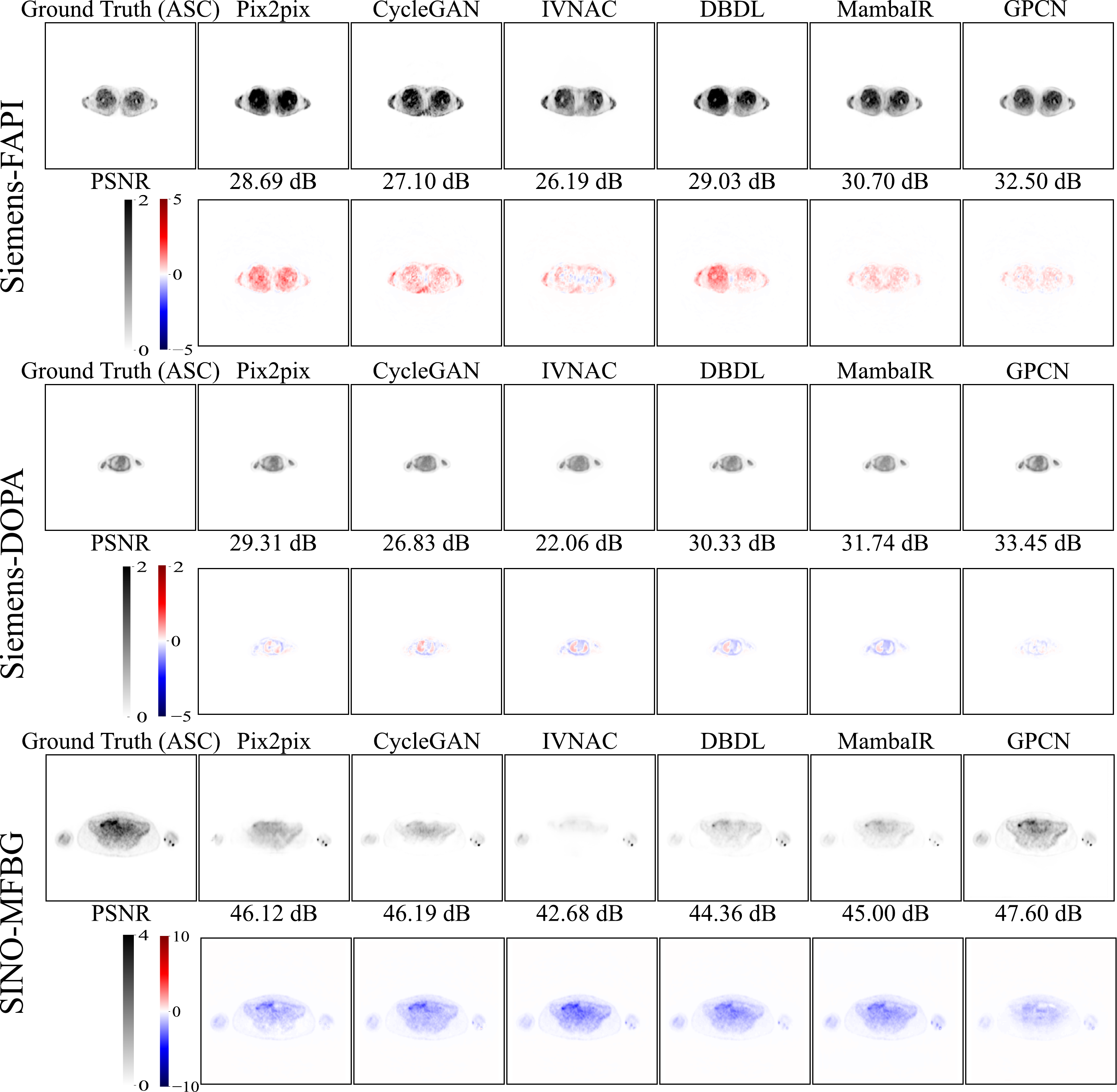}
\caption{Qualitative visual comparison of generalization performance on external scanners and tracers. Each row corresponds to a representative case from a specific test domain.}
\label{fig:visualcross}
\end{figure}
\subsection{Generalization Analysis on Unseen Scanners and Radiotracers}
The clinical landscape of PET imaging is inherently heterogeneous, encompassing a diverse array of radiotracers and numerous scanners across different institutions. Therefore, to rigorously assess the generalizability of our proposed GPCN against competing methods, we designed a demanding cross-domain evaluation. In this setup, all models were trained exclusively on data from a single domain: $^{18}$F-FDG scans acquired on a Siemens Biograph mCT scanner. Subsequently, these trained models were directly tested on six entirely unseen domains: Siemens-DOTATATE, Siemens-MFBG, Siemens-FAPI, Siemens-DOPA, SinoUnion-FDG, and SinoUnion-MFBG. This strict separation ensures a fair and explicit comparison of each model's robustness against both scanner-specific and tracer-specific domain shifts.

The results of this evaluation are summarized in Fig.~\ref{fig:crosspsnr}, which reports four key metrics across all test domains, including the in-domain Siemens–FDG data and six out-of-domain datasets from scanners unseen during training. GPCN not only achieves the best performance on the in-domain data but also maintains high PSNR/SSIM and low MAE/NMSE across every out-of-domain set. In contrast, competing methods, while reasonable on Siemens–FDG, degrade markedly on unseen data, with methods like CycleGAN and IVNAC showing large errors and variances. This indicates that they overfit to the training distribution, whereas GPCN learns a more fundamental and robust mapping for PET correction. To rigorously substantiate our GPCN's robustness on unseen scanners and radiotracers, we extended the cross-domain evaluation beyond conventional image similarity metrics to include critical clinical and radiomic features. Fig. \ref{fig:suv_cross} presents the relative error for VOI-based $\text{SUV}_{\text{max}}$, $\text{SUV}_{\text{mean}}$, total lesion glycolysis (TLG), and GLCM features across the external test domains. While baseline models such as CycleGAN and IVNAC suffer from severe systematic underestimation and high variance, GPCN consistently yields the lowest median bias and the tightest interquartile ranges across all evaluated metrics. Fig.~\ref{fig:visualcross} shows that GPCN reconstructions demonstrate improved structural fidelity and reduced visible artifacts across the evaluated domains, while competing methods introduce noticeable artifacts and structural distortions, especially in the Siemens–FAPI case, where only GPCN preserves anatomically plausible uptake patterns.
\subsection{Ablation Study of Core GPCN Components}
\label{sec:ablation}
To validate the contribution of each core component in our proposed dual-domain architecture, we conducted a comprehensive ablation study. We evaluated three model configurations: (1) the full GPCN model, (2) GPCN without the multi-band contextual refinement (w/o MBCR), where the image-domain branch was removed, and (3) GPCN without the frequency-aware spectral decoupling (w/o FASD), where the Fourier-domain branch was removed. All configurations were trained and evaluated under identical conditions on the Siemens $^{18}$F-FDG dataset.
\begin{table}[!t]
  \centering
  \setlength{\tabcolsep}{1.6pt}
      \renewcommand\arraystretch{1}
  \caption{Ablation study of the core GPCN components. All configurations were trained on the Siemens $^{18}$F-FDG dataset.}
  \label{tab:ablation_study}
  \begin{tabular}{lcccc}
    \toprule
    {Model} & {PSNR (dB) $\uparrow$} & {SSIM $\uparrow$} & {nMAE $\downarrow$} & {NMSE $\downarrow$} \\
    \midrule
w/o MBCR  & 32.49 $\pm$ 2.15 & 0.876 $\pm$ 0.01 & 10.57 $\pm$ 4.94 & 0.07 $\pm$ 0.04 \\
    w/o FASD  & 36.59 $\pm$ 4.06 & 0.992 $\pm$ 0.01 & 2.10 $\pm$ 1.82 & 0.03 $\pm$ 0.02 \\
    GPCN      & \textbf{38.61 $\pm$ 4.02} & \textbf{0.993 $\pm$ 0.01} & \textbf{1.88 $\pm$ 1.75} & \textbf{0.02 $\pm$ 0.02} \\
    \bottomrule
  \end{tabular}
\end{table}
\begin{figure}[t]
\centering
\setlength{\tabcolsep}{2pt} 
\renewcommand{\arraystretch}{0.5} 
\begin{tabular}{ccccc}
    & {Ground Truth} & {w/o MBCR} & {w/o FASD} & {GPCN} \\
     &
    \includegraphics[width=0.23\linewidth]{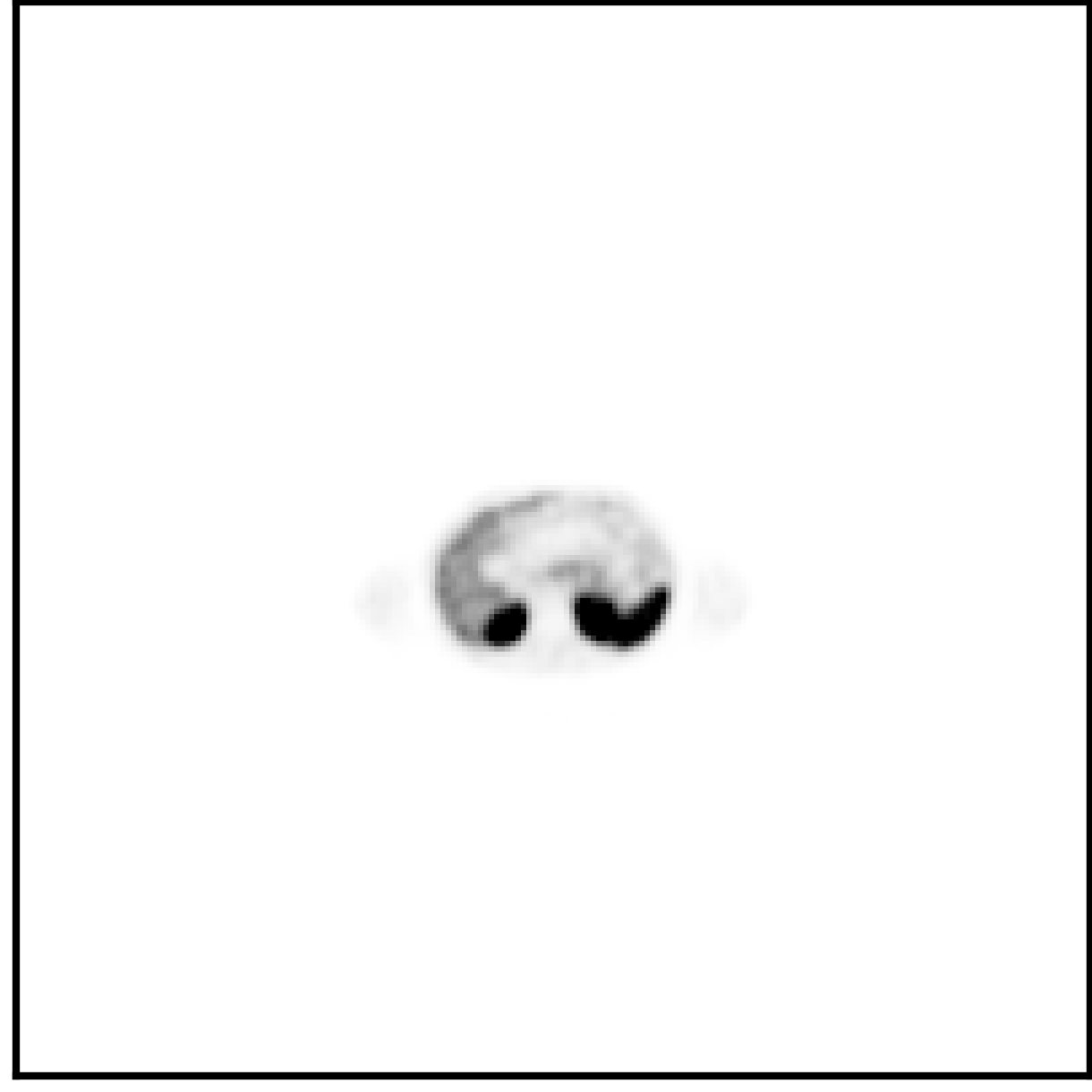} &
    \includegraphics[width=0.23\linewidth]{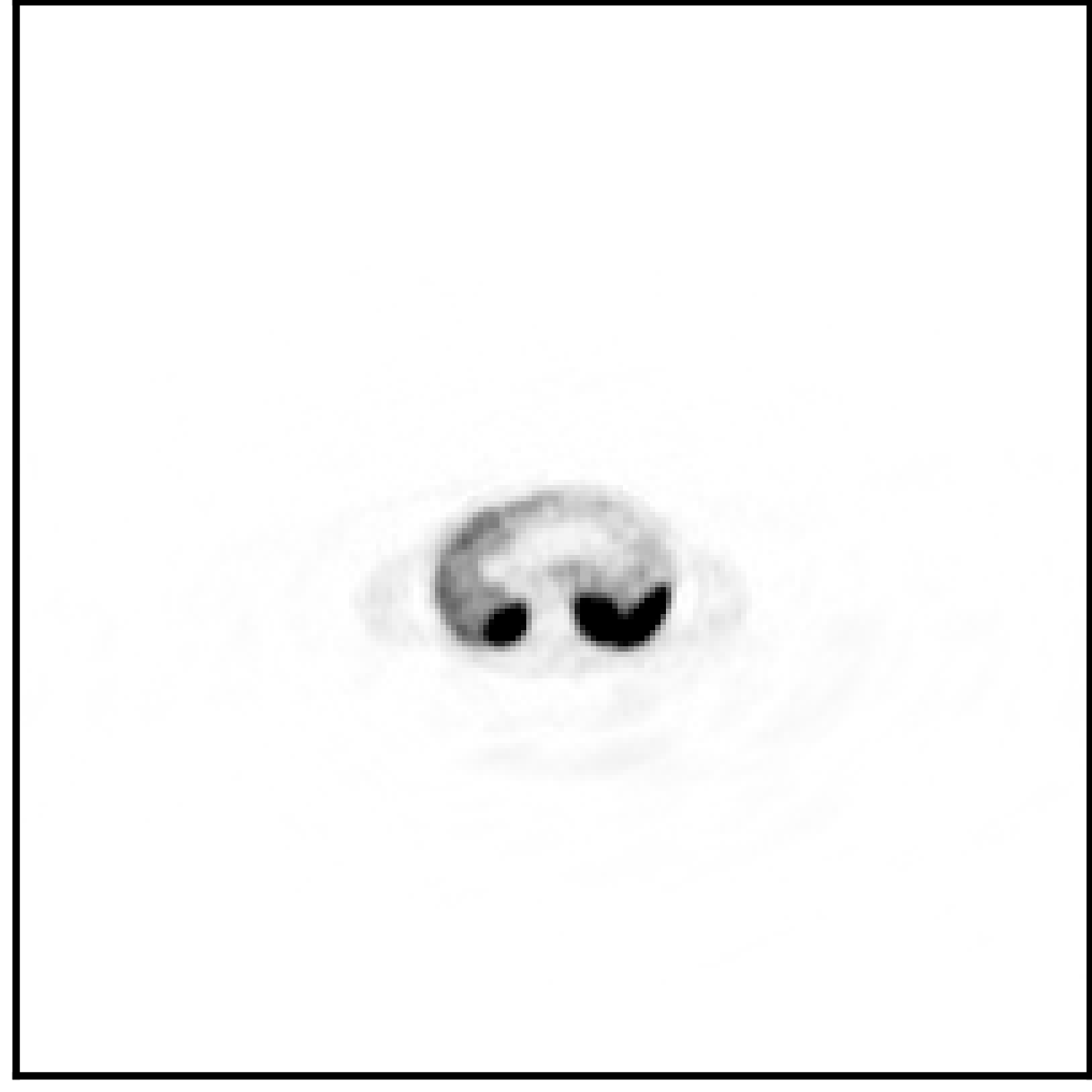} &
    \includegraphics[width=0.23\linewidth]{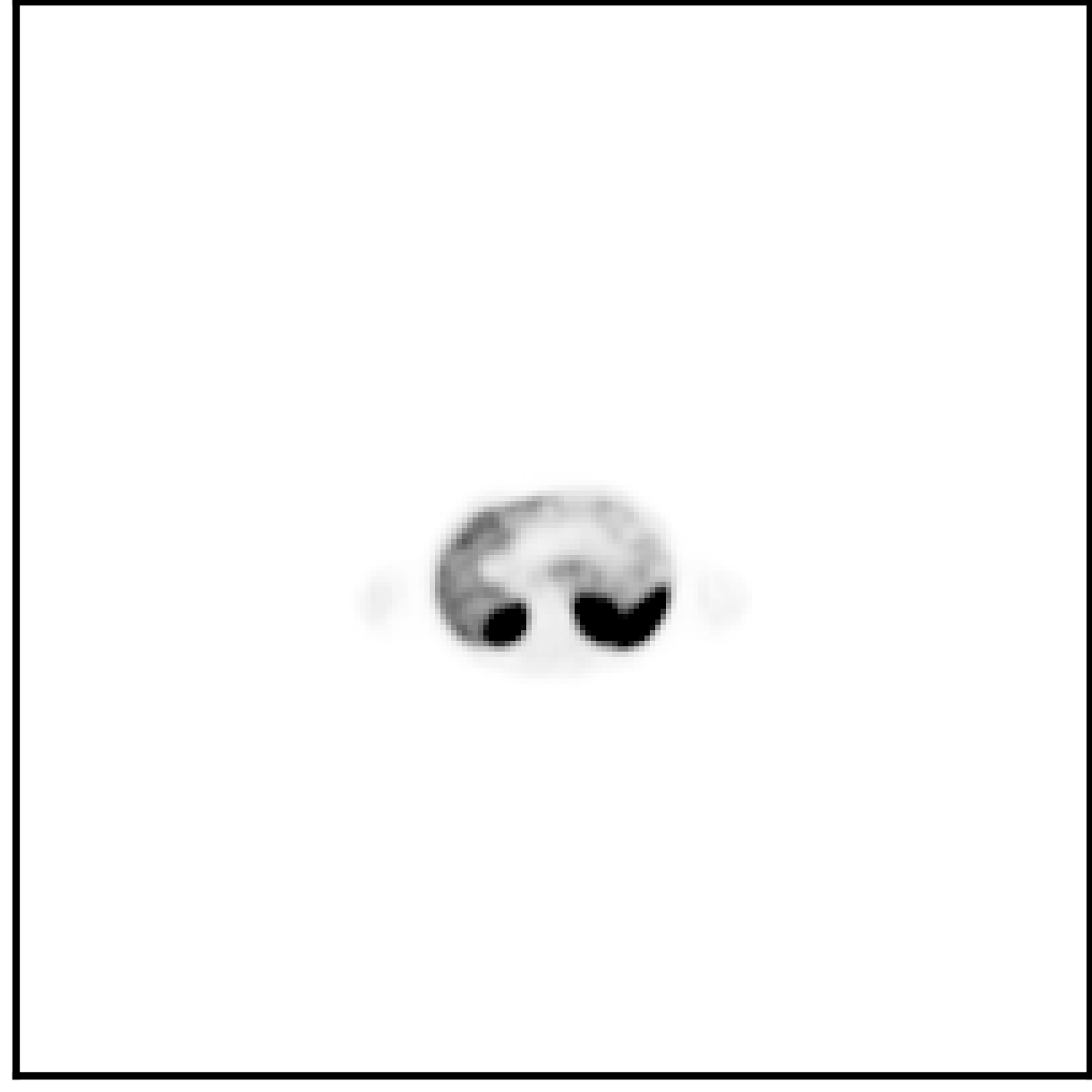} &
    \includegraphics[width=0.23\linewidth]{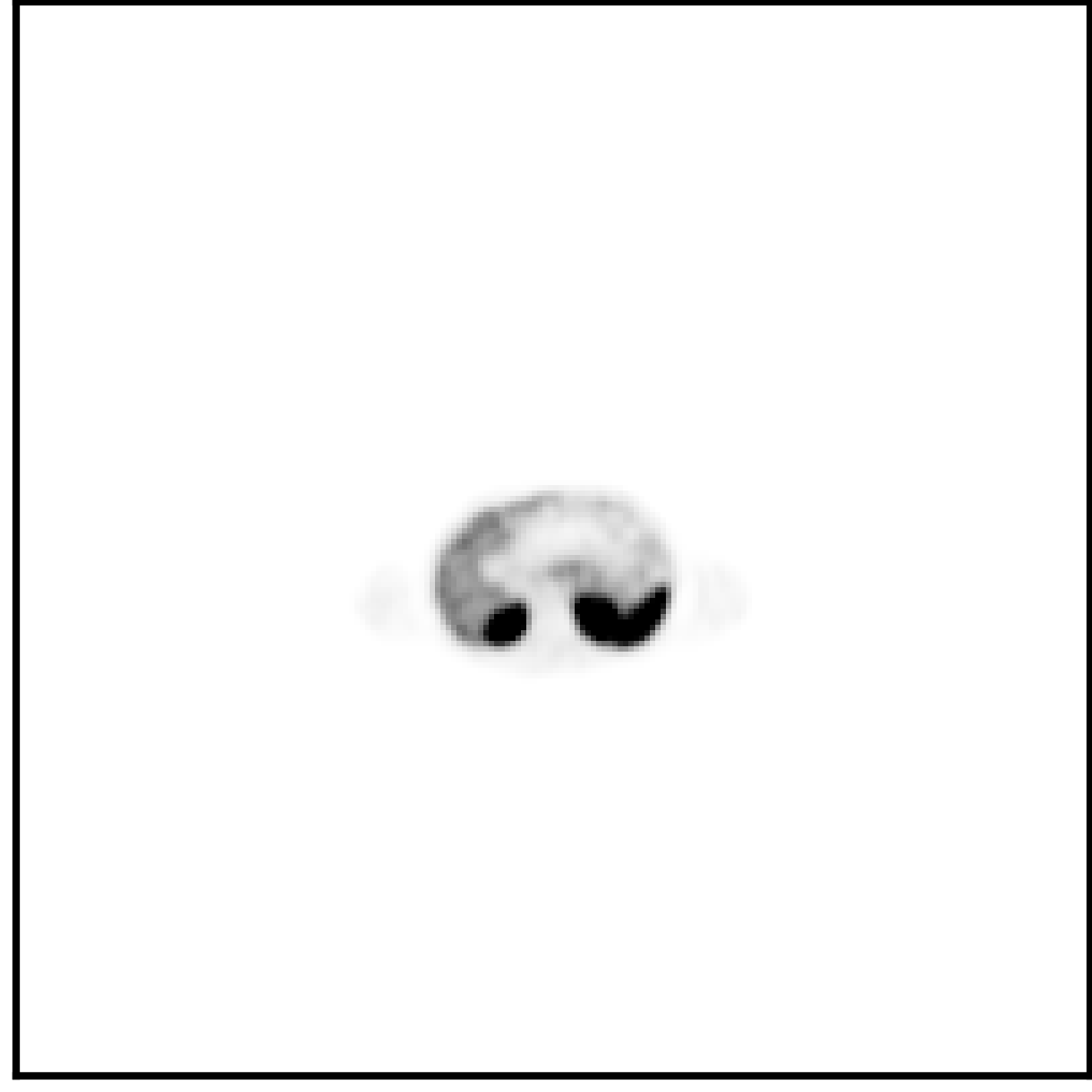} \\
    
     &
    \begin{minipage}[b]{0.23\linewidth} 
        \hfill 
        \includegraphics[height=1.0\linewidth]{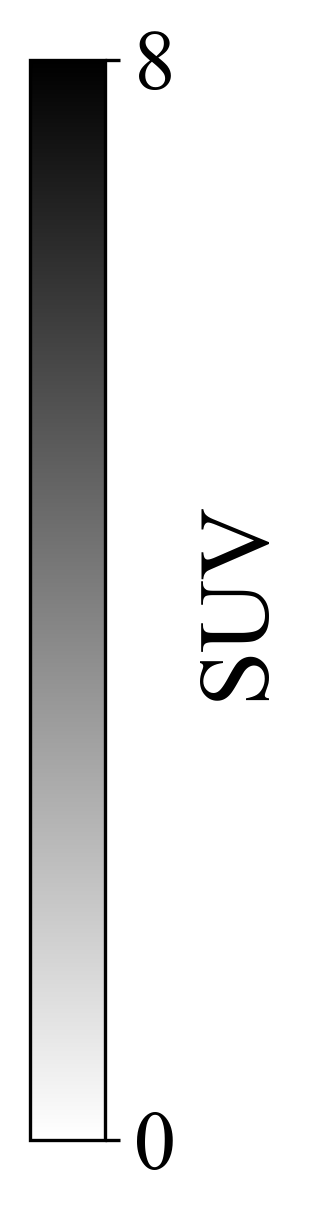}
        \hspace{1mm}
        \includegraphics[height=1.0\linewidth]{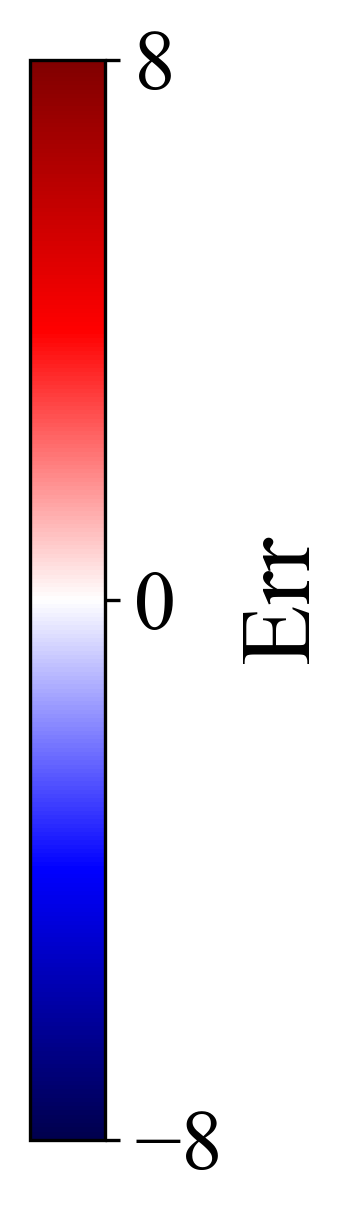}
    \end{minipage} &
    \includegraphics[width=0.23\linewidth]{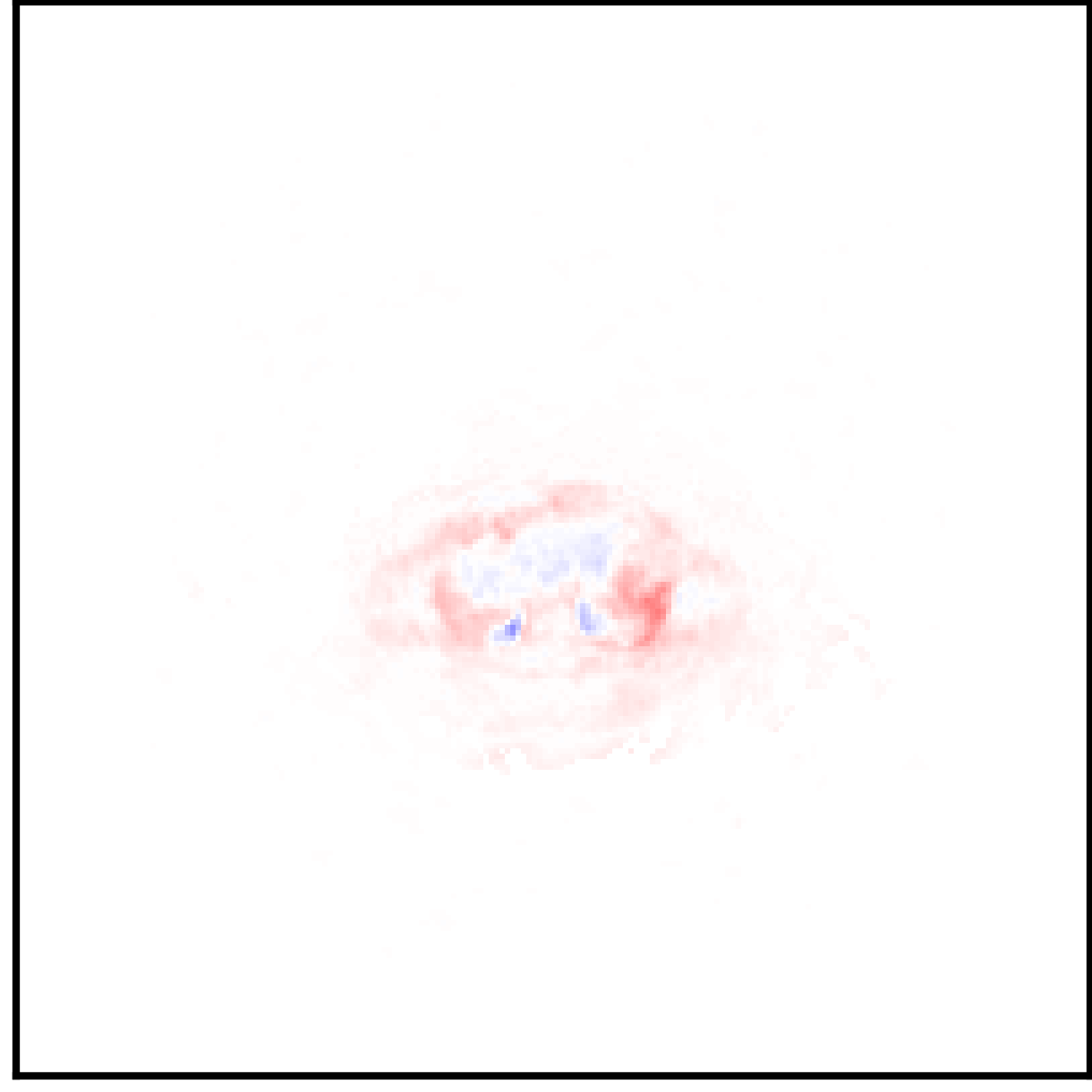} &
    \includegraphics[width=0.23\linewidth]{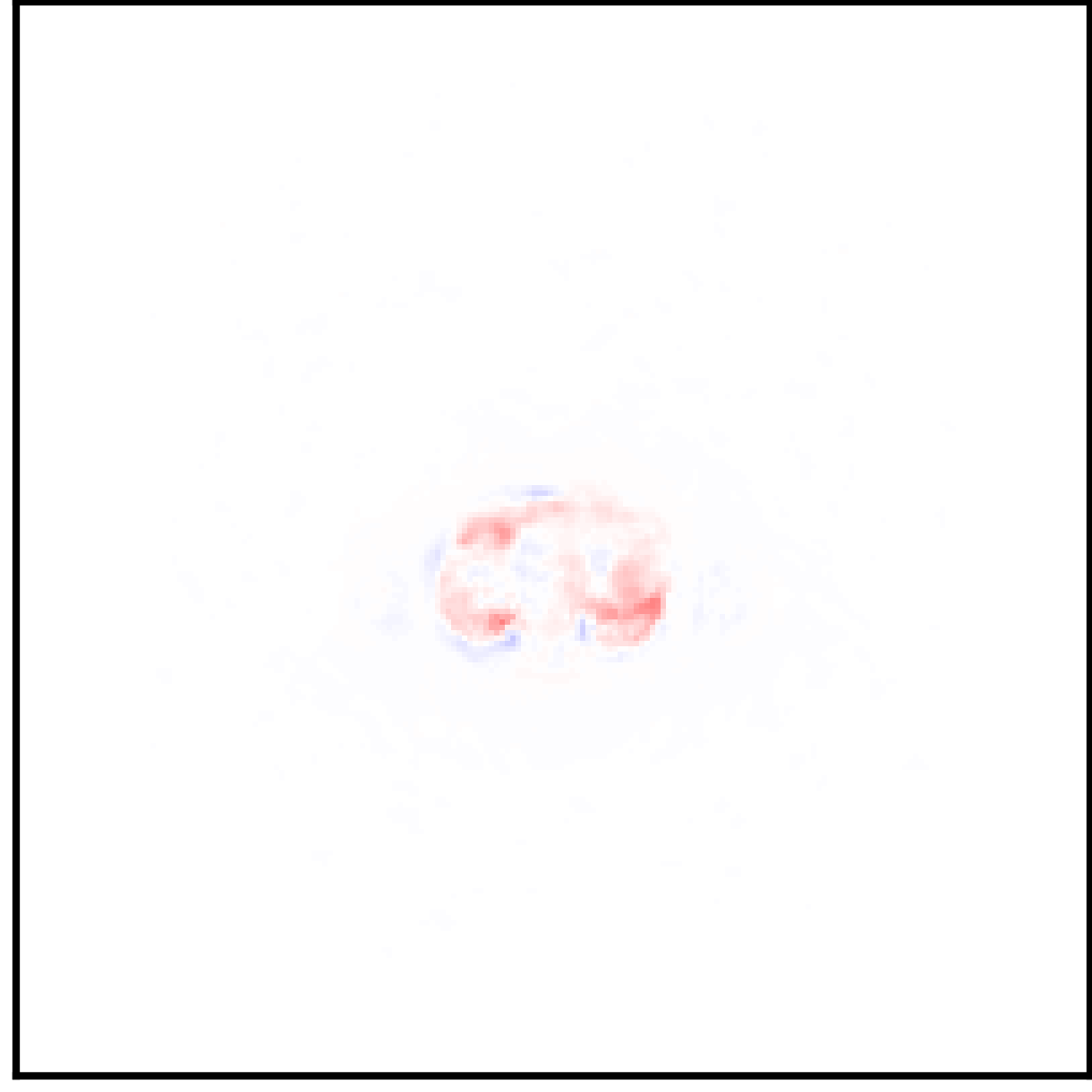} &
    \includegraphics[width=0.23\linewidth]{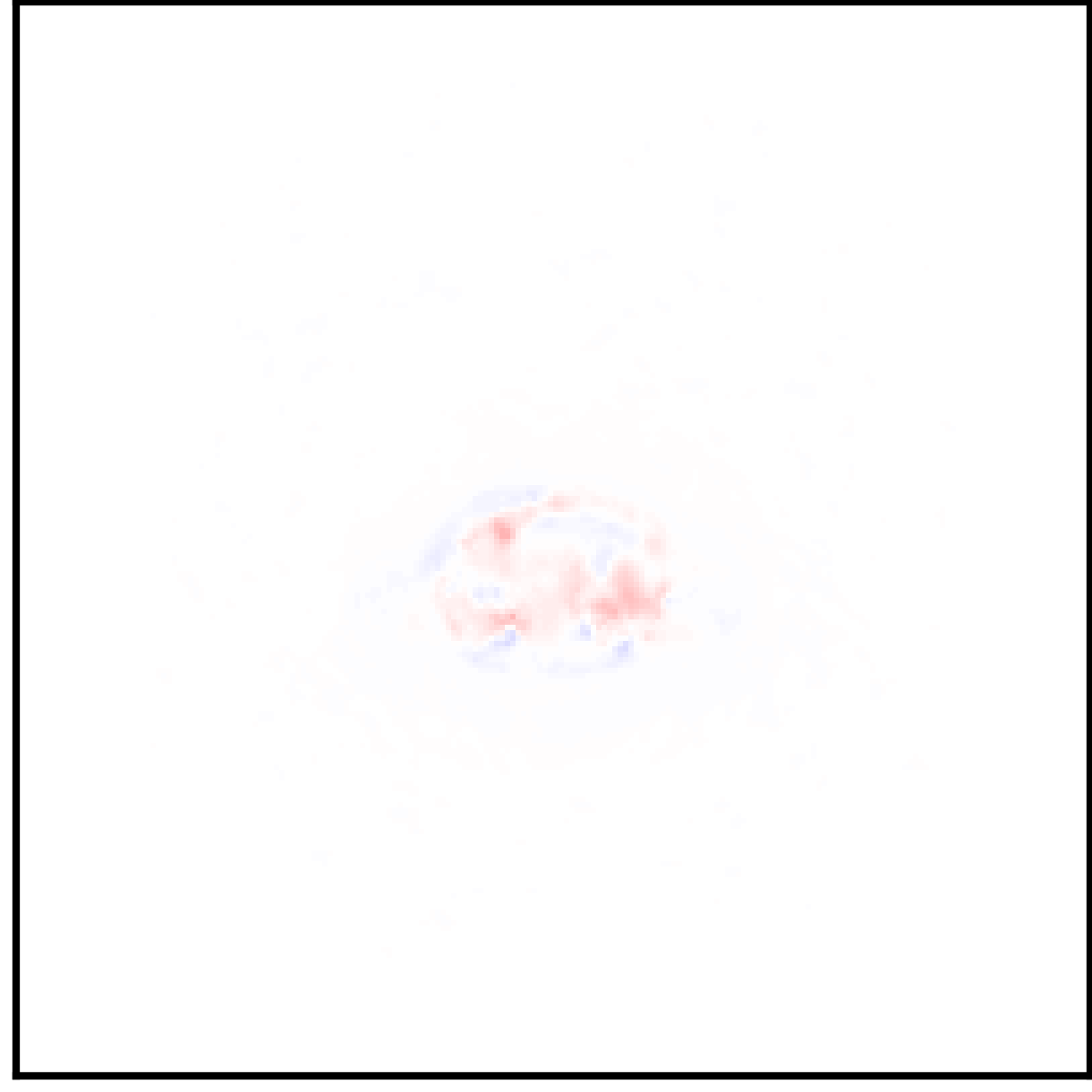} \\

     &
    \includegraphics[width=0.23\linewidth]{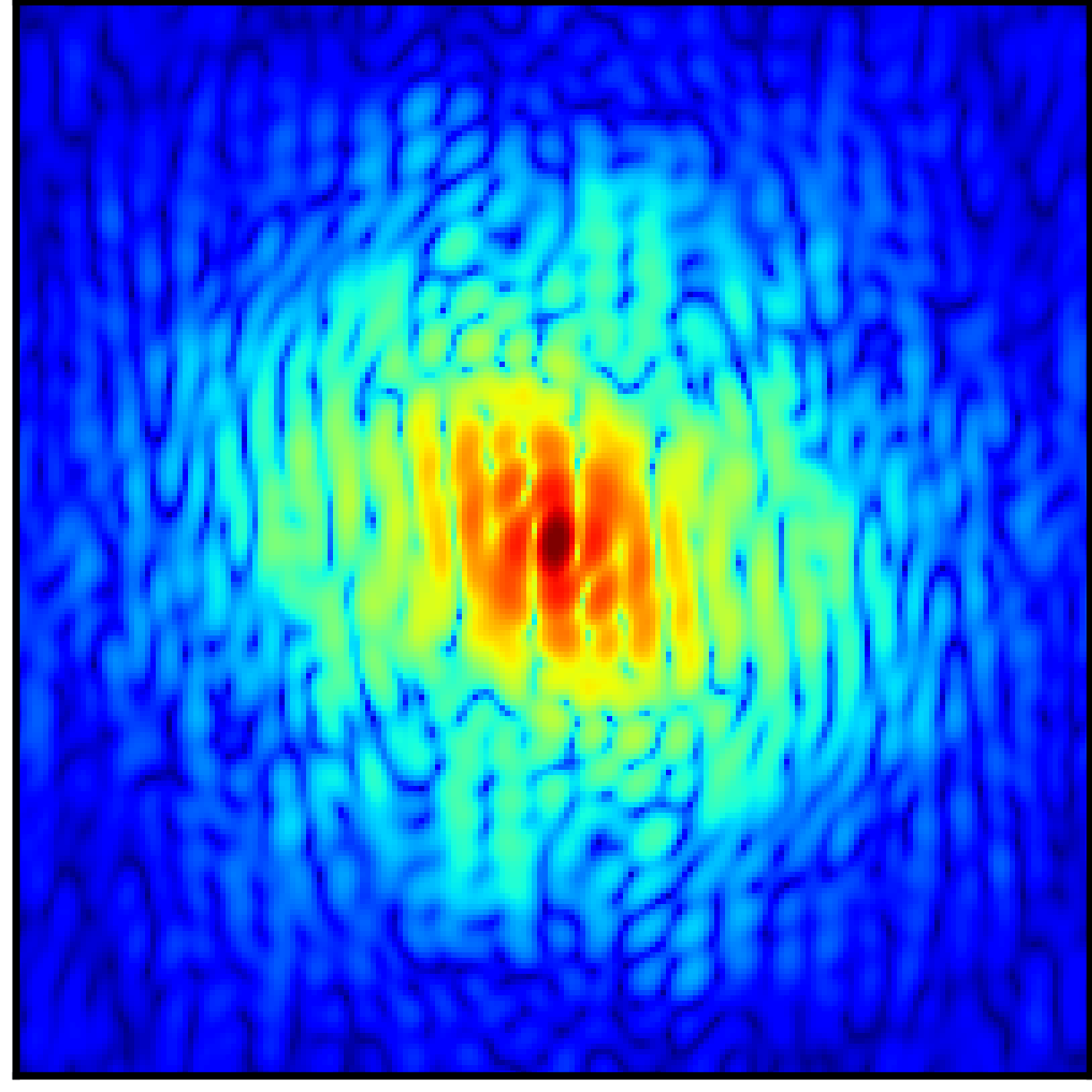} &
    \includegraphics[width=0.23\linewidth]{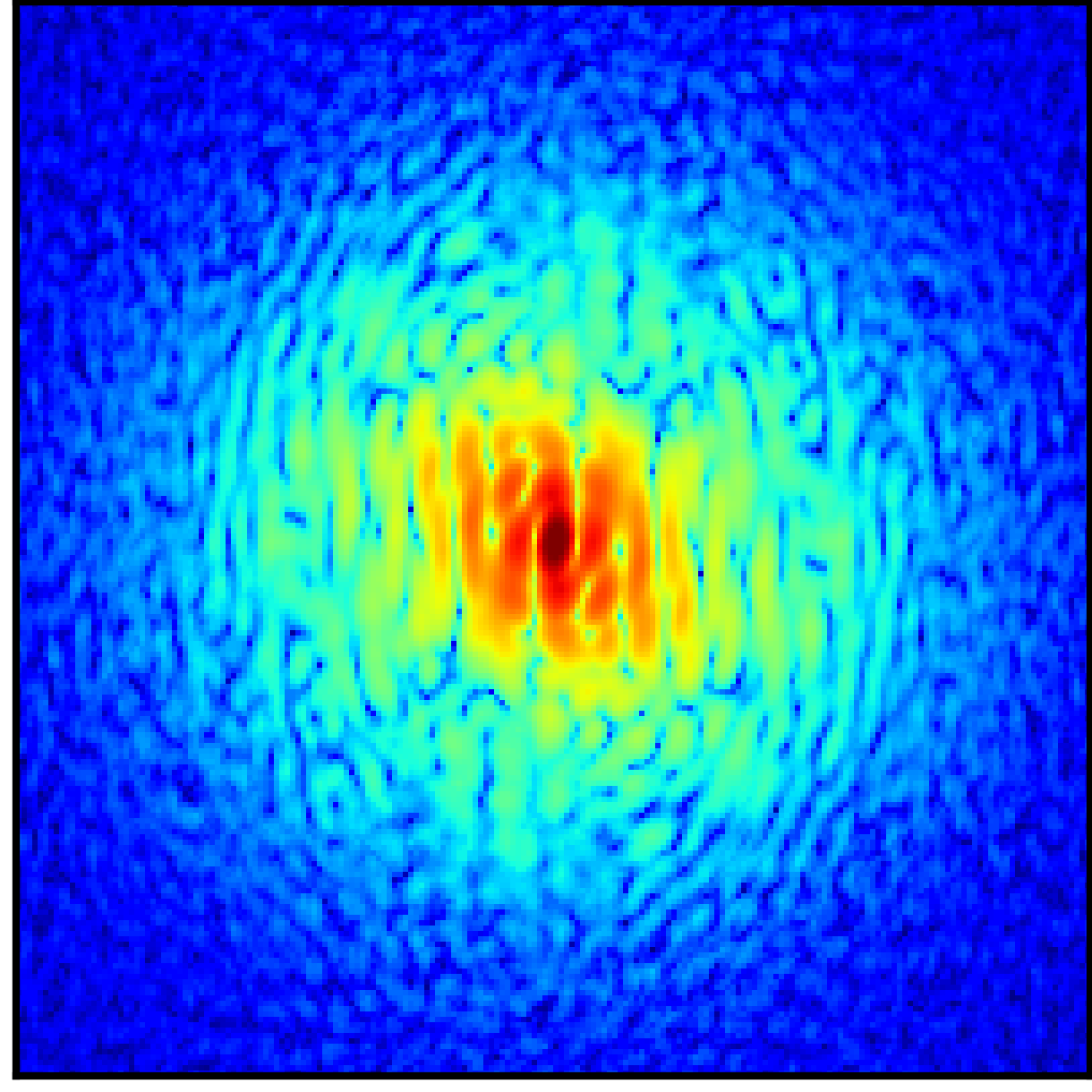} &
    \includegraphics[width=0.23\linewidth]{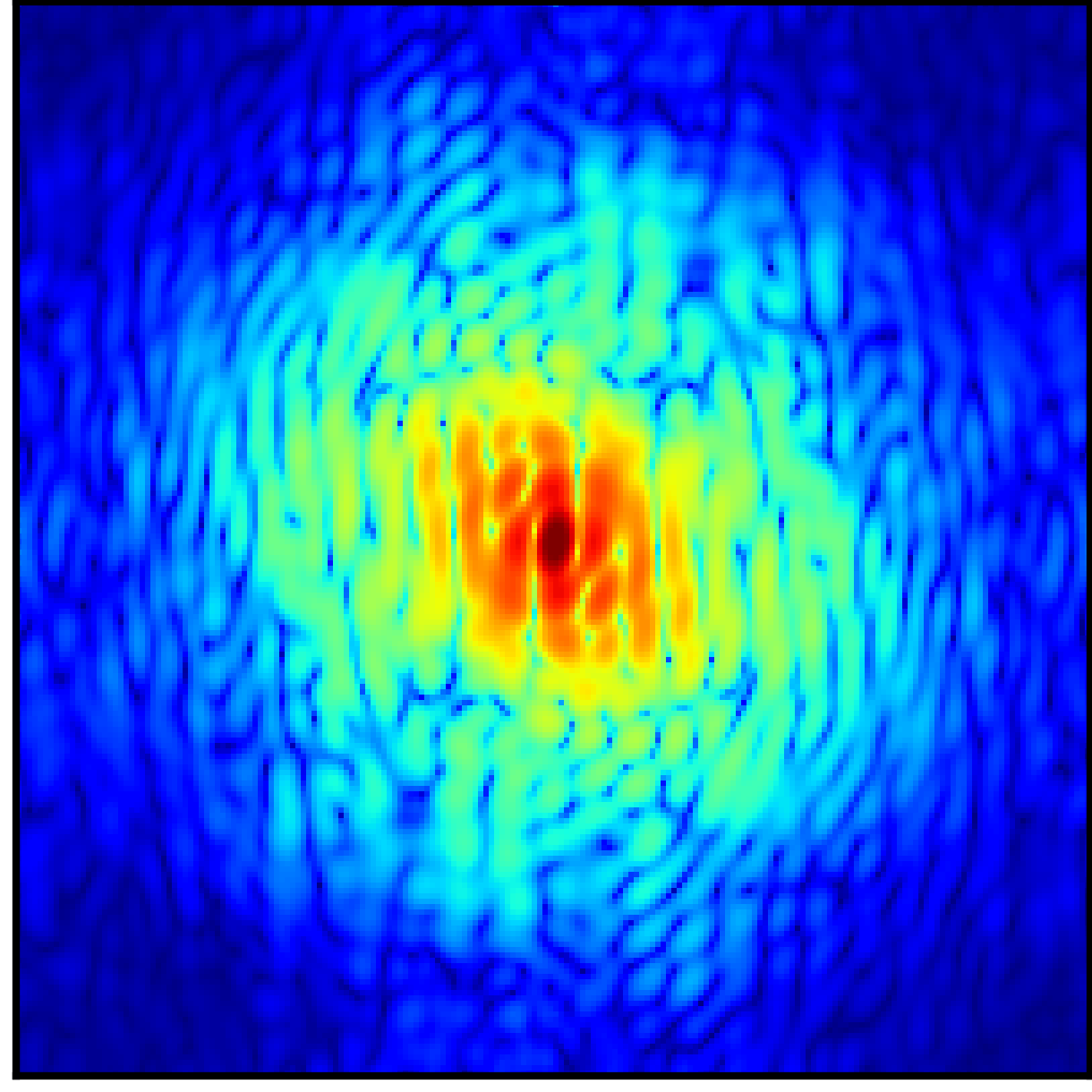} &
    \includegraphics[width=0.23\linewidth]{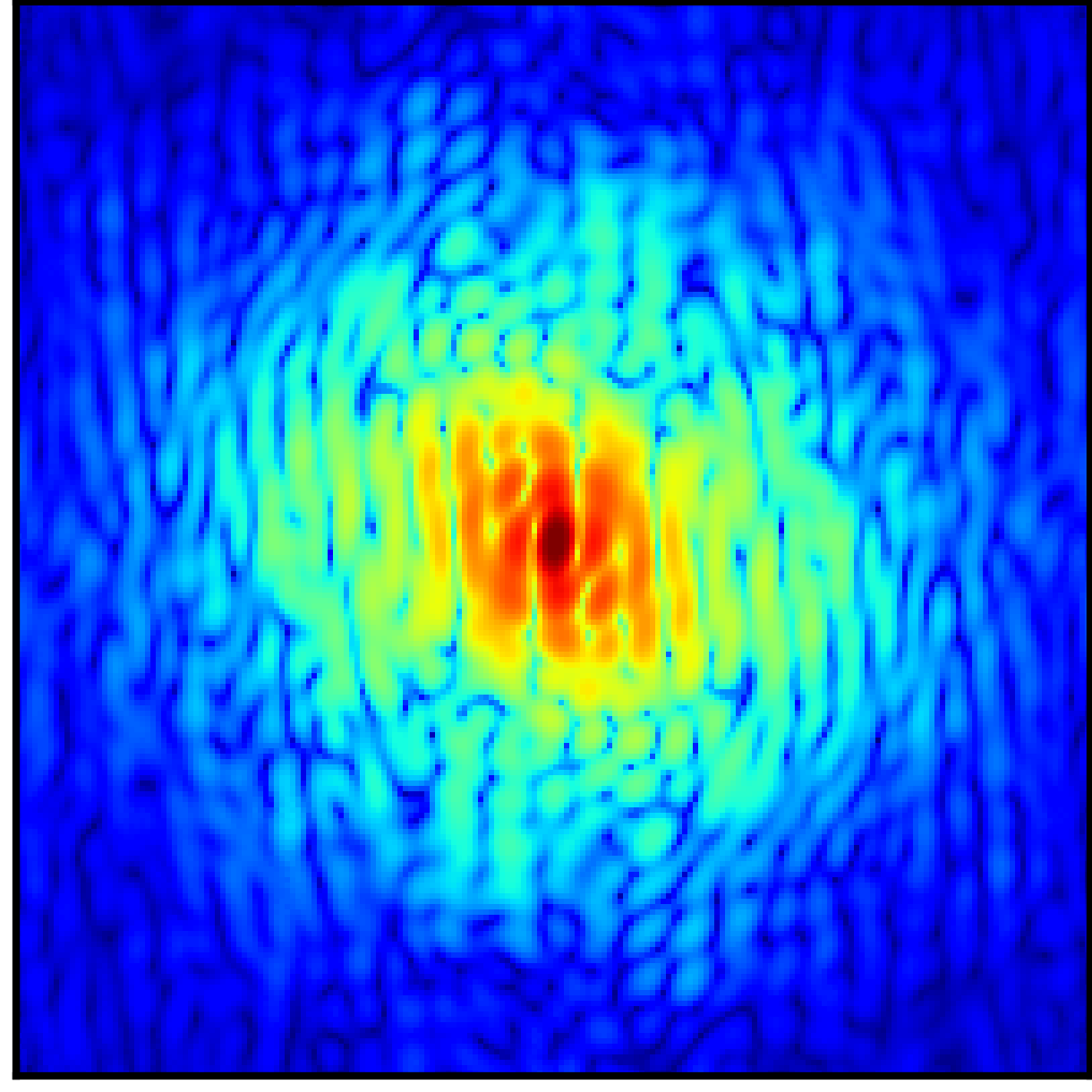} \\
    
     &
    \begin{minipage}[b]{0.23\linewidth} 
        \hfill 
        \includegraphics[height=1.0\linewidth]{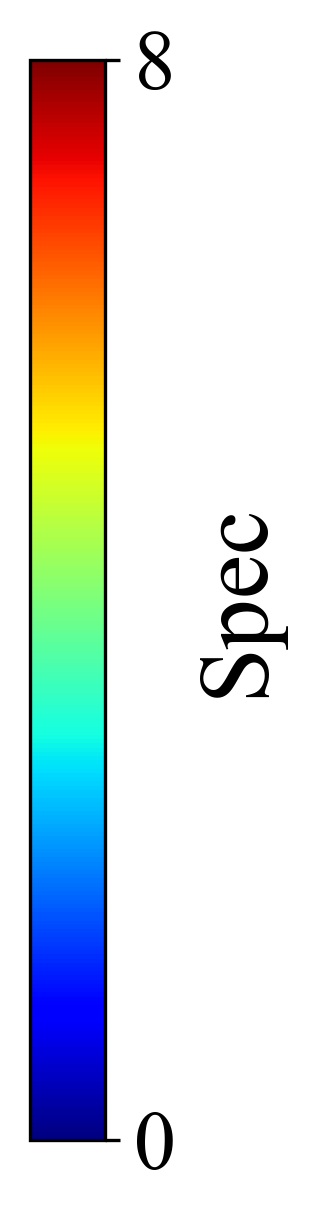}
        \hspace{1mm}
        \includegraphics[height=1.0\linewidth]{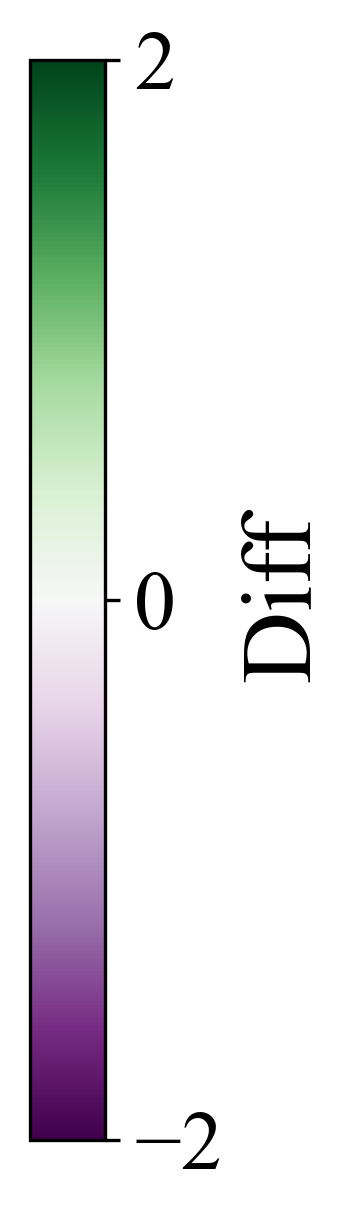}
    \end{minipage} &
    \includegraphics[width=0.23\linewidth]{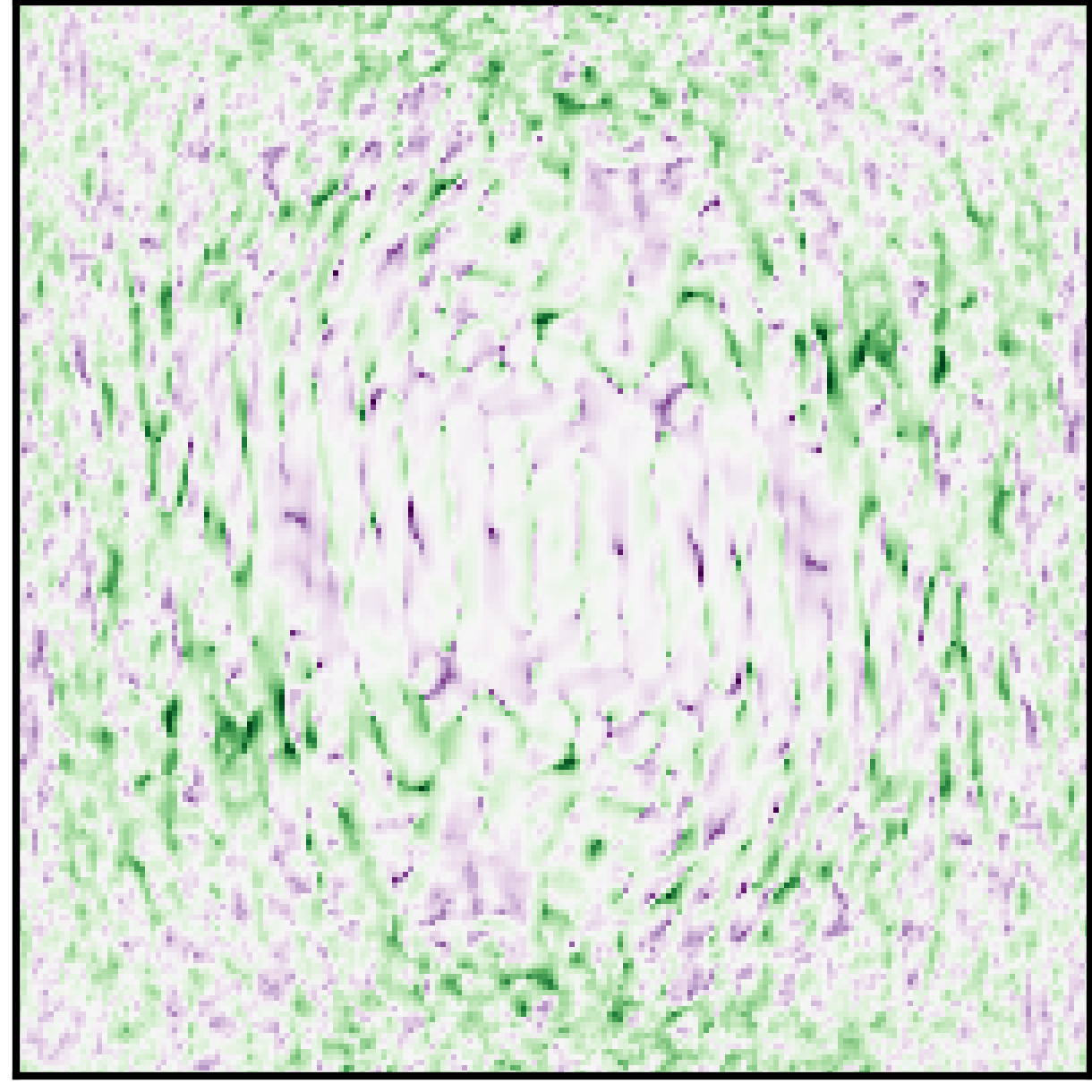} &
    \includegraphics[width=0.23\linewidth]{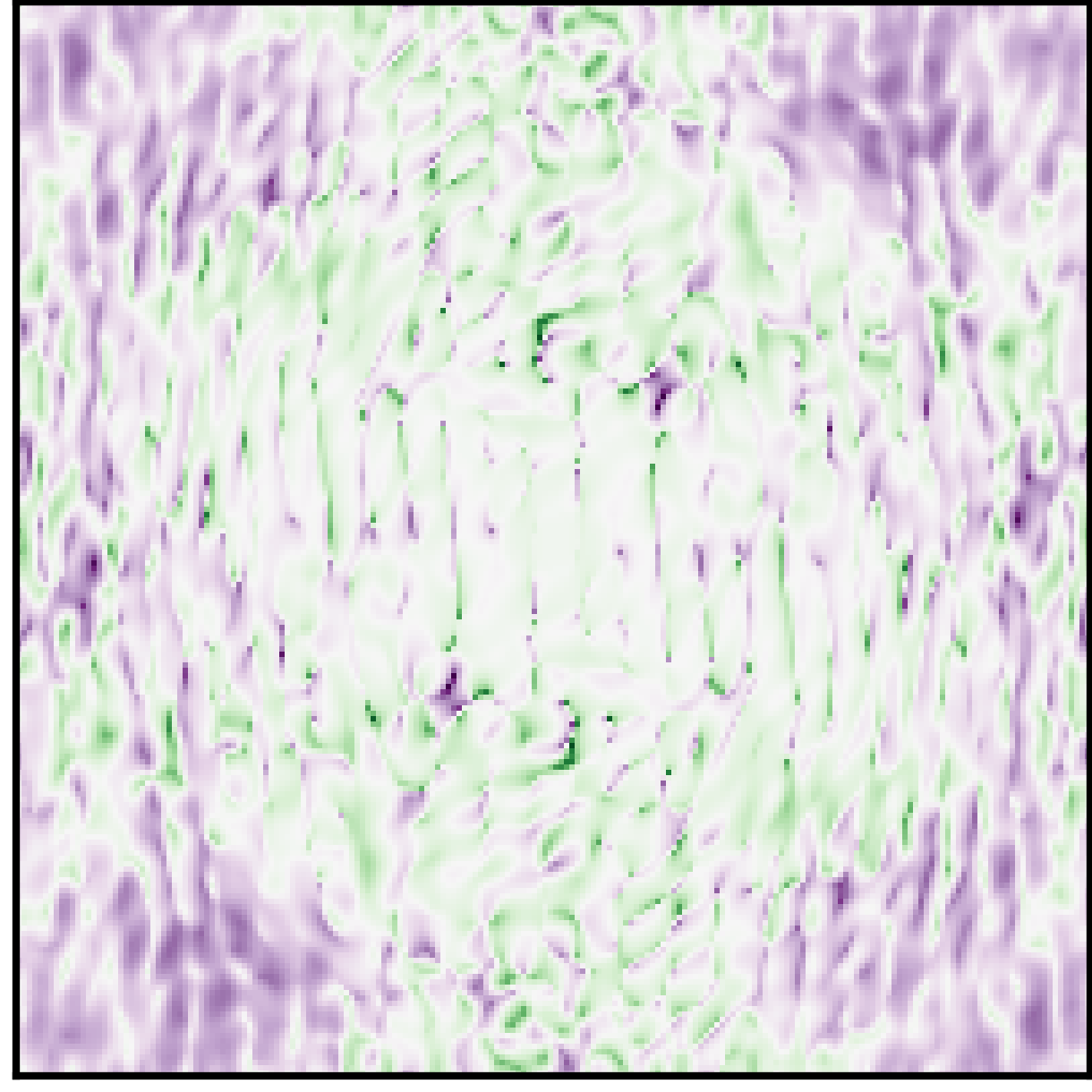} &
    \includegraphics[width=0.23\linewidth]{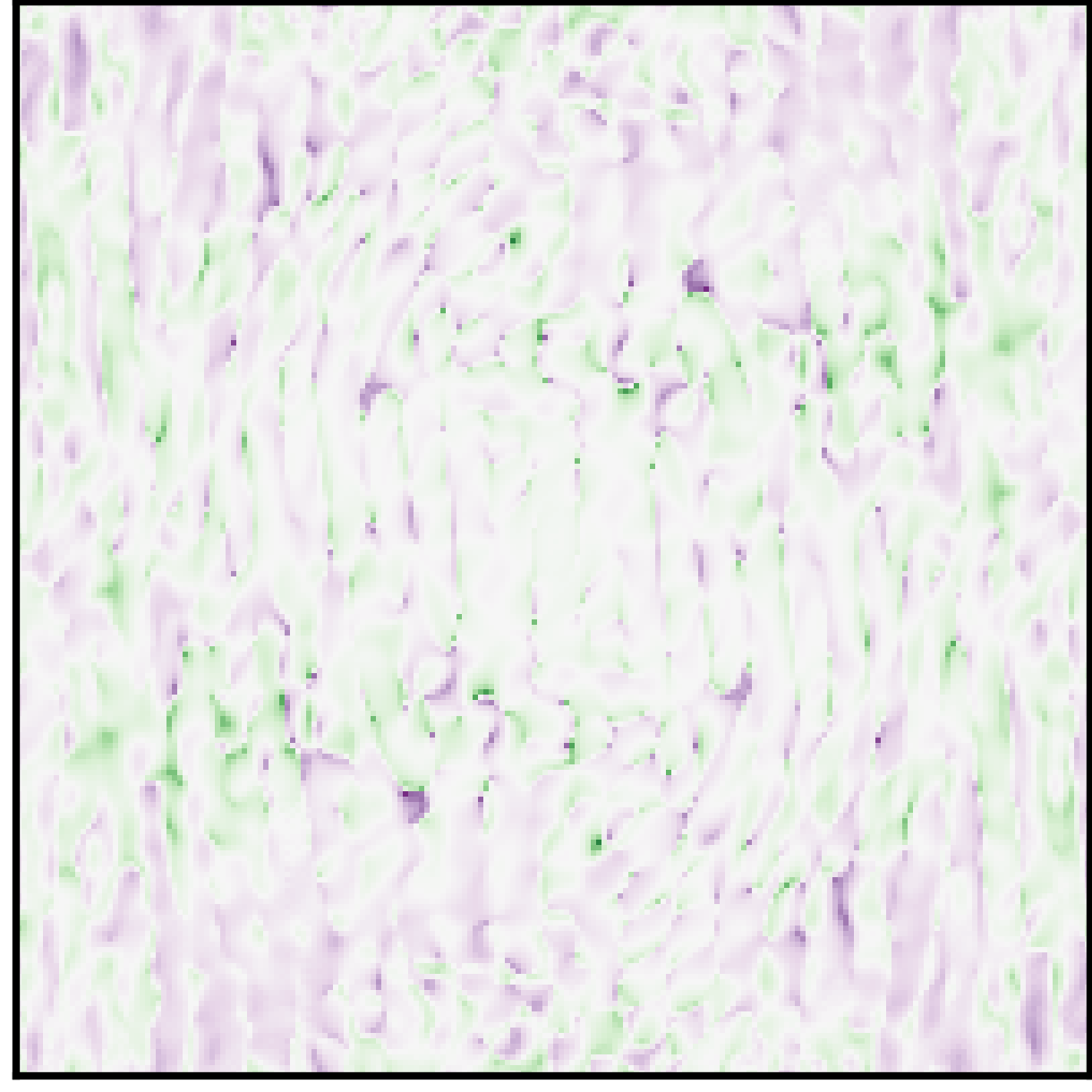} \\
\end{tabular}
\caption{Visual results of the ablation study for different modules. From top to bottom, the rows display the predicted SUV images, spatial error maps, frequency spectra, and frequency domain errors, respectively.}
\label{fig:ablation_axial_12555888}
\end{figure}
\begin{figure}[htbp]
    \centering
    \includegraphics[width=0.8\linewidth]{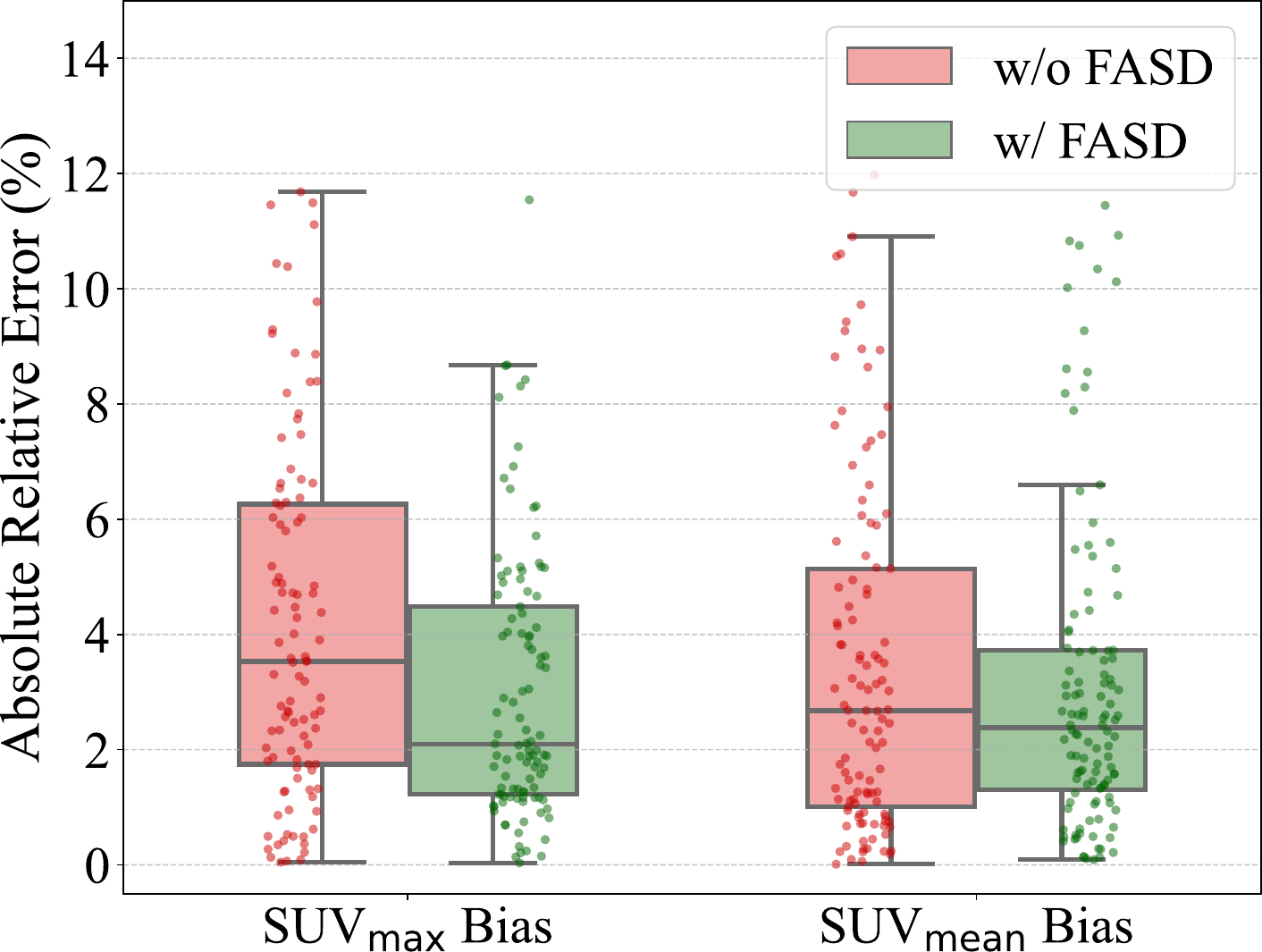}
    \caption{Lesion-specific quantitative ablation analysis. The bar chart compares the absolute relative error (\%) of tumor $\text{SUV}_{\text{max}}$ and $\text{SUV}_{\text{mean}}$ between the model without the frequency-domain branch (w/o FASD) and the full model (w/ FASD). }
    \label{fig:ablation_lesion}
\end{figure}

\begin{table}[htbp]
\centering
\caption{Region-wise absolute SUV Bias (\%) within different regions.}
\label{tab:tissue_bias}
\resizebox{\linewidth}{!}{
\begin{tabular}{lcccccc}
\toprule
Tissue Type & Pix2pix & CycleGAN & IVNAC & DBDL & MambaIR & GPCN (Ours) \\
\midrule
Bone        & 7.52  & 17.10 & 17.26 & 8.25 & 7.36  & \textbf{6.49} \\
Lung        & 10.50 & 15.43 & 56.85 & 7.02 & 13.02 & \textbf{2.00} \\
Soft Tissue & 2.88  & 10.09 & 5.53  & 3.57 & 2.26  & \textbf{0.65} \\
\bottomrule
\end{tabular}
}
\end{table}

\begin{figure}[htbp]
    \centering
    \includegraphics[width=\linewidth]{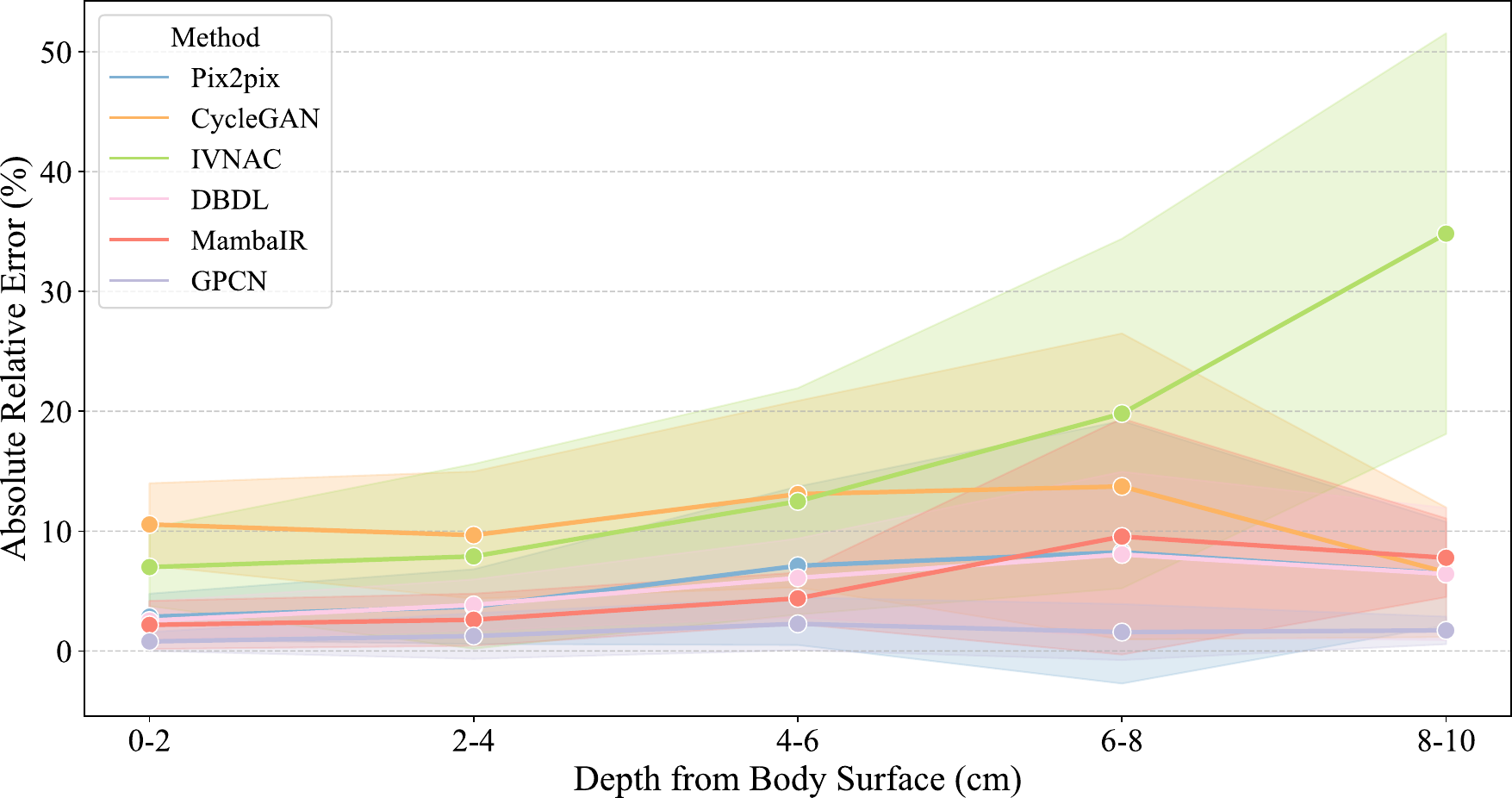}
    \caption{Absolute relative SUV error as a function of voxel depth from the patient surface. Solid lines indicate the mean error, while shaded regions represent the variance across the test set.}
    \label{fig:depth_bias}
\end{figure}

The quantitative results in Table~\ref{tab:ablation_study} demonstrate the necessity of both modules. Removing MBCR substantially worsened all metrics, especially MAE and NMSE, indicating that the Fourier-domain branch alone cannot model the complex correction mapping. Likewise, removing FASD reduced PSNR and increased errors, underscoring the importance of position-guided, physically grounded frequency-domain correction. The complete GPCN consistently outperformed both ablated variants, highlighting the complementary roles of both modules.

Fig.~\ref{fig:ablation_axial_12555888} visually corroborates these quantitative findings. The variant without MBCR suffers from severe spatial distortions and anatomically implausible artifacts. Its highly structured error map confirms the indispensability of multi-band contextual modeling for preserving long-range topological coherence. Conversely, while the model without FASD appears macroscopically acceptable at first glance in the spatial domain, its spatial error map reveals clear residual inaccuracies and degraded local contrast. More importantly, its corresponding frequency spectrum residual maps explicitly expose pronounced errors in the high-frequency regions of both amplitude and phase. In contrast, the full GPCN produces remarkably clean error profiles in both the spatial and spectral domains. This directly evidences the effectiveness of our synergistic dual-domain design, which strictly enforces both spatial structural fidelity and spectral quantitative precision.

To explicitly validate the rationale for the frequency-domain branch, we performed a targeted lesion-specific analysis. Focal tumors exhibit high-frequency characteristics that are easily over-smoothed by purely spatial networks. As shown in Fig. \ref{fig:ablation_lesion}, the model lacking FASD (w/o FASD) systematically underestimates tracer uptake, yielding high relative errors for tumor $\text{SUV}_{\text{max}}$ and $\text{SUV}_{\text{mean}}$. Incorporating the FASD module drastically reduces these errors. This confirms that while the spatial branch captures the anatomical backbone, the explicit phase and amplitude recovery in FASD is indispensable for preserving the quantitative integrity of high-frequency focal lesions.

\subsection{Physical Validation of Attenuation Behavior}
To rigorously substantiate that our proposed framework learns a fundamental physical mapping rather than a purely data-driven appearance translation, we conducted a dedicated physical validation assessing tissue-specific and path-length-dependent attenuation behaviors. 

In PET imaging, photon attenuation is inherently density-dependent. Thus, we first extracted regions of interest (ROIs) across three distinct anatomical areas representing extremes of density variations: lung (low density), soft tissue (water equivalent), and bone (high density). As summarized in Table \ref{tab:tissue_bias}, conventional generative baselines struggle significantly to adapt to varying physical densities. Models such as CycleGAN and IVNAC exhibit severe density-dependent deviations, with IVNAC failing catastrophically in the low-density lung region and CycleGAN yielding high errors in highly attenuating bone structures. While recent methods like DBDL and Pix2pix show relative improvements, they still suffer from large fluctuations across different tissues. 

In stark contrast, GPCN achieves the absolute lowest bias across all evaluated tissue types. Remarkably, our framework constrains the relative error to merely 0.65\% in soft tissue and 2.00\% in the lung, while also successfully managing the extreme attenuation of bone with a minimal bias of 6.49\%. This consistent quantitative accuracy across tissue types suggests that GPCN captures attenuation-related regularities more reliably than the compared baselines. While these results do not prove that the network explicitly models PET physics, they indicate that its learned correction is more compatible with known attenuation behavior across different densities.

Furthermore, following the Beer-Lambert law, physical attenuation increases exponentially with the photon path length through tissue. To evaluate whether the networks internalized this principle, we analyzed the absolute relative SUV error as a function of voxel depth from the patient surface. As illustrated in Fig. \ref{fig:depth_bias}, traditional DL methods such as CycleGAN and IVNAC demonstrate a strong depth-dependent bias. While their errors are marginal near the surface, they escalate exponentially in deep-seated regions, indicating a failure to account for path-length attenuation. In stark contrast, GPCN yields a highly stable, flat error profile that remains robust regardless of depth. The narrow variance (shaded region) further confirms that GPCN consistently and accurately compensates for cumulative photon attenuation in deep tissues, a hallmark of a highly robust quantitative correction model.

\subsection{Validation on Downstream Segmentation Task}
To further assess whether the corrected PET volumes preserve anatomically relevant information, we performed a downstream organ and tumor segmentation experiment using SegAnyPET~\cite{zhang2025seganypet}, a recent state-of-the-art foundation model for universal promptable segmentation in molecular imaging. Pre-trained on the massive PETS-5k dataset (5,700 whole-body PET scans), SegAnyPET effectively captures the characteristic noise properties, resolution constraints, and physiological uptake patterns of PET, and its 3D encoder–decoder architecture fully exploits volumetric spatial context. In our evaluation, we used SegAnyPET in prompt-guided mode to delineate key organs (liver and spleen) and tumor lesions. To quantitatively evaluate the segmentation accuracy, we employed three metrics: the dice similarity coefficient (Dice) to measure volumetric overlap, alongside the 95th percentile hausdorff distance (HD95) and average symmetric surface distance (ASSD) to rigorously assess boundary preservation and surface geometric fidelity. 
\paragraph{Organ Segmentation}
\begin{table}[!t] 
\centering
\scriptsize 
\setlength{\tabcolsep}{2pt} 
\renewcommand\arraystretch{1.1}
\caption{Quantitative comparison of segmentation performance on Liver, Spleen, and Tumor.}
\label{tab:seg}
\begin{tabular}{lccccccccc}
\toprule
\multirow{2}{*}{Method} & \multicolumn{3}{c}{Liver} & \multicolumn{3}{c}{Spleen} & \multicolumn{3}{c}{Tumor} \\
\cmidrule(lr){2-4} \cmidrule(lr){5-7} \cmidrule(lr){8-10}
 & Dice$\uparrow$ & HD95$\downarrow$ & ASSD$\downarrow$ & Dice$\uparrow$ & HD95$\downarrow$ & ASSD$\downarrow$ & Dice$\uparrow$ & HD95$\downarrow$ & ASSD$\downarrow$ \\
\midrule
Pix2Pix  & 0.789 & 22.289 & 7.398 & 0.195 & 6.356 & 3.241 & 0.904 & 4.073 & 2.232 \\
CycleGAN & 0.885 & 15.788 & 4.565 & 0.370 & 32.876 & 10.578 & 0.833 & 8.146 & 3.328 \\
IVNAC    & 0.686 & 23.264 & 8.986 & 0.167 & 16.291 & 4.487 & 0.859 & 5.760 & 2.984 \\
DBDL     & 0.885 & 12.280 & 3.944 & 0.274 & 10.095 & 4.680 & 0.871 & 4.073 & 2.787 \\
MambaIR  & 0.916 & 8.690 & 3.080 & 0.528 & 13.519 & 3.659 & 0.919 & 4.073 & 1.930 \\
GPCN  & \textbf{0.940} & \textbf{6.531} & \textbf{2.154} & \textbf{0.808} & \textbf{5.720} & \textbf{2.208} & \textbf{0.924} & \textbf{4.073} & \textbf{1.745} \\
\bottomrule
\end{tabular}
\end{table}
\begin{figure}[!t]
    \centering
    \scriptsize
    \setlength{\tabcolsep}{1pt}

    \begin{tabular}{cccccccc}
        Prompts & Pix2Pix & CycleGAN & IVNAC & DBDL& MambaIR & GPCN & ASC \\
        \includegraphics[width=0.115\linewidth]{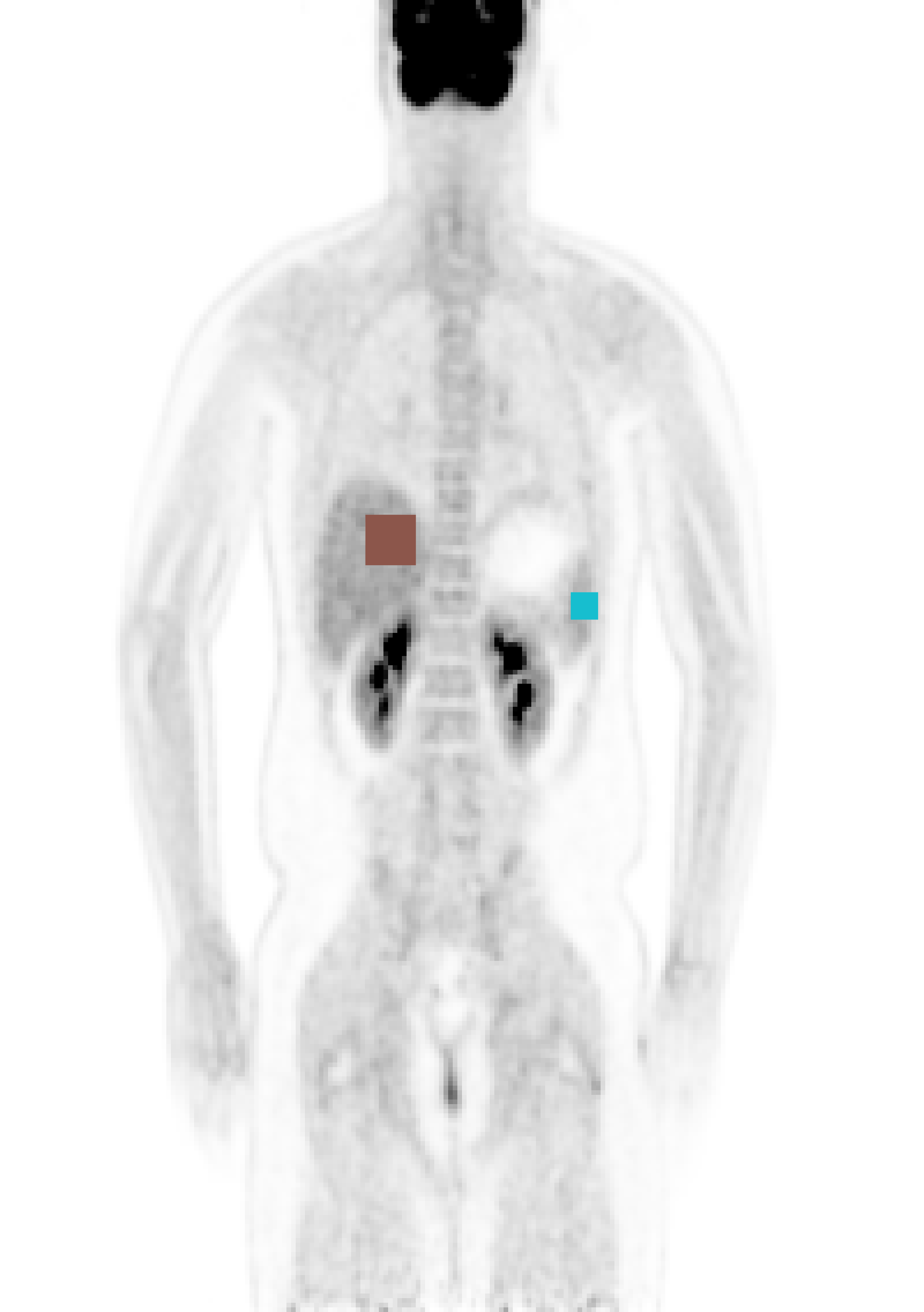} &
        \includegraphics[width=0.115\linewidth]{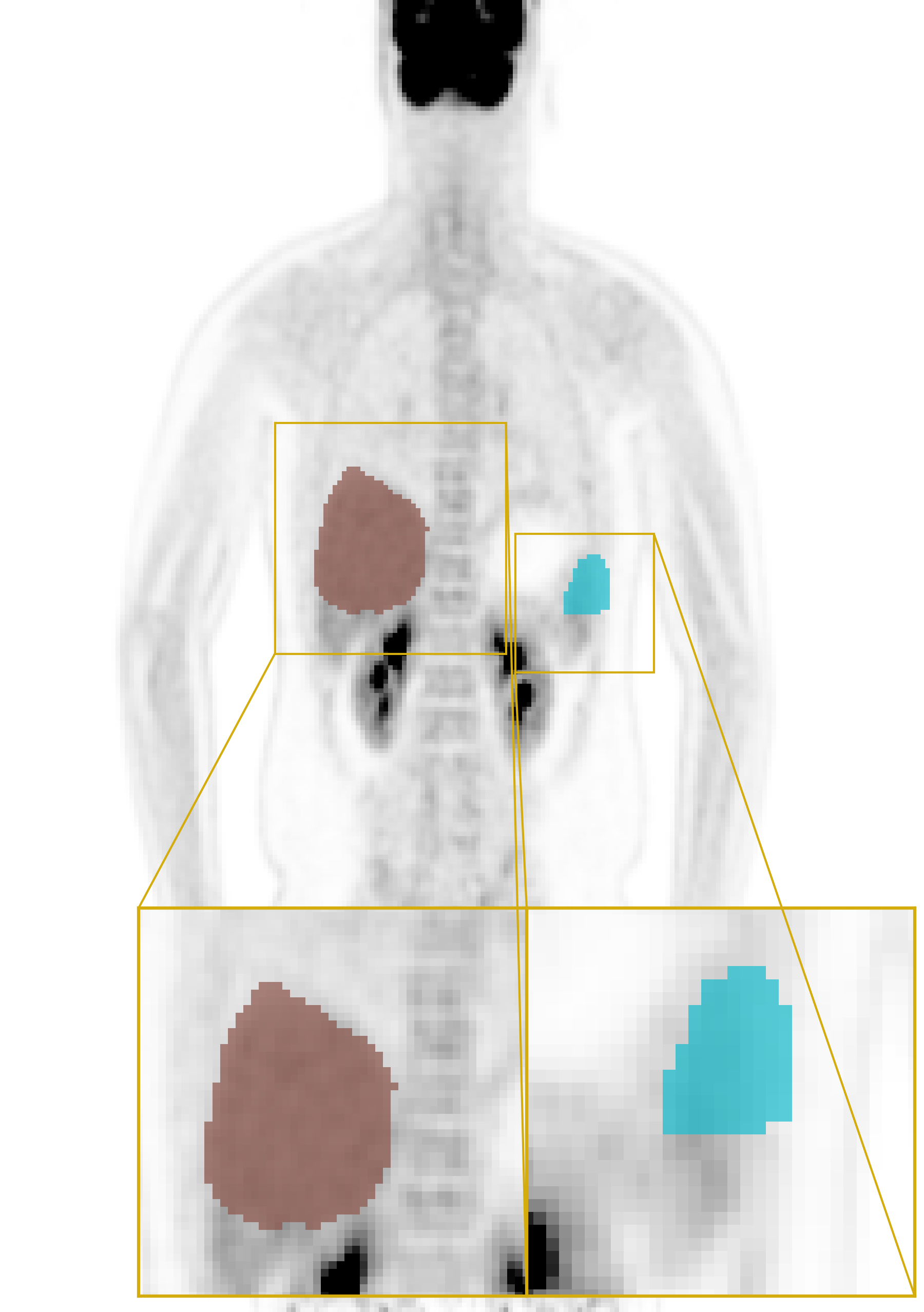} &
        \includegraphics[width=0.115\linewidth]{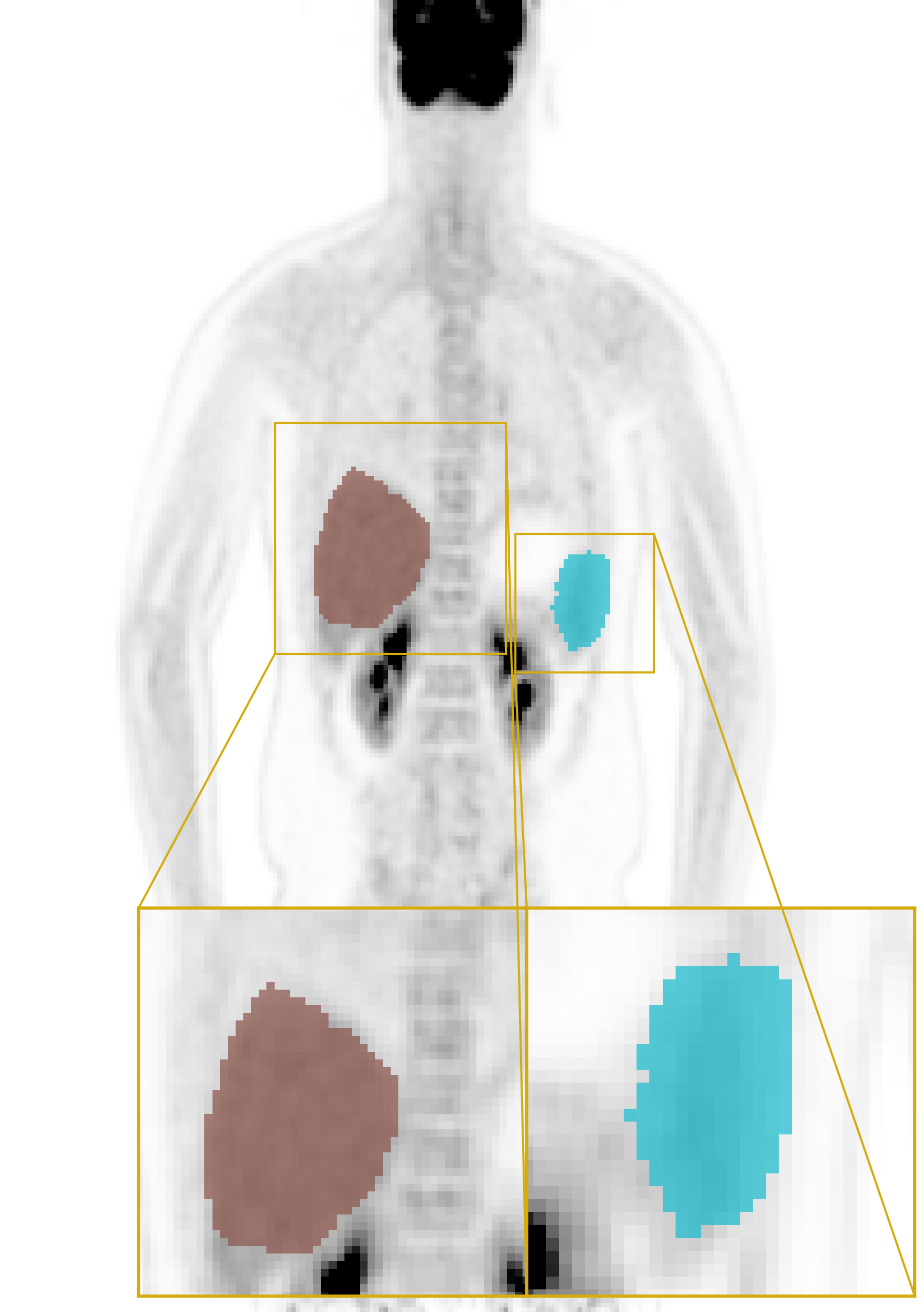} &
        \includegraphics[width=0.115\linewidth]{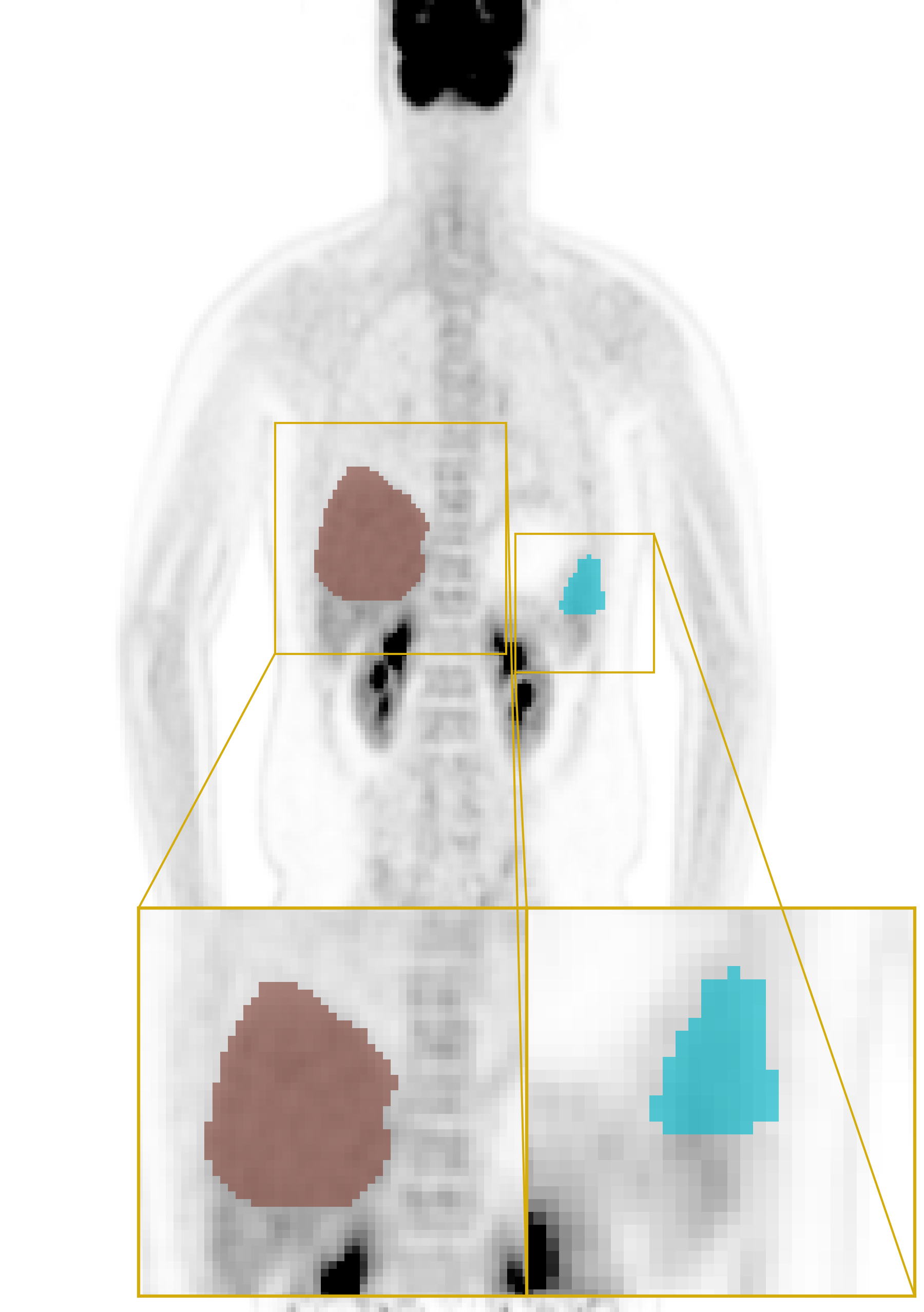} &
        \includegraphics[width=0.115\linewidth]{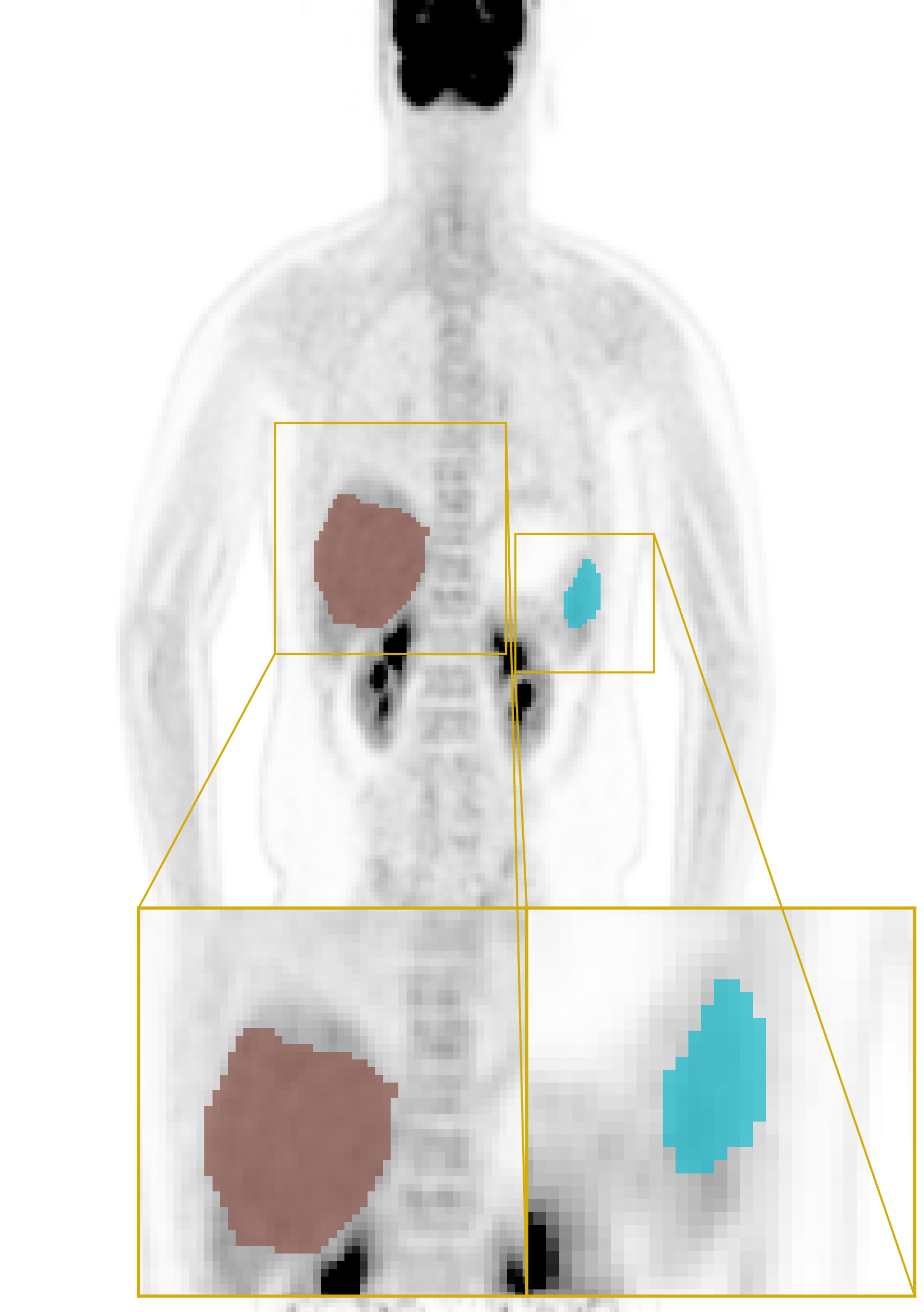} &
        \includegraphics[width=0.115\linewidth]{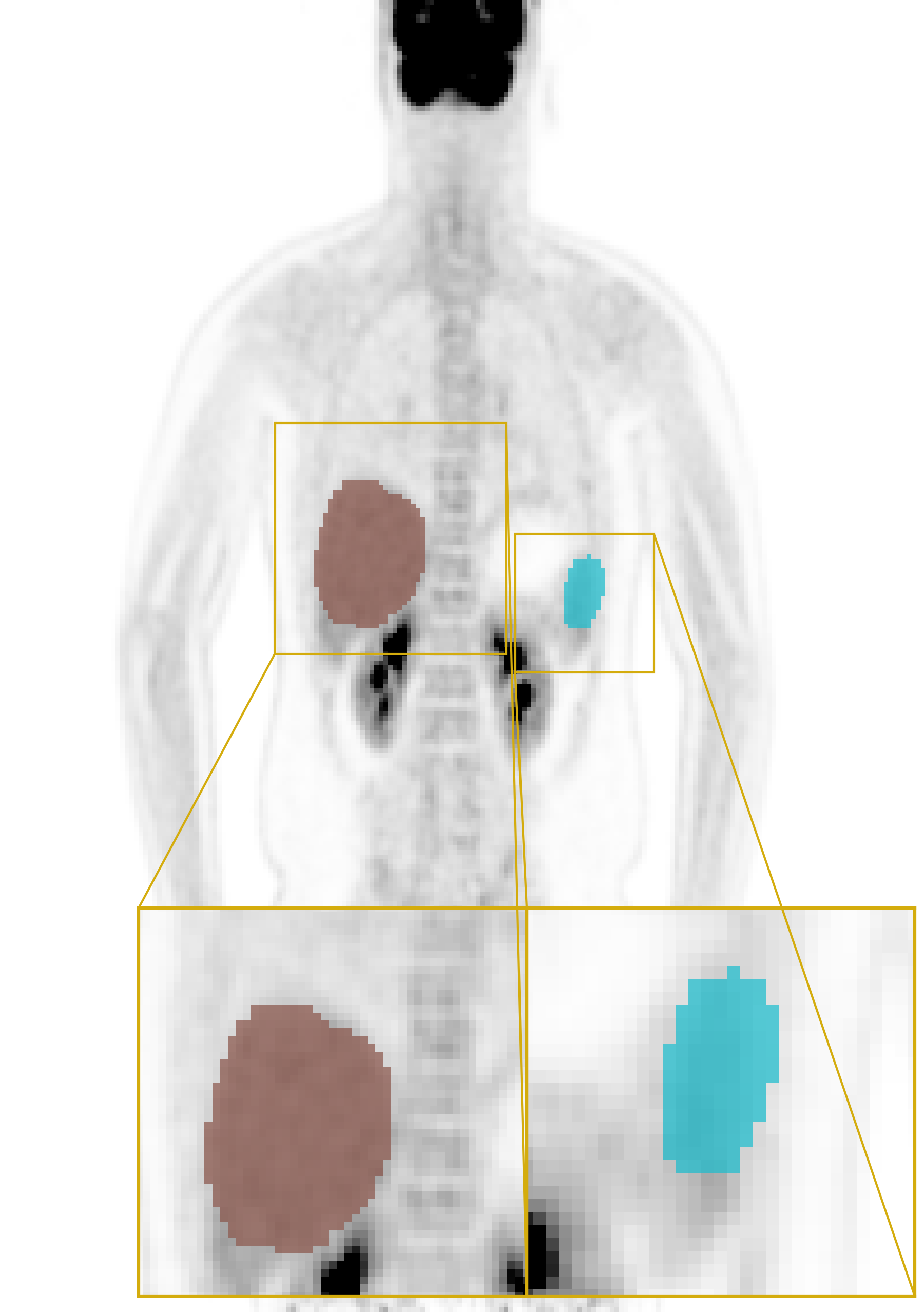} &
        \includegraphics[width=0.115\linewidth]{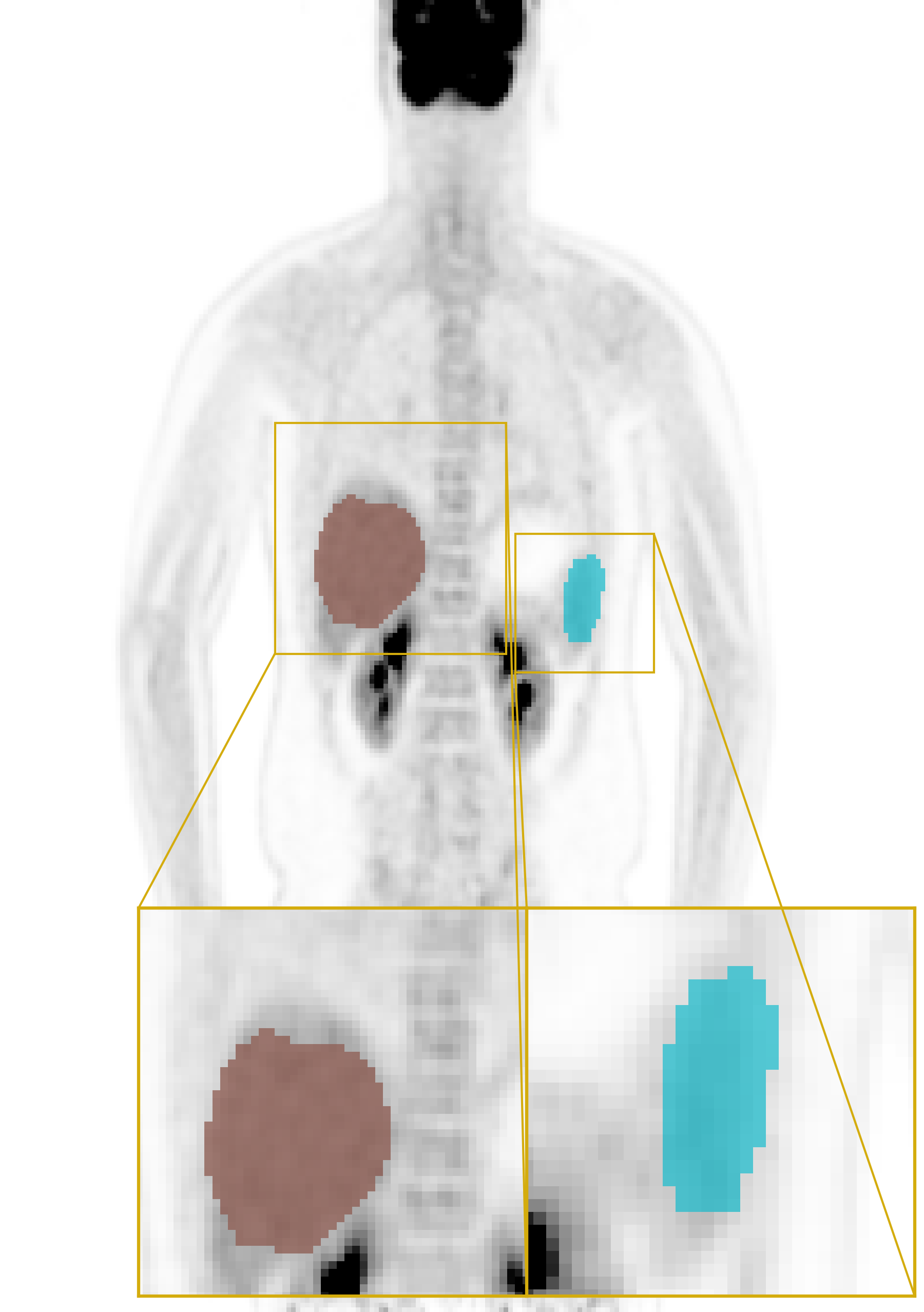} &
        \includegraphics[width=0.115\linewidth]{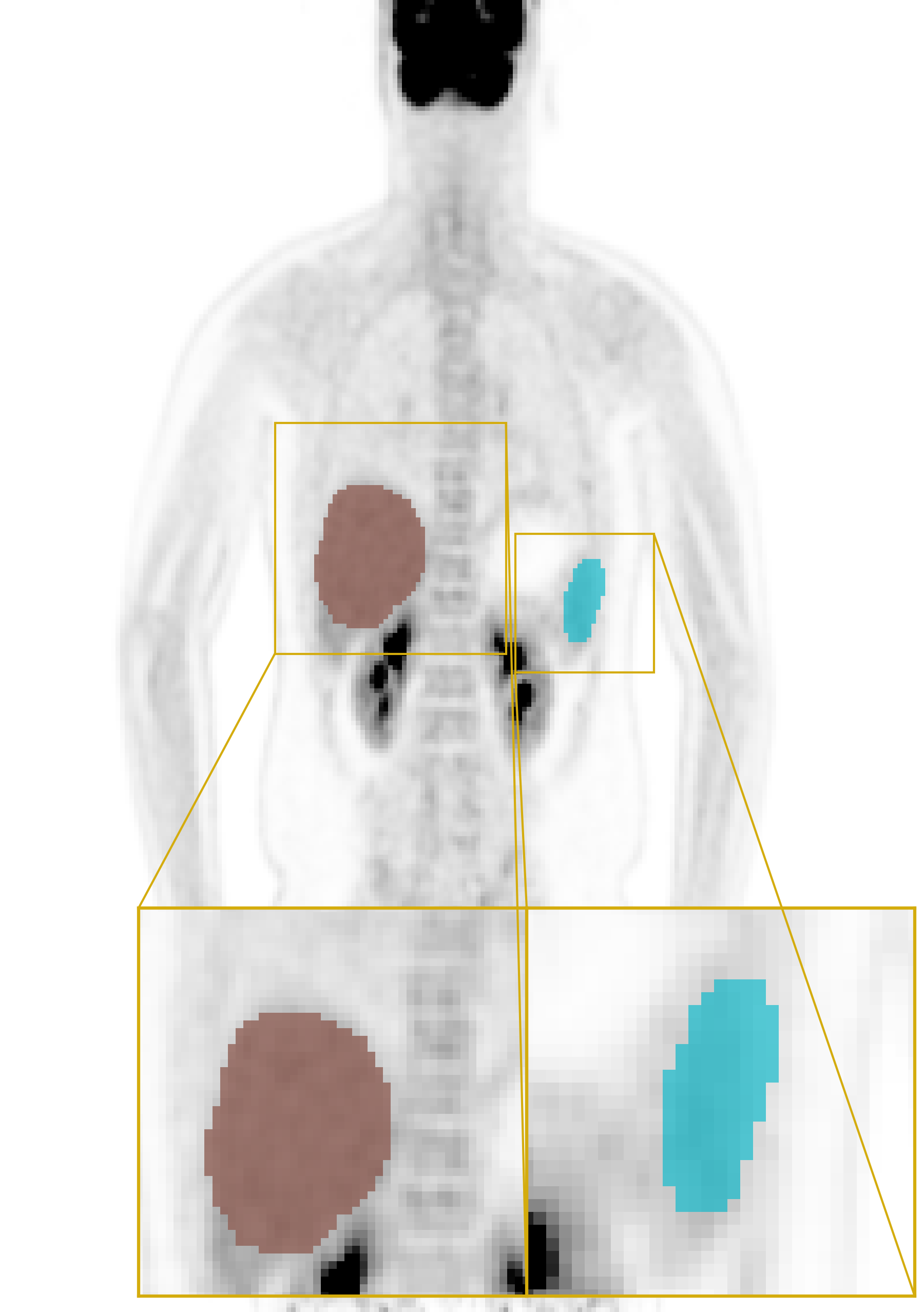} \\
    \end{tabular}
    \begin{tabular}{cccccccc}
        \includegraphics[width=0.115\linewidth]{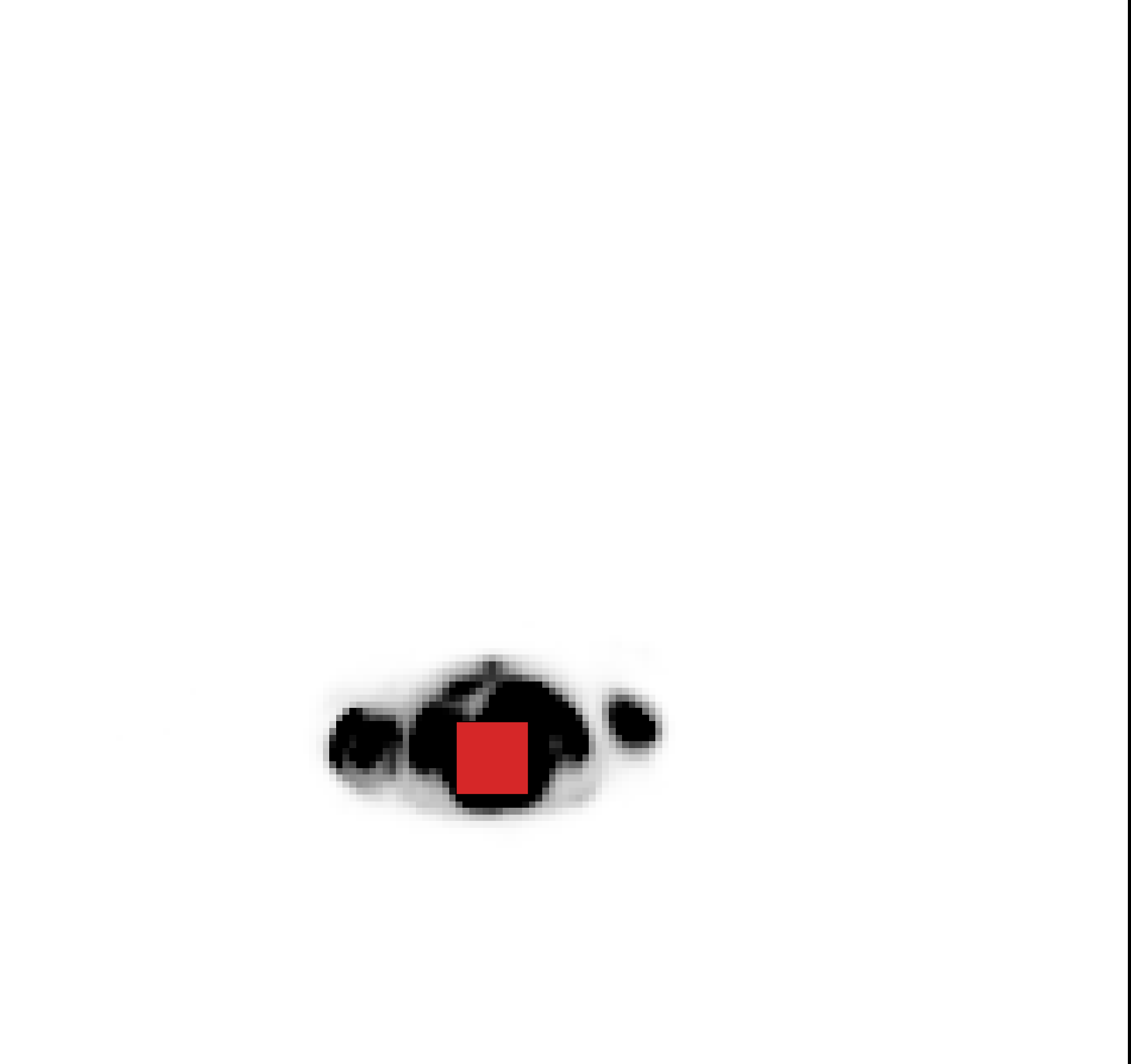} &
        \includegraphics[width=0.115\linewidth]{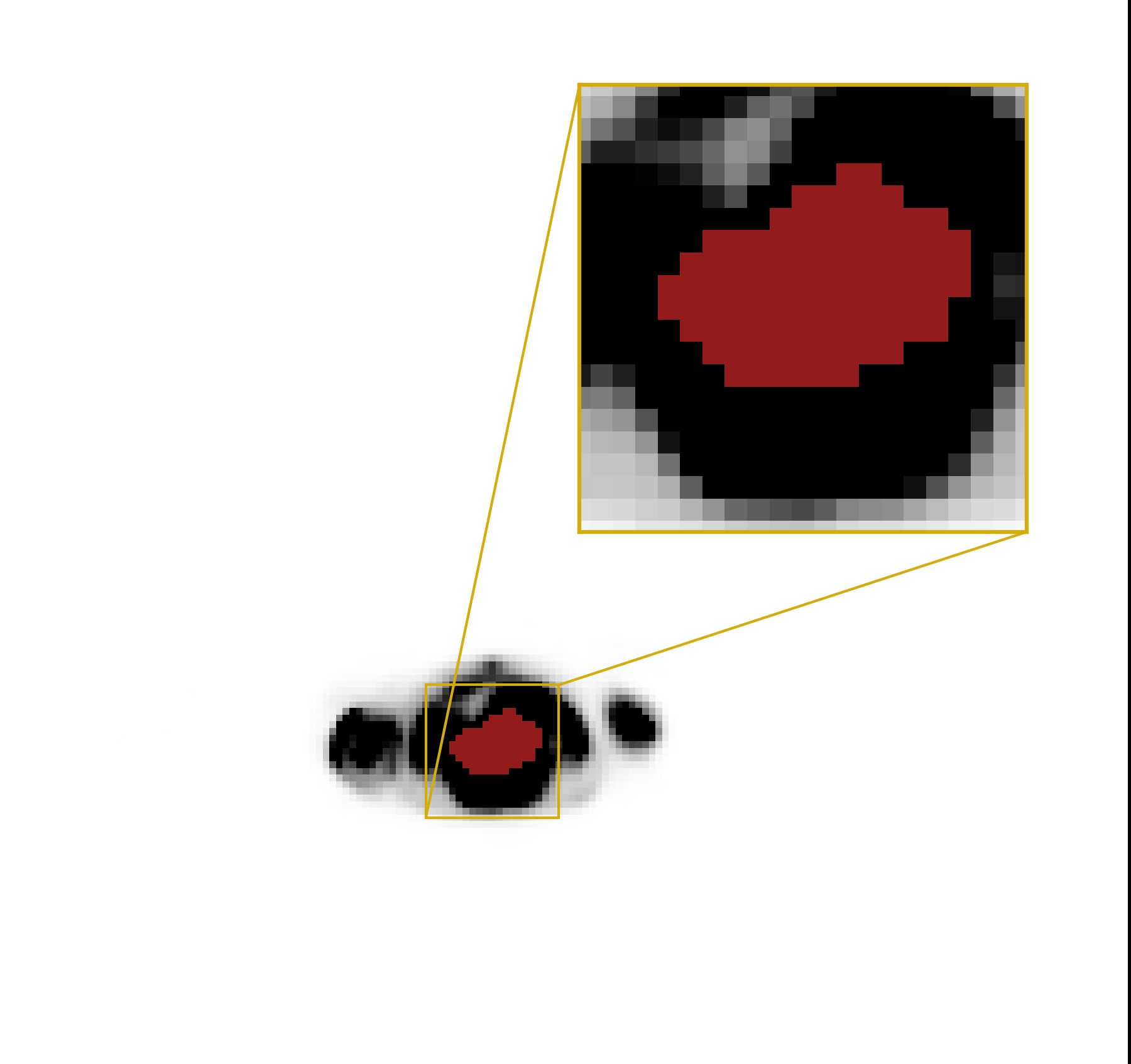} &
        \includegraphics[width=0.115\linewidth]{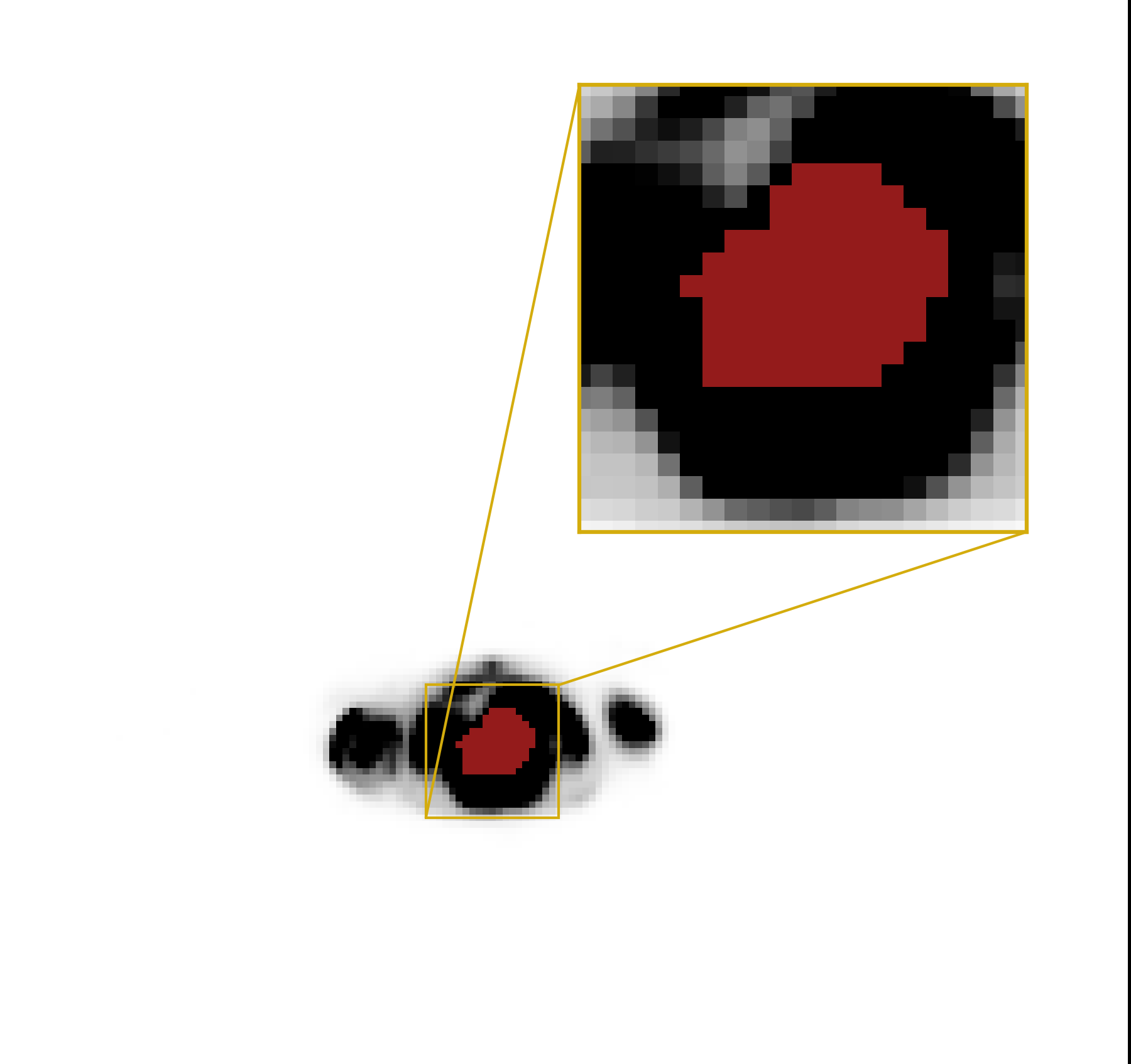} &
        \includegraphics[width=0.115\linewidth]{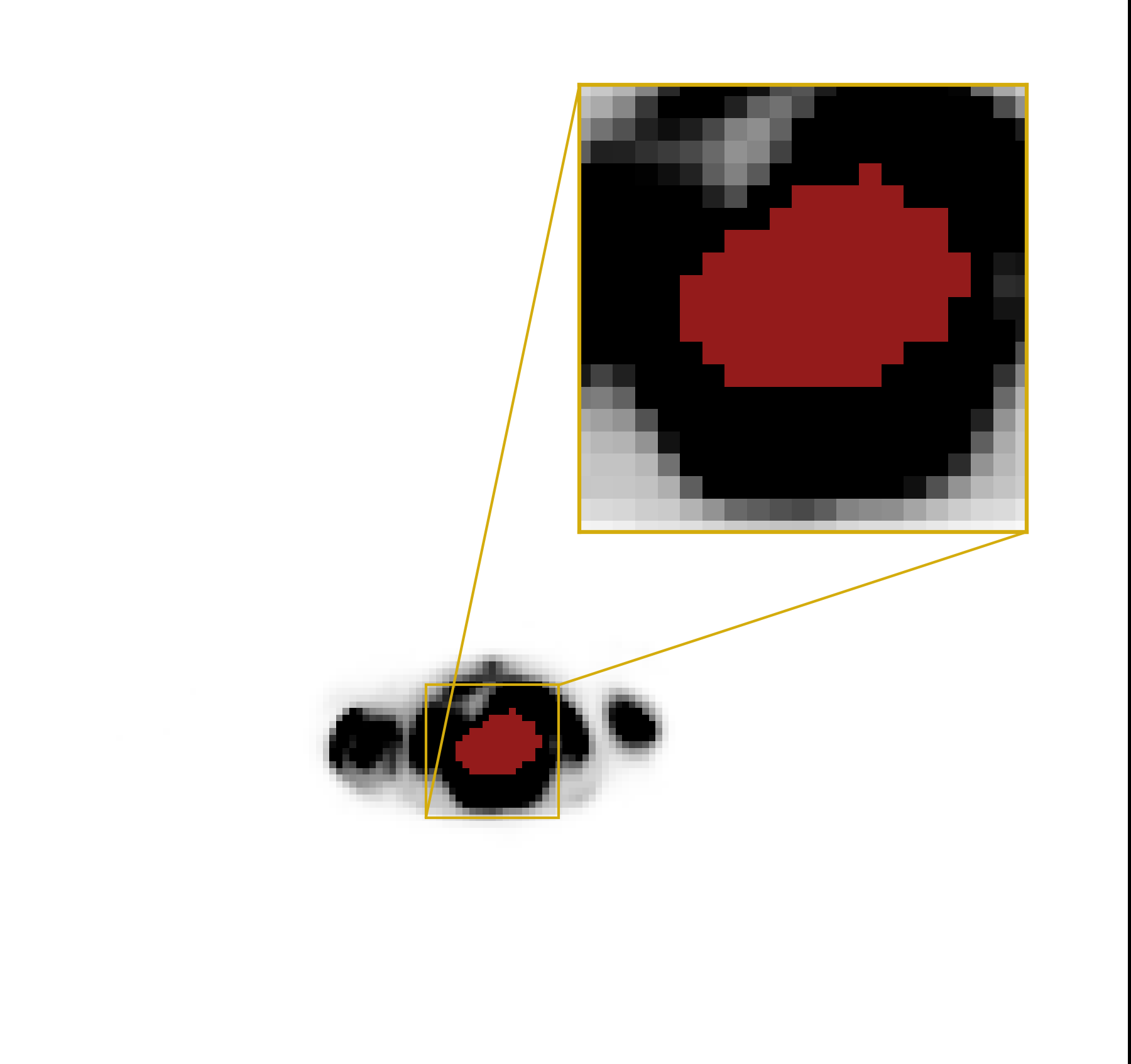} &
        \includegraphics[width=0.115\linewidth]{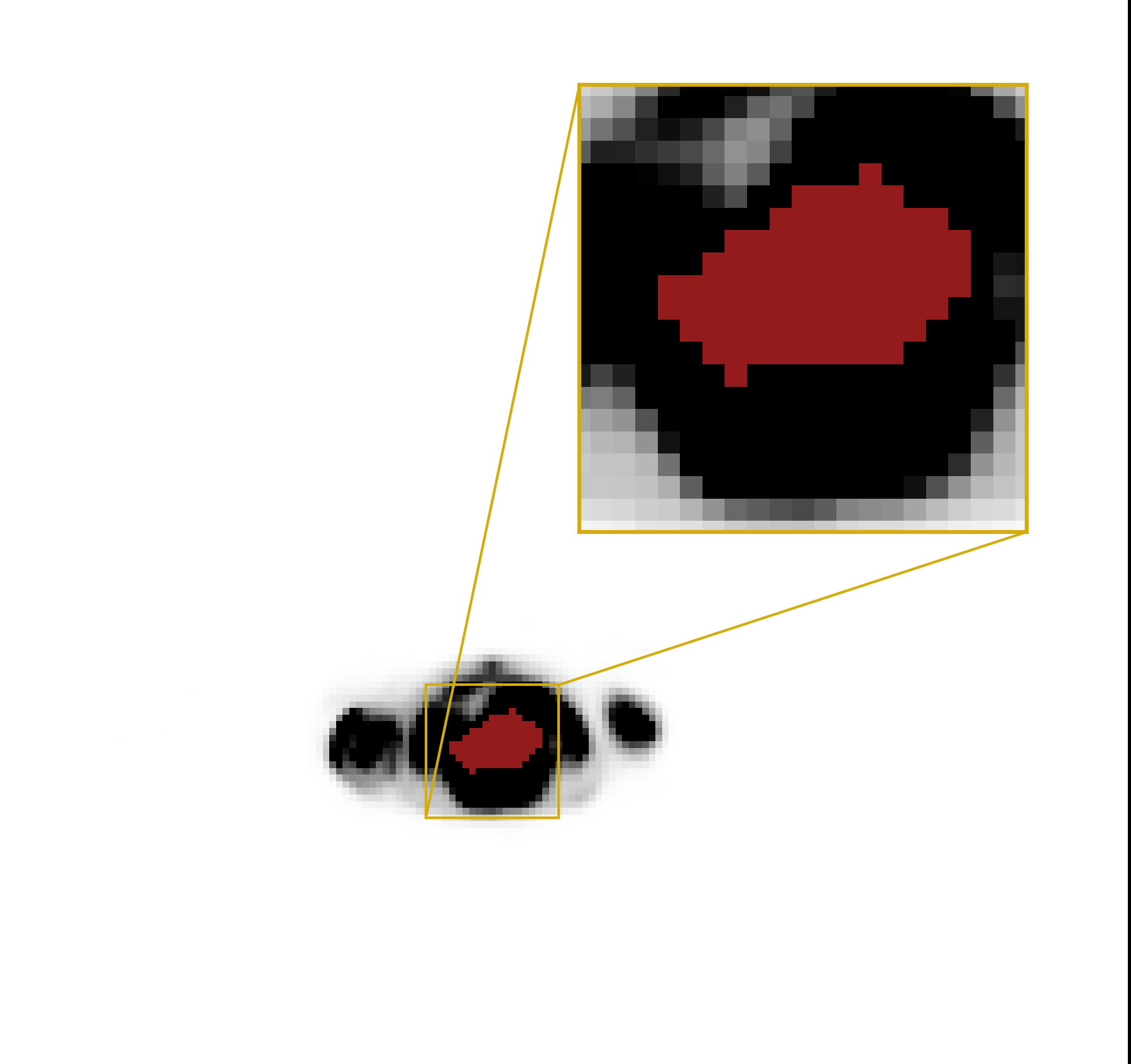} &
         \includegraphics[width=0.115\linewidth]{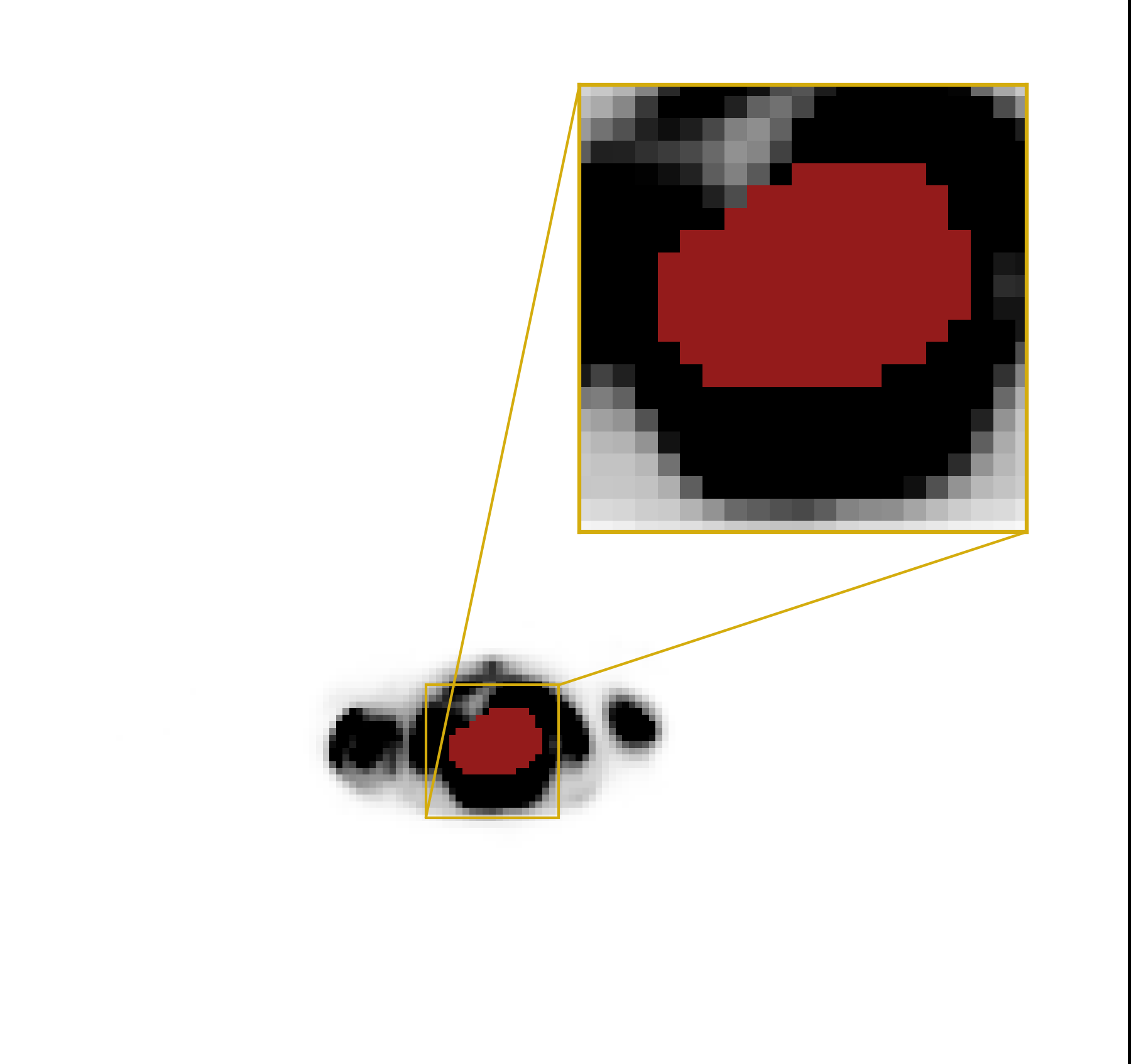} &
      \includegraphics[width=0.115\linewidth]{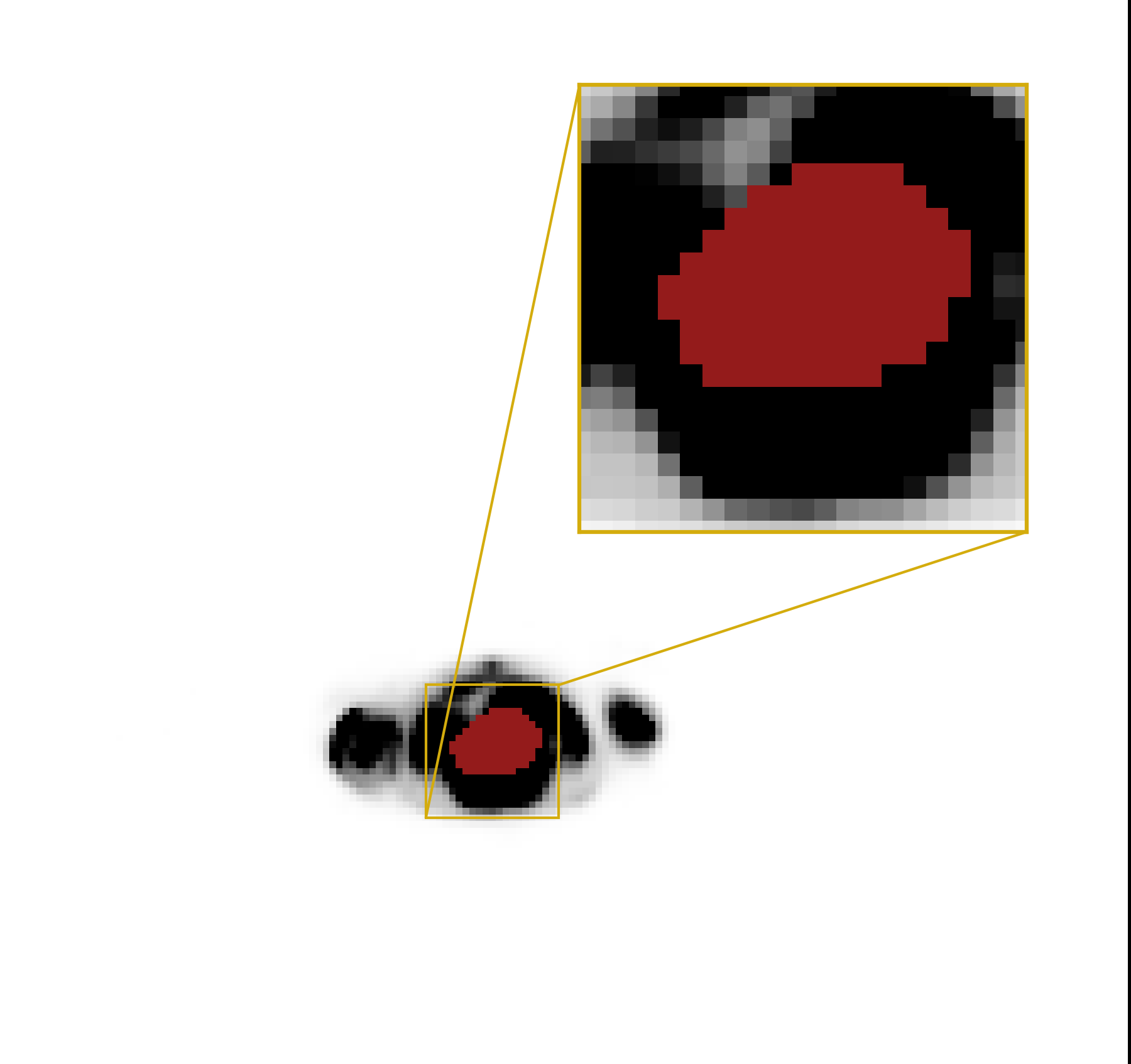} &
        \includegraphics[width=0.115\linewidth]{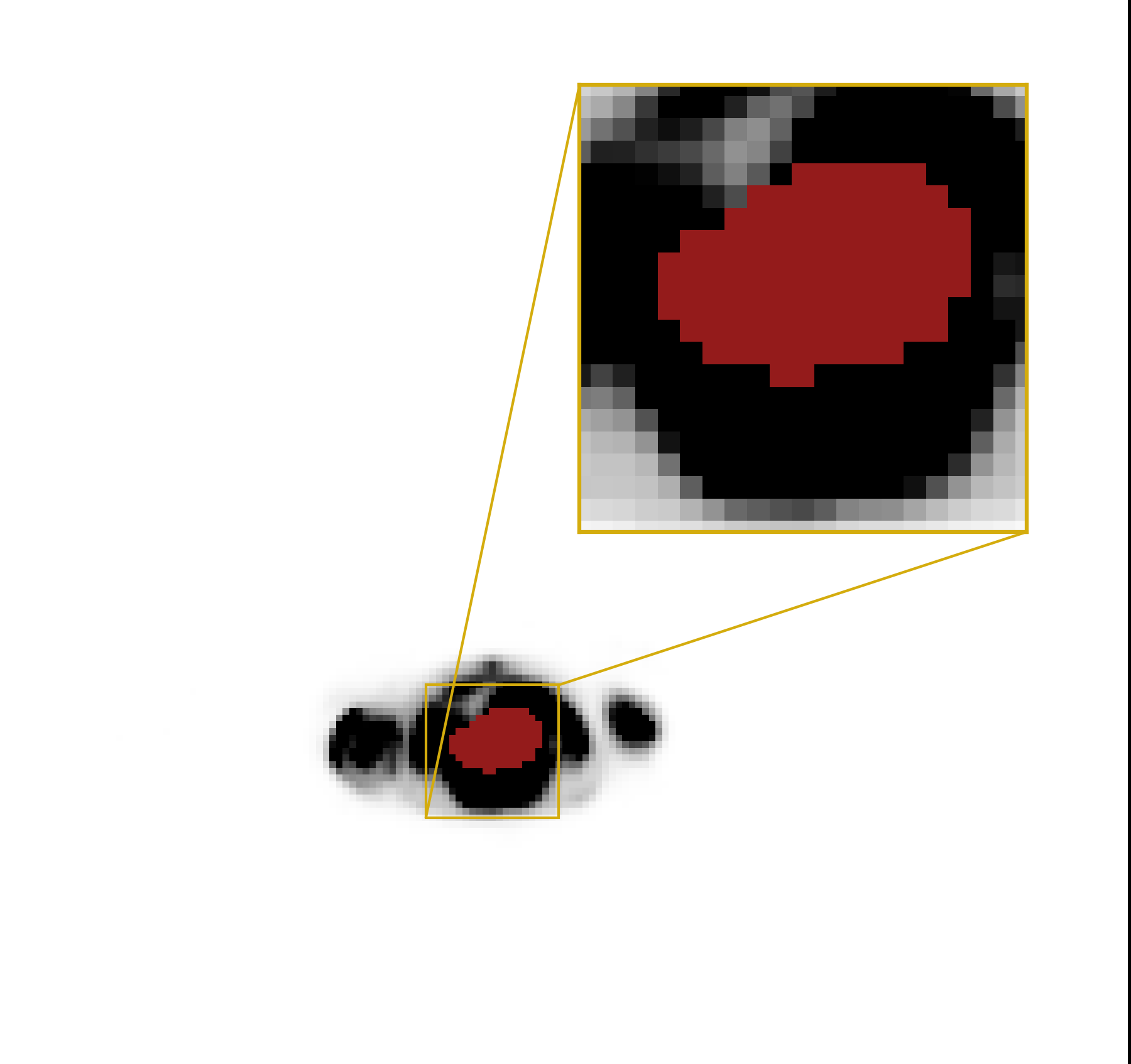} \\
    \end{tabular}

    \caption{Visual comparison of the downstream segmentation tasks. {Top row:} Organ segmentation, where Brown represents the liver and Cyan represents the spleen. {Bottom row:} Tumor segmentation.}
    \label{fig:visual_seg_combined}
\end{figure}
While conventional metrics such as PSNR and SSIM provide a global assessment of reconstruction quality, they do not fully capture the preservation of local anatomical boundaries critical for clinical decision-making. To more directly validate the clinical utility of our images, we performed a downstream automated segmentation task for the liver and spleen, using a prompt-guided framework to delineate these organs on PET images corrected by GPCN and competing methods, with segmentations from ground-truth ASC-PET serving as the reference standard.

As shown in Table~\ref{tab:seg}, GPCN achieves state-of-the-art performance across all metrics, demonstrating superior structural preservation over baseline methods. For the liver, GPCN attains the highest Dice score of 0.940 and the lowest 95\% Hausdorff distance (HD95) of 6.531 mm. For the spleen, an inherently challenging organ due to its variable size and complex boundaries, GPCN reduces HD95 to 5.720 mm, markedly lower than competing approaches such as CycleGAN and IVNAC, which exhibit substantially larger boundary errors.

Fig.~\ref{fig:visual_seg_combined} shows organ segmentation results for all methods. In the zoomed-in views, baselines struggle with boundary definition: Pix2Pix exhibits severe under-segmentation, missing the tip of the spleen, and CycleGAN introduces artifacts that distort organ shape. By contrast, GPCN segmentations align closely with the ASC image references, effectively mitigating spatial distortions and edge blurring seen in other CT-free correction methods and preserving high geometric fidelity suitable for downstream clinical applications.
\paragraph{Tumor Segmentation}
Table~\ref{tab:seg} presents the quantitative results for all compared methods. GPCN achieves outstanding performance on this challenging task, with a state-of-the-art Dice score of 0.924, representing a clear improvement over competing methods such as CycleGAN and IVNAC. Fig.~\ref{fig:visual_seg_combined} shows the corresponding tumor segmentation results: competing methods often fail to accurately delineate tumor margins, leading to under-segmentation or irregular boundary artifacts, whereas the segmentation produced by GPCN closely matches the ASC ground truth. This high fidelity in preserving lesion morphology indicates that GPCN can recover critical diagnostic information without the structural distortion commonly seen in conventional generative approaches, providing a reliable foundation for tumor detection and analysis.
\subsection{Quantification of avoided radiation dose with GPCN}
\begin{table}[!t]
  \centering
  \setlength{\tabcolsep}{4pt}
  \caption{Summary of effective dose savings provided by the GPCN framework across different radiotracers.}
  \label{tab:dose_savings}
  \begin{tabular}{lcccc}
    \toprule
    & {Mean} (mSv) & {Median} (mSv) & {Min} (mSv) & {Max} (mSv) \\
    \midrule
    \textbf{Average} & 10.8 & 11.46 & 3.47 & 18.56 \\
    \bottomrule
  \end{tabular}
\end{table}
\label{sec:dose_calculation}
To quantify the radiation burden eliminated by our GPCN framework, the effective dose ($E$) from the CT component was calculated for patients scanned on the Siemens Biograph mCT system. The calculation followed the standardized, internationally recognized methodology \cite{shrimpton2004assessment}.

The process involves two primary steps. First, the dose-length product (DLP) was determined by multiplying the volume CT dose index (CTDI$_{\text{vol}}$) by the total scan length ($L$). Both parameters were derived from the scan's DICOM data.
\begin{equation}
\text{DLP (mGy}\cdot\text{cm)} = \text{CTDI}_{\text{vol}} \text{ (mGy)} \times L \text{ (cm)}
\end{equation}
Second, the DLP, a measure of absorbed dose, was converted to the effective dose ($E$), which accounts for whole-body radiation risk. This was achieved using age-specific conversion factors ($k$), reported in milliSieverts (mSv).
\begin{equation}
E \text{ (mSv)} = \text{DLP (mGy}\cdot\text{cm)} \times k \text{ (mSv/(mGy}\cdot\text{cm))}
\end{equation}
The methodology for calculating effective dose from CT parameters is consistent with the framework established by international bodies such as the ICRP and applied in practice-oriented guidelines like AAPM Report No. 204 \cite{boone2011size}. The specific age- and region-dependent conversion factors ($k$) used in this study were adopted from the widely recognized values for pediatric examinations published in the literature \cite{shrimpton2006national}.

The calculated radiation savings are summarized in Table~\ref{tab:dose_savings}. By eliminating the need for a CT scan for attenuation correction, GPCN avoids an average effective dose of 10.8~mSv per scan. This substantial reduction highlights the profound clinical value of our method in adhering to the ALARA principle, offering a safer molecular imaging pathway for pediatric patients.
\section{Discussion}
The lack of generalizability has long hampered the clinical translation of deep learning for CT-free pediatric PET \cite{finlayson2021clinician,guo2022using}. We addressed this by validating our generalizable PET correction network (GPCN) on an unprecedentedly large and diverse dataset. GPCN not only achieves state-of-the-art accuracy but also exhibits consistent robustness across multiple scanners and unseen radiotracers. By eliminating an average effective dose of 10.8 mSv per scan (Table~\ref{tab:dose_savings}), our framework offers a safer, generalizable solution for longitudinal monitoring in pediatric oncology, strictly adhering to the ALARA principle.

The robustness and generalizability of GPCN opens the door to several novel and expanded applications of PET that are currently impractical. For instance, in pediatric inflammatory or infectious diseases where the risk-benefit ratio is debatable, a CT-free approach would make PET a much more accessible diagnostic tool. Furthermore, its utility in performing multiple scans in a short timeframe is transformative for pharmacokinetic studies, early therapy response evaluation, and the rapid, low-dose assessment of new agents in drug development. Finally, for multi-modal imaging like PET/MRI where a CT scan is not natively available, a robust DL-based correction method like GPCN provides an elegant and accurate solution, further enhancing the synergy between functional and anatomical imaging modalities.

The superior performance and generalizability of GPCN stem from its unique, structure-aware architecture. Unlike conventional methods that treat the task as a generic image-to-image translation problem \cite{isola2017image,zhu2017unpaired}, GPCN's dual-domain approach effectively disentangles tracer emission from tissue attenuation. The multi-band contextual refinement (MBCR) isolates the stable, low-frequency anatomical information from variable, high-frequency tracer-specific textures \cite{gu2023mamba}. Concurrently, the frequency-aware spectral decoupling (FASD), explicitly guided by a spectral coordinate map, performs a principled, frequency-adaptive correction in the Fourier domain. This synergistic design, guided by explicit frequency coordinates, prevents anatomically implausible errors when faced with out-of-domain data and allows GPCN to learn a fundamental intrinsic mapping rather than simply overfitting to a specific data distribution.

Our work builds upon prior deep learning efforts for PET correction \cite{gong2018iterative,reader2020deep,shiri2023decentralized}, but represents a significant advancement over studies like Guo et al. \cite{guo2022using} in three key aspects. Firstly, we validate our framework on a pediatric dataset of unprecedented scale and heterogeneity. Secondly, our architecture synergizes three complementary concepts (multi-scale analysis, phase-amplitude separation, and positional awareness) to enhance robustness. Thirdly, we introduce a shift in training strategy—using a heterogeneous dataset to learn an intrinsically robust mapping, a more powerful approach than generalizing from homogeneous data alone \cite{sun2024artificial}.

We acknowledge several limitations. First, our supervised training fundamentally relies on CT-AC PET as the ground truth. However, CT-AC itself is not a flawless reference, particularly in pediatric imaging where respiratory motion mismatch between the fast CT and slow PET acquisitions can cause misalignment, and beam hardening or HU-to-$\mu$ conversion uncertainties can introduce artifacts \cite{cao2018region}. Our model may inadvertently inherit some of these inherent CT-based biases. Second, while our method demonstrates robust intra-institution, cross-protocol generalization, the data originates from a single center; multi-center external validation is necessary. Finally, although our objective metrics and downstream segmentation tasks demonstrate high physical fidelity, formal blinded reader studies by experienced nuclear medicine physicians are required to definitively confirm the diagnostic equivalency of the generated images, especially for detecting sub-centimeter lesions.

\section{Conclusion}
This study presents GPCN, a dual-domain framework for CT-free pediatric PET correction. By synergizing multi-scale contextual refinement with frequency-aware spectral decoupling, GPCN achieves state-of-the-art accuracy and robust generalization across diverse scanners and radiotracers.

The elimination of the CT component saves an average of 10.8 mSv per scan, offering a safer, ALARA-compliant pathway for pediatric molecular imaging. Future work will focus on prospective multi-center validation.

\bibliographystyle{ieeetr}
\bibliography{References}
\begin{IEEEbiography}[{\includegraphics[width=1in,height=1.25in,clip,keepaspectratio]{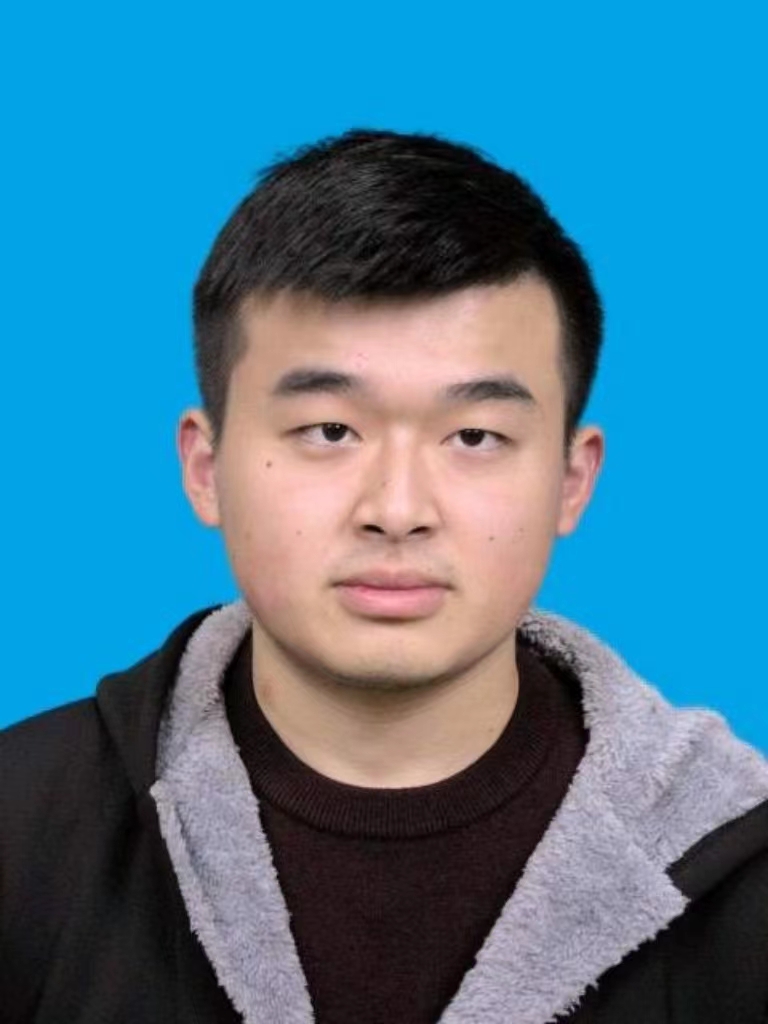}}]{Jia-Mian Wu} received the B.E. degree from Chongqing University of Science and Technology, Chongqing, China, in 2020, and the M.E. degree from Southwestern University of Finance and Economics (SWUFE), Chengdu, China, in 2024. He is currently a Ph.D. student with the School of Computing and Artificial Intelligence, Southwestern University of Finance and Economics. His current research interests include medical imaging via tensor modeling and deep learning.
\end{IEEEbiography}
\begin{IEEEbiography}[{\includegraphics[width=1in,height=1.25in,clip,keepaspectratio]{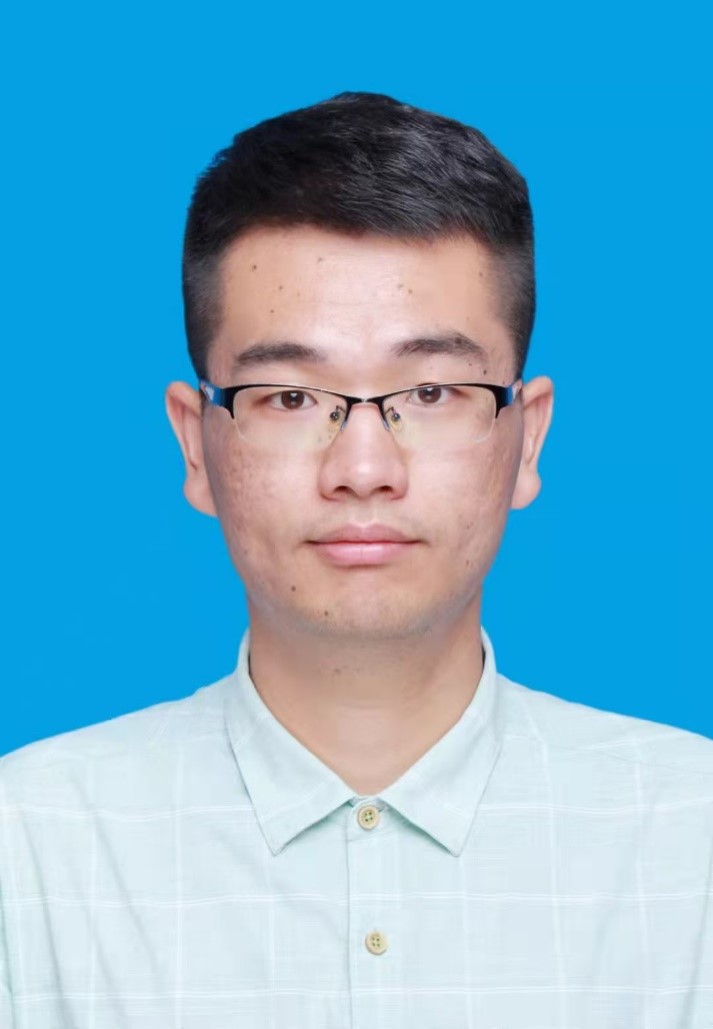}}]{Jun Liu} received his doctorate from Capital Medical University in 2022. He is currently an attending physician in the Department of Nuclear Medicine at Beijing Friendship Hospital, Capital Medical University. His research primarily focuses on pediatric nuclear medicine, with a specialized emphasis on the application of artificial intelligence (AI) within the field. 
\end{IEEEbiography}
\begin{IEEEbiography}[{\includegraphics[width=1in,height=1.25in,clip,keepaspectratio]{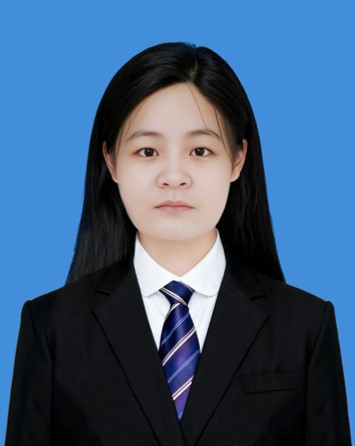}}]{Siqi Li} is a doctoral candidate in the Department of Nuclear Medicine at Beijing Friendship Hospital, Capital Medical University. She earned her bachelor’s degree from Hebei Medical University in 2021 and her master’s degree from Capital Medical University in 2024. After obtaining her master’s degree, she continued her studies toward a Ph.D. Her current research focuses on the application of multimodal molecular nuclear medicine imaging in the diagnosis and treatment of pediatric neuroblastoma.
\end{IEEEbiography}
\begin{IEEEbiography}[{\includegraphics[width=1in,height=1.25in,clip,keepaspectratio]{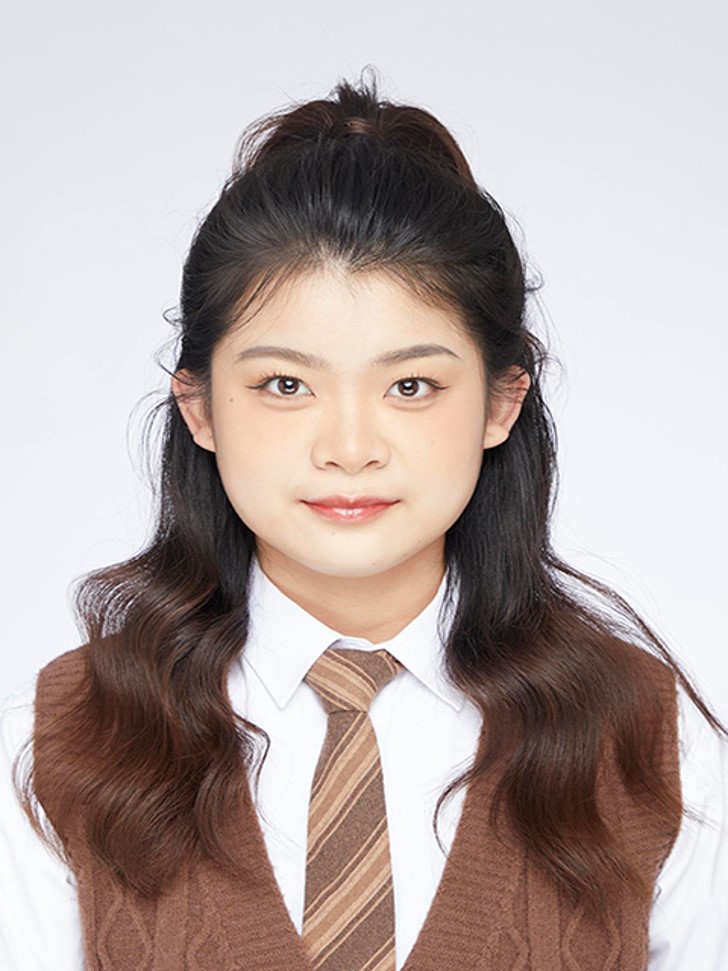}}]{Xiaoya Wang} received the B.M. degree from Chongqing Medical University, Chongqing, China, in 2017, and the M.M. degree from Capital Medical University, Beijing, China, in 2025. She is currently a Ph.D. student with the Department of Nuclear Medicine, Beijing Friendship Hospital, Capital Medical University. Her current research interests include the diagnostic and therapeutic value of multimodal imaging in neuroblastoma.
\end{IEEEbiography}
\begin{IEEEbiography}[{\includegraphics[width=1in,height=1.25in,clip,keepaspectratio]{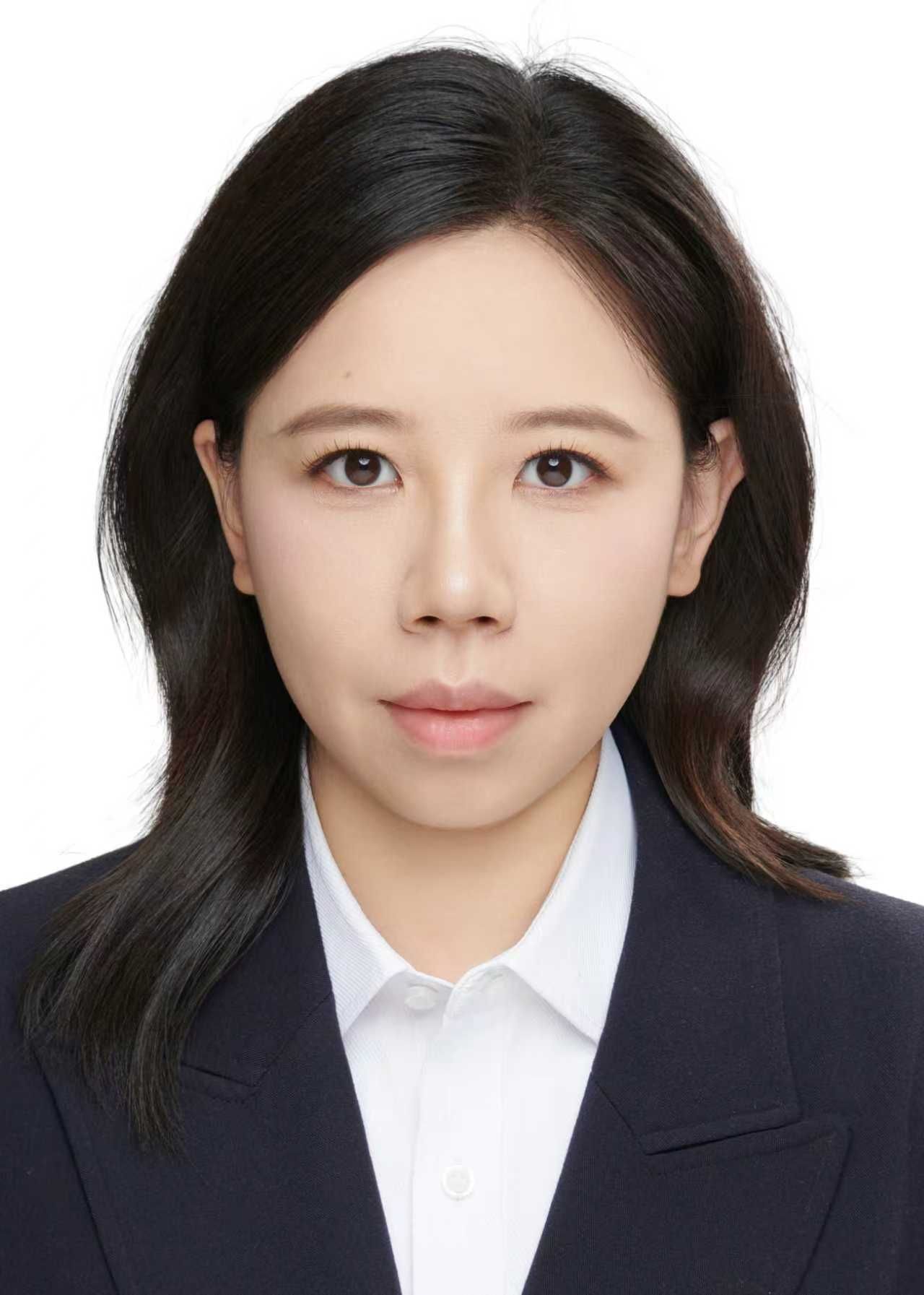}}]{Shibai Yin} received the Ph.D. degree in Control Science and Engineering from Chang'an University in 2013. She is currently a Professor with the School of Computing and Artificial Intelligence, Southwestern University of Finance and Economics. She was a visiting scholar at Memorial University of Newfoundland and the University of Alberta. Her research interests include computer vision, deep learning, and multimodal retrieval. She has published over 30 high-quality academic papers in top journals and conferences, such as Pattern Recognition, Neural Networks, and IEEE Transactions on Circuits and Systems for Video Technology. Her homepage is https://nicelab.swufe.edu.cn/info/1013/1167.htm.
\end{IEEEbiography}
\begin{IEEEbiography}[{\includegraphics[width=1in,height=1.25in,clip,keepaspectratio]{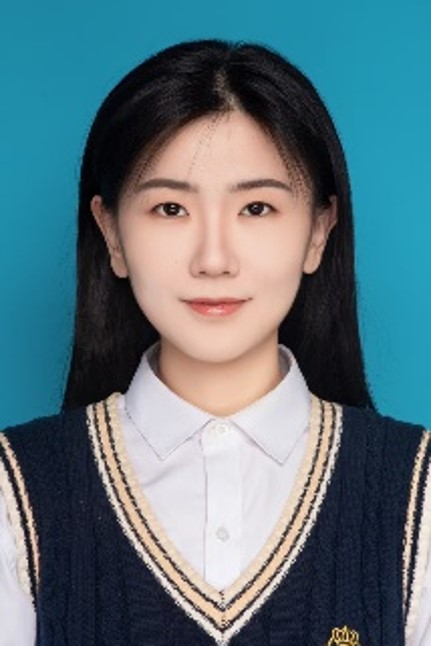}}]{Huanyu Luo} obtained his PhD from Capital Medical University in 2024. He is currently a physician in the Department of Radiology at Beijing Children's Hospital, Capital Medical University. His primary research focus lies in pediatric radiology.
\end{IEEEbiography}
\begin{IEEEbiography}[{\includegraphics[width=1in,height=1.25in,clip,keepaspectratio]{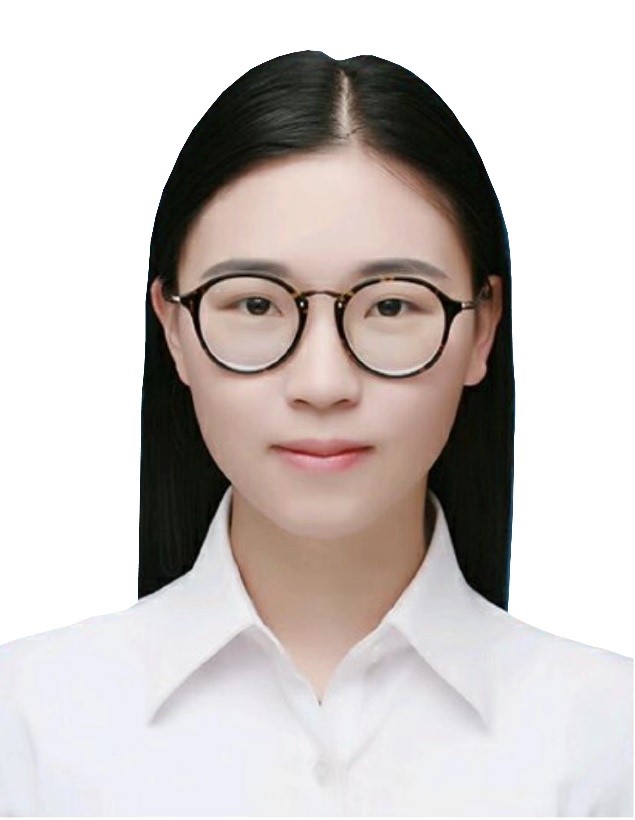}}]{Lingling Zheng} is a Ph.D. candidate in the Department of Nuclear Medicine at Capital Medical University, affiliated with the Department of Nuclear Medicine at Beijing Friendship Hospital. She received her Master’s degree in Diagnostic Radiology and Nuclear Medicine from Fudan University in 2021. Her research focuses on the development of antibody-based radiotracers for precision imaging and targeted therapy of tumors. 
\end{IEEEbiography}
\begin{IEEEbiography}[{\includegraphics[width=1in,height=1.25in,clip,keepaspectratio]{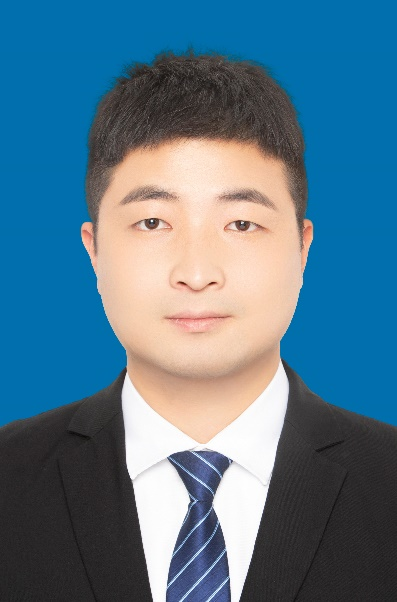}}]{Qiang Gao} received the B.S. degree in management and the Ph.D. degree in software engineering from the University of Electronic Science and Technology of China (UESTC). From 2019 to 2020, he was a joint Ph.D. student with Northwestern University, USA. He is currently a Professor and a Ph.D. Supervisor with the School of Computing and Artificial Intelligence, Southwestern University of Finance and Economics. He has authored or coauthored nearly 60 papers in top-tier journals and conferences, including NeurIPS, KDD, IJCAI, AAAI, WWW, SIGIR, IEEE TKDE, and IEEE TNNLS. His research interests include human mobility representation learning, spatio-temporal data processing and deep learning, and regional economics. He serves as a PC member or reviewer for several leading international conferences and journals, and served as a Session Chair for IJCAI 2024. His homepage is https://qianggao.xyz.
\end{IEEEbiography}
\begin{IEEEbiography}[{\includegraphics[width=1in,height=1.25in,clip,keepaspectratio]{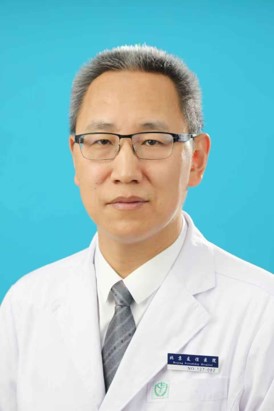}}]{Jigang Yang} obtained his doctorate from Capital Medical University in 2009. He currently serves as the Director of the Department of Nuclear Medicine at Beijing Friendship Hospital, Capital Medical University. Professor Yang is a distinguished expert in pediatric nuclear medicine and the application of artificial intelligence (AI) in nuclear medicine. To date, he has authored over 100 high-impact research articles and has been the principal investigator for numerous national, provincial, and ministerial-level research grants.
\end{IEEEbiography}
\begin{IEEEbiography}[{\includegraphics[width=1in,height=1.25in,clip,keepaspectratio]{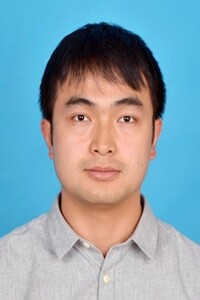}}]{Tai-Xiang Jiang} received the Ph.D. degree in mathematics from the University of Electronic Science and
Technology of China (UESTC), in 2019. He was a
co-training Ph.D. student in the University of Lisbon
supervised by Prof. Jose M. Bioucas-Dias from 2017
to 2018. He was a research assistant in the Hong
Kong Baptist University supported by Prof. Michael
K. Ng in 2019. He is currently a Professor with
the School of Computing and Artificial Intelligence,
Southwestern University of Finance and Economics.
His research interests include tensor decomposition and low-rank modeling for multi-dimensional image processing, especially on the low-level inverse problems for multi-dimensional images. His homepage is https://taixiangjiang.github.io/.
\end{IEEEbiography}
\vfill

\end{document}